\documentclass[fleqn,opre,nonblindrev]{informs_draft} 
\OneAndAHalfSpacedXI

\usepackage{eqndefns-left} 
\RequirePackage{tgtermes}
\RequirePackage{newtxtext}
\RequirePackage{newtxmath}
\RequirePackage{bm}
\RequirePackage{endnotes}

\usepackage{algorithm}
\usepackage{algpseudocode}
\usepackage{tikz}
\newcommand{\wTheta}{{\widetilde{\Theta}}}
\newcommand{\OSARkrr}{\text{OSAR}^\text{+}}
\newcommand{\OSARkrrPP}{\text{OSAR}^\text{++}}
\newcommand{\OSARkrrFD}{\text{OSAR}^\text{+FD}}
\newcommand{\OSARkrrPS}{\text{OSAR}^\text{+PS}}
\usepackage{blkarray, xcolor, subfiles, enumitem, subfigure}
\usepackage{multirow, thmtools, comment}
\usepackage{pgfplots}
\usepackage{pgfplotstable}
\usetikzlibrary{calc, positioning, intersections, arrows, automata, patterns}
\usepgfplotslibrary{patchplots}

\usepackage{mathtools, accents}
\mathtoolsset{showonlyrefs}

\newcommand{\lt}{\left}
\newcommand{\rt}{\right}
\newcommand{\Real}{\mathbb{R}}

\newcommand{\expec}{\mathrm{E}}
\newcommand{\prob}{\mathrm{P}}

\newcommand{\Tas}{{\Theta_{\text{a.s.}}}}

\makeatletter
\def\thickhline{%
  \noalign{\ifnum0=`}\fi\hrule \@height \thickarrayrulewidth \futurelet
   \reserved@a\@xthickhline}
\def\@xthickhline{\ifx\reserved@a\thickhline
               \vskip\doublerulesep
               \vskip-\thickarrayrulewidth
             \fi
      \ifnum0=`{\fi}}
\makeatother

\newlength{\thickarrayrulewidth}
\newcommand\norm[1]{\left\lVert#1\right\rVert}
\newcommand\inner[1]{\left\langle#1\right\rangle}

\newcommand{\hBFtheta}{\hat{\BFtheta}}

\makeatletter
\newcommand{\doublehat}[1]{%
\begingroup%
  \let\macc@kerna\z@%
  \let\macc@kernb\z@%
  \let\macc@nucleus\@empty%
  \hat{\raisebox{.3ex}{\vphantom{\ensuremath{#1}}}\smash{\hat{#1}}}%
\endgroup%
}
\makeatother

\makeatletter
\newcommand{\doubletilde}[1]{%
\begingroup%
  \let\macc@kerna\z@%
  \let\macc@kernb\z@%
  \let\macc@nucleus\@empty%
  \tilde{\raisebox{.3ex}{\vphantom{\ensuremath{#1}}}\smash{\tilde{#1}}}%
\endgroup%
}
\makeatother
\newcommand{\dKL}{\textsf{D}^\textup{KL}}

\usepackage{natbib}
\bibpunct[, ]{(}{)}{,}{a}{}{,}%
\def\bibfont{\fontsize{8}{9.5}\selectfont}%
\TheoremsNumberedThrough     
\ECRepeatTheorems

\EquationsNumberedThrough    

\gdef\AQ#1{}
\gdef\CQ#1{}
\begin{document}
	
\def\COPYRIGHTHOLDER{INFORMS}%
\def\COPYRIGHTYEAR{2017}%
\def\DOI{\fontsize{7.5}{9.5}\selectfont\sf\bfseries\noindent https://doi.org/10.1287/opre.2017.1714\CQ{Word count = 9740}}


        \RUNAUTHOR{Authors' names not included for peer review}
	\RUNTITLE{Optimizing Input Data Collection for Ranking and Selection}

\TITLE{Optimizing Input Data Collection for Ranking and Selection}



\ARTICLEAUTHORS{%
\AUTHOR{Eunhye Song}
\AFF{School of Industrial and Systems Engineering, Georgia Institute of Technology, GA 30332, \EMAIL{eunhye.song@isye.gatech.edu}}
\AUTHOR{Taeho Kim}
\AFF{HKUST Business School, The Hong Kong University of Science and Technology, Clear Water Bay, Hong Kong, \EMAIL{thk5594@gmail.com}} 
} 


	
	\ABSTRACT{ 
We study a ranking and selection (R\&S) problem when all solutions share common parametric Bayesian input models updated with the data collected from multiple independent data-generating sources. Our objective is to identify the best system by designing a sequential sampling algorithm that collects input and simulation data given a budget. We adopt the most probable best (MPB) as the estimator of the optimum and show that its posterior probability of optimality converges to one at an exponential rate as the sampling budget increases. 
Assuming that the input parameters belong to a finite set, we characterize the $\epsilon$-optimal static sampling ratios for input and simulation data that maximize the convergence rate. Using these ratios as guidance, we propose the optimal sampling algorithm for R\&S (OSAR) that achieves the $\epsilon$-optimal ratios almost surely in the limit. We further extend OSAR by adopting the kernel ridge regression to improve the simulation output mean prediction. This not only improves OSAR's finite-sample performance, but also lets us tackle the case where the input parameters lie in a continuous space with a strong consistency guarantee for finding the optimum. We numerically demonstrate that OSAR outperforms a state-of-the-art competitor.  

 }



\AREAOFREVIEW{Simulation.}


	
	%
	
\maketitle
	
\section{Introduction} \label{sec:intro}
\noindent In many applications, decision-makers build simulation models to optimize the designs of complex real-world stochastic systems. The input models that generate random variates fed into these simulators are typically estimated from data observed from the systems. Due to finiteness of the data, the estimated input models are subject to error. When simulations are run with the estimated input models, the estimation error is propagated to the simulation output, inflating the simulation output variance compared to when the exact system input distributions are known. Such inflation is known as \emph{input uncertainty}. 

This paper investigates a ranking and selection (R\&S) problem under input uncertainty where all competing solutions' are simulated with a collection of common estimated input models. In particular, we consider the case when the real-world system has multiple independent input-generating processes from which additional data can be acquired at a cost to improve the input models. The objective of the R\&S problem is to find the solution that has the optimal mean performance in the system. However,  input uncertainty makes it challenging to provide a high probability guarantee of selecting the optimum with finite input data~\citep{songnelsonhong2015}. Nevertheless, as more input data are collected and the estimated input models become closer to the input-generating distributions, one might expect the probability guarantee to improve.  

We study a unified framework to sequentially decide how to allocate a finite sampling budget between input data collection and simulation replication to discover the optimum. In practice, the parties that own the data and utilize it for decision making do not necessarily coincide. Then, the decision-maker must purchase the input data to improve their simulation models. Meanwhile, a complex simulation model tends to require heavy computing resources, which can be purchased by hours from a third-party provider such as Amazon Web Services. Hence, there is a trade-off between purchasing more data versus computing hours. Another example is when the decision maker has a large database of raw data, but it requires non-negligible computing effort to be transformed so that it can be fed to the distribution fitting software to improve the input models.  Then, the computation budget must be split between data processing and simulation.

To find the true optimum efficiently, it is important to carefully choose an estimator of the optimum and design an algorithm to allocate the sampling budget so that the estimator converges to the optimum fastest in some probabilistic sense. In this paper, we adopt parametric Bayesian input models by imposing a prior distribution on the parameter vector of the input models and updating it to the corresponding posterior given the input data. We utilize the \emph{most probable best} (MPB), the maximum a posteriori estimator (MAP) of the optimum given the posterior on the input parameter~\citep{kimkimsong}, to infer the optimum. We refer to the posterior probability that each solution is optimal as the solution's \emph{posterior preference}. We show that the MPB is not only a strongly consistent estimator of the optimum but also its posterior preference converges exponentially fast to one if each input source attains a nonzero fraction of the budget. In addition, the convergence rate can be characterized as a function of the sampling ratios of all input data sources. By maximizing the rate function, input data sampling can be optimized.

However, there are two challenges in solving the optimization problem.  
First, the constants in the optimization problem are functions of the simulation output means of the solutions, which are to be estimated with error. Second, without assuming a functional form of the simulation output mean of each solution in the parameters, the problem is difficult to solve when the prior distribution imposed on the input model parameters is continuous. 

To tackle these challenges, we first consider the case where the parameter vector of the input models has a finite support, and establish an estimator of the MPB by replacing the unknown means with simulation sample means. Then, we analyze the rate at which the estimated MPB's posterior preference converges to one while the MPB is correctly identified by the sample means. The rate is a function of the sampling ratios of the data sources as well as the simulation sampling ratios of all solution-parameter vector combinations. We formulate an asymptotic budget allocation problem to optimize the rate and show that it can be reformulated into a linear program (LP) that has the classical R\&S computing budget allocation problem as subproblems. 

Based on the analyses, we propose the \emph{optimal sampling algorithm for R\&S (OSAR)} that sequentially allocates the budget in batches, updates input models and simulation sample means, and resolves the LP to decide the next batch allocation. 
Two strengths of OSAR are i) the LP can be solved fast, and ii) the R\&S subproblems can be solved by any fixed budget algorithm known to achieve the optimal sampling ratios for the classical R\&S problem in the limit. We show that the cumulative sampling ratios almost surely converge to a set of static sampling ratios whose convergence rate can be made arbitrarily close to the optimal convergence rate. 

However, sequential algorithms based on the optimal static sampling ratios computed from plug-in estimates may suffer from performance degradation due to the noisy initial estimates~\citep{wu2019}. To overcome this, we sharpen each solution's mean estimates by adopting the kernel ridge regression (KRR) to pool the simulation outputs made at all simulated parameter vectors. We not only demonstrate that the KRR significantly improves the empirical performance of OSAR, but also show that the same asymptotic performance guarantee for OSAR can be achieved with the KRR in some cases.

The KRR also lets us extend OSAR to the case where the input parameter has a continuous support. While the simulations are still only run at a finite design points within the continuous support,~$\OSARkrrFD$ and~$\OSARkrrPS$  predict the means at any parameter by interpolating the samples means by KRR. These procedures are shown to achieve strong consistency of finding the optimum for the continuous-support problem as the data collection budget increases.

We benchmark OSAR against a state-of-the-art Bayesian optimization (BO) algorithm designed to solve a general simulation optimization problem under input uncertainty \citep{ungredda2022}. In both discrete- and continuous-support problems, OSAR exhibits superior performances in identifying the true optimum than the benchmark at a cheaper computational overhead. Moreover, the empirical sampling ratios allocated by OSAR become closer to the optimal static allocation ratios, whereas those from the benchmark fail to. 

\noindent \textbf{Related Literature.} Several approaches have been proposed to account for input uncertainty in simulation optimization in recent years; see~\cite{hesong2024tutorial} for a comprehensive review of the literature. Many of these concern the case where additional input data collection is infeasible. A common approach in this case is to adopt an appropriate statistical measure to reflect the risk of input uncertainty and focus on finding the ``optimal'' solution under the measure. For instance,~\cite{corlu2015} and \cite{pearce2017} take the mean function as a risk measure (risk-neutral) and define the optimum as the solution with the best performance averaged with respect to input uncertainty, while~\cite{xie2015} and \cite{helin2020} consider risk measures such as conditional value at risk or value at risk. Alternatively, one can assume that there is an uncertainty set of distributions containing the true input distribution, and find a solution whose worst-case performance within the set outperforms others'~\citep{gao2017robust,fan2020distributionally}.

Some recent work incorporates streaming input data into simulation optimization~\citep{liu2021bayesian, HeSongShangbag24:streamingSO, di2024}. In this problem context, the input models are continuously improved, however, the decision maker passively collects the incoming data rather than deciding which input data to collect. 

The work most closely related to ours is~\cite{ungredda2022}. In addition to adopting Bayesian parametric input models, they impose the Gaussian process (GP) prior on the simulation mean function; together, these two models let them evaluate the value of information of additional data/simulation sampling and utilize it to guide sequential sampling. \cite{wang2023} study the same R\&S problem as ours, however, their scope of input models are limited (the point estimator of the true parameter vector must be in a sample average form). Another distinction is that their sequential sampling rule is guided by a central limit theorem rather than the large deviations theory.

\noindent \textbf{Our contribution.} We are the \emph{first} to analyze the \emph{convergence rate} of the probability that an estimated optimum of R\&S converges to the optimum considering \emph{both} input uncertainty and simulation estimation error from the large deviations theory perspectives. We show that the effect of input uncertainty in identifying the optimum can be measured by the smallest Kullbeck-Leibler (KL) divergence between the true input distributions and the input models parameterized by the parameters at which a suboptimal solution becomes the optimum. Furthermore, our analysis reveals that the asymptotically optimal sampling ratios prescribe zero simulation ratios to some parameters, which helps OSAR concentrate the simulation effort to those parameters that matter the most for identifying the optimum. 

Adopting the KRR in OSAR enables it to tackle continuous-parameter problems while achieving strong consistency of finding the optimum. However, the design matrix for the KRR increases in size as new points in the parameter space are simulated, which poses computational and theoretical challenges. We tackle this issue by reducing the effective dimension of the KRR by projecting the prediction function onto a subspace spanned by a finite number of design points. The same technique can be applied to other R\&S problems under input uncertainty or contextual R\&S to address continuous-parameter problems.

\noindent {\bf Paper Organization.} Section~\ref{sec:prob.form} introduces Bayesian input models and defines the MPB. Focusing on a finite parameter set, Section~\ref{sec:MPB_asymp} discusses optimization of input data collection based on the large-deviation analysis of the posterior preference of the MPB. This framework is combined with the simulation budget allocation problem in Section~\ref{sec:simulation_error}. Section~\ref{sec:seq.sampling} introduces OSAR and discusses how its finite-sample performance can be improved by adopting the KRR. OSAR is further extended in Section~\ref{sec:continuous.case.extension} to tackle a continuous parameter space. Section~\ref{sec:num.exp} demonstrates OSAR's empirical performances against benchmarks. Proofs of all theorems are presented in the Electronic Companions.

\section{Problem Formulation}\label{sec:prob.form}

Suppose there exist $L > 1$ independent input data sources in the system. Let $f_\ell$ and $\theta^\ell$ represent the distribution function and the parameter vector of the $\ell$-th input. We assume that $f_1,f_2,\ldots,f_L$ are known, however, the true parameter vector, $\BFtheta_0 = (\theta_0^1, \theta_0^2, \ldots, \theta_0^L)$, is unknown and must be estimated from input data observed from the $L$ sources in the system.

The R\&S problem we concern has $k$ solutions in comparison, all of which are simulated with the same joint input model. For each $1 \leq i\leq k$ and parameter vector $\BFtheta$, $Y_i(\BFtheta)$ represents the simulation output of Solution $i$ when inputs are generated from the input model parameterized by $\BFtheta$. Let  $\eta_i(\BFtheta) := \expec[Y_i(\BFtheta)|\BFtheta]$.
The ultimate goal of the problem is to find the optimum, 
\begin{oldequation}\label{eq:true.opt}
    i_0 := \argmin\nolimits_{1 \leq i\leq k} \eta_i(\BFtheta_0).
\end{oldequation}

We consider the problem setting when we can collect additional input data from any of the $L$ sources at a cost. 
To model the uncertainty about $\BFtheta_0$ given the current input data set, we adopt the Bayesian estimator, $\BFtheta = (\theta^1, \theta^2, \ldots, \theta^L)$. Let $\mathcal{Z}_{m_\ell}^\ell := \{Z_1^\ell, Z_2^\ell, \ldots, Z_{m_\ell}^\ell\}$ be the size-$m_\ell$ independent and identically distributed (i.i.d.) observations from the $\ell$-th input distribution, and $\BFm := (m_1,m_2,\ldots,m_L)$. 
The likelihood process for $\mathcal{Z}_{\BFm} := \cup_{\ell = 1}^L \mathcal{Z}^\ell_{m_\ell}$ is $f_{\bm{\theta}}(\mathcal{Z}_{\BFm}) := \prod\nolimits_{\ell = 1}^L f^\ell_{\theta^\ell}(\mathcal{Z}^\ell_{m_\ell})$.
Furthermore, we denote the prior {distribution} of $\theta^\ell$ and its support by $\BFpi^\ell_0(\theta^\ell)$ and  $\Theta^\ell$, respectively. Then, for each $1 \leq \ell \leq L$, the posterior distribution of $\theta^\ell$ can be written as
    $\BFpi^\ell_{m_\ell}(\theta^\ell) = \frac{\BFpi_0^\ell(\theta^\ell)f^\ell_{\theta^\ell}(\mathcal{Z}_{m_\ell}^\ell)}{\int_{\Theta^\ell} \BFpi_0^\ell(\theta_1^\ell)f^\ell_{\theta_1^\ell}(\mathcal{Z}_{m_\ell}^\ell)d\theta_1},$
and the joint posterior distribution of $\BFtheta\in\Theta:=\bigtimes_{\ell = 1}^L \Theta^\ell$ is $\BFpi_\BFm = \prod_{\ell=1}^L \BFpi^\ell_{m_\ell}(\theta^\ell)$, where $\bigtimes$ is the Cartesian product.
We assume that the $\ell$th data source has associated constant cost $c_\ell$ for collecting one additional observation.

To analyze the effect of input uncertainty on the R\&S problem, we adopt the notion, \emph{posterior preference}, first proposed by~\cite{kimkimsong2022paper}. We define the \emph{favorable set} of Solution $i$ as $\Theta_i = \lt\{\BFtheta: \eta_i(\BFtheta) \leq \min_{j \neq i} \eta_j(\BFtheta) \rt\}$ and its complement as the \emph{adversarial set} of Solution $i$. The favorable and adversarial sets for each solution are deterministic regardless of $\BFpi_\BFm$.
The posterior preference of Solution~$i$ is  the probability assigned to its favorable set by $\BFpi_\BFm, \prob_{\BFpi_\BFm}(\Theta_i),$
which quantifies \textit{how likely Solution~$i$ is optimal given  $\BFpi_\BFm$}. Under the following assumption, the posterior preferences of all solutions sum to one.
\begin{assumption}\label{asmp:uniq.cond.opt}
The optimum, $i_0$, is unique. Moreover, $\argmin_{1 \leq i\leq k} \eta_i(\BFtheta)$ is unique almost everywhere under $\BFpi_0$.
\end{assumption}

Because $\prob_{\BFpi_\BFm}(\Theta_i)$ depends on the $\BFm$ observed data, its value changes when additional input data are collected. In Section~\ref{sec:MPB_asymp}, we show that $\prob_{\BFpi_\BFm}(\Theta_{i_0})$ converges to one almost surely with an exponential rate as $m$ grows under  mild assumptions.  This result combined with  Assumption~\ref{asmp:uniq.cond.opt} implies that $i_0$ has the largest posterior preference for sufficiently large $m$. Hence, it is sensible to adopt the solution with the largest posterior preference as an estimator for $i_0$; \cite{kimkimsong} refer to this estimator as the \emph{most probable best} (MPB). Mathematically, the MPB given $\BFpi_\BFm$ is defined as
\begin{oldequation}\label{eq:MPB_def}
    i^*(\BFpi_\BFm):= \argmax\nolimits_{1 \leq j \leq k} \prob_{\BFpi_\BFm}(\Theta_j). 
\end{oldequation}
In Section~\ref{sec:MPB_asymp}, we show that  $i^*(\BFpi_\BFm)$ is indeed a strongly consistent estimator for $i_0$ under some assumptions.

In addition to learning $\BFtheta_0$, since $\eta_i(\BFtheta)$ has no known analytical expression in general, it must be estimated for each $(i, \BFtheta)$ by simulation to find $i^*(\BFpi_\BFm)$. Thus, the decision maker has two types of sampling choices to make: i) how much additional data to collect from each of $L$ input sources?; ii) which $(i, \BFtheta)$ to simulate so that $i^*(\BFpi_\BFm)$ can be correctly estimated? We make the following simplifying assumption about the relative sampling costs of input and simulation data.

\begin{assumption}\label{asmp:cost}
    The total sampling budget, $T$, is allocated to input data collection and simulation. For any $(i, \BFtheta)$, the cost of simulation is one. For $1\leq \ell\leq L$, the cost of the one additional observation from the $\ell$-th input data source relative to the simulation cost is $c_\ell$. 
\end{assumption}

Under Assumption~\ref{asmp:cost}, we propose a sequential algorithm to allocate $T$, where both input and simulation data collection decisions are made  considering their relative costs as well as their significance in identifying $i_0$. For this, it is crucial to analyze how fast the simulation estimator of $i^*(\BFpi_\BFm)$ approaches $i_0$  as the $T$ increases. 
For finite $T$, this problem is difficult even for a classical R\&S problem without input uncertainty. Instead, we first analyze the asymptotic properties of $i^*(\BFpi_\BFm)$ when $T\to\infty$ assuming each $\eta_i(\BFtheta)$ is known in Section~\ref{sec:MPB_asymp}, and utilize this to propose an input and simulation budget allocation framework in Section~\ref{sec:simulation_error}.

\section{Asymptotic Properties of the Most Probable Best}\label{sec:MPB_asymp}

We begin this section with Assumptions~\ref{asmp:continuity_y} and~\ref{asmp:input_LDR} below, which summarize the additional conditions we impose on $f_{\theta^\ell}$ and $\BFpi_0(\theta^\ell)$ to study the asymptotic behavior of $i^*(\BFpi_\BFm)$.

\begin{assumption}\label{asmp:continuity_y}
For each $i$, the mean function $\eta_i(\BFtheta)$ is continuous in $\BFtheta$.
\end{assumption} 
The definition of the favorable set and Assumption~\ref{asmp:continuity_y} together imply that  $\Theta_i$ is a closed set, whereas $\Theta_i^c$ is open for all $i$.
\begin{assumption}\label{asmp:input_LDR}
 For each $1\leq \ell \leq L$, $\mathcal{F}_{\Theta^\ell}$ and $\mathcal{F}_\ell$ are $\sigma$-algebras of prior $\BFpi_0^\ell$ and $Z \sim f^\ell_{\theta_0^\ell}$, respectively, and the following holds:
 \begin{itemize}
     \item[(a)] $\Theta^\ell$ is compact;
     \item[(b)] For all $m_\ell$, the product likelihood, $f^\ell_{\theta^\ell}(\mathcal{Z}_{m_\ell}^\ell)$, is $\mathcal{F}_{\Theta^\ell} \times \mathcal{F}_\ell^{m_\ell}$-measurable, where $\mathcal{F}_\ell^{m_\ell}$ is a product $\sigma$-algebra with respect to $\mathcal{Z}_{m_\ell}^\ell$;
     \item[(c)]\label{asmp:GC_class} The class of functions, $\mathcal{H} : = \lt\{\log({f^\ell_{\theta^\ell}}/{f^\ell_{\theta^\ell_0}}) : \theta^\ell \in \Theta^\ell\rt\}$, is $\mathrm{P}_{\theta^\ell_0}$-Glivenko-Cantelli;
    \item[(d)]\label{asmp:continuity} $\dKL(\theta^\ell_0||\theta^\ell)$, is continuous in both $\theta^\ell$ and $\theta_0^\ell$, and $0 < \inf_{\theta^\ell \in \Theta^\ell} \BFpi^\ell_0(\theta^\ell)\leq \sup_{\theta^\ell \in \Theta^\ell} \BFpi^\ell_0(\theta) < \infty$, where 
 $\dKL(\theta^\ell_0||\theta^\ell) := \expec_{Z \sim f^\ell_{\theta^\ell_0}}[\log ({f^\ell_{\theta^\ell_0}(Z)}/{f^\ell_{\theta^\ell}(Z)})]$ is the KL divergence of $f^\ell_{\theta^\ell_0}$ from $f^\ell_{\theta^\ell}.$
 \end{itemize}
\end{assumption}

Given probability measure $\prob$, functional class $\mathcal{H}$ is said to be a \emph{$\prob$-Glivenko-Cantelli} (GC) class if $ \norm{\prob_m - \prob}_{\mathcal{H}} : = \sup_{f \in \mathcal{H}}\lt|\prob_m f - \prob f\rt| \rightarrow 0, \prob\text{-a.s}$ where $\prob_m$ is an empirical measure consisting of $m$ i.i.d observations from $\prob$ and $\prob f = \int f d\prob$. In other words, the strong law of large numbers holds uniformly over the GC class. Proposition~\ref{prop:exp.fam} below presents a sufficient condition for Assumption~\ref{asmp:input_LDR}(c) to hold when the input model belongs to an exponential family, which includes a wide range of parametric distributions such as exponential, Gamma, Gaussian, and Poisson.

\begin{proposition}\label{prop:exp.fam}
For each $1\leq \ell \leq L$, let us consider an exponential family whose density $f^\ell_{\theta^\ell}(\BFx)$ is given as $f^\ell_{\theta^\ell}(\BFx) = h^\ell(\BFx)\exp\lt(\inner{\theta^\ell, S^\ell(\BFx)} - A^\ell(\theta^\ell)\rt)$. Further, assume $\Theta^\ell$ is compact and $A^\ell(\theta^\ell)$ is twice differentiable at $\theta_0^\ell$ for all $1\leq \ell \leq L$. Then, Assumption~\ref{asmp:GC_class}(c) holds.
\end{proposition}

Assumption~\ref{asmp:data_size} is made to investigate asymptotic consistency of $i^*(\BFpi_\BFm)$ as $T$ increases.

\begin{assumption}\label{asmp:data_size}
For each $\ell$, $m_\ell\rightarrow \infty$ as $T\to\infty$, and $\lim_{T \rightarrow \infty} \frac{c_\ell m_\ell}{T} = \beta_\ell$ for some $\beta_\ell > 0$ and $\sum_{\ell=1}^L \beta_\ell \leq 1$. 
\end{assumption}

In words, $\beta_\ell$ is the asymptotic fraction of the sampling budget spent on the $\ell$th input source. Since we assume $\eta_i(\BFtheta)$ is known in this section, one may regard $\sum_{\ell=1}^L \beta_\ell = 1$ in the remainder of the section. However, the results presented in this section hold even if $\sum_{\ell=1}^L \beta_\ell < 1$. 

We begin by stating the strong consistency of $i^*(\BFpi_\BFm)$ and the convergence rate of its posterior preference.

\begin{theorem}\label{thm:LDR_input}
 Suppose Assumptions~\ref{asmp:uniq.cond.opt}--\ref{asmp:data_size} hold. Then, with probability 1, (a) $i^*(\BFpi_\BFm)$ converges to $i_0$ as $T\to\infty$; (b) we have
 \begin{oldequation}\label{eq:ldr_mpb}
    \lim_{T \rightarrow \infty} -\frac{1}{T}\log (1- \prob_{\BFpi_\BFm}(\Theta_{i^*(\BFpi_\BFm)})) = \inf_{\BFtheta \in \Theta^c_{i_0}}\sum\nolimits_{\ell = 1}^L \frac{\beta_\ell}{c_\ell} \dKL(\theta^\ell_0||\theta^\ell). 
\end{oldequation}
\end{theorem}

Theorem~\ref{thm:LDR_input}(a) establishes that $i^*(\BFpi_\BFm)$ is a strongly consistent estimator of $i_0$ while its posterior preference $\prob_{\BFpi_\BFm}(\Theta_{i^*(\BFpi_\BFm)})$ converges to one as $T\to\infty$. 
Thus, it is sensible to allocate $T$ so that $\prob_{\BFpi_\BFm}(\Theta_{i^*(\BFpi_\BFm)})$ converges at the fastest rate to quickly identify $i_0$.
Theorem~\ref{thm:LDR_input}(b) states that $\prob_{\BFpi_\BFm}(\Theta_{i^*(\BFpi_\BFm)})$ converges to one exponentially in $T$ at the rate characterized by the smallest limiting weighted sum of the KL divergences of $f^\ell_{\theta^\ell_0}$ from  $f^\ell_{\theta^\ell}$ for $\ell=1,2,\ldots,L$, within $\BFtheta\in\Theta_{i_0}^c$. This measures how ``far'' $\BFtheta_0$ is from the set of $\BFtheta$s that make $i_0$ suboptimal in the input distribution space---this is a loose statement as the KL divergence is not a distance metric. The $\ell$th weight, $\beta_\ell/c_\ell$, factors in how much of the total sampling effort is to be spent on the $\ell$th input data relative to its sampling cost. 

The limiting $\BFtheta$ where the infimum of~\eqref{eq:ldr_mpb} is attained at can be interpreted as the point at which the posterior probability assigned to the adversarial set of $i_0$ concentrates. Such  $\BFtheta$ depends on 
$\BFbeta := \{\beta_\ell\}_{\ell=1}^L$, which determines how $\BFpi_\BFm(\BFtheta)$ concentrates at $\BFtheta_0$ as $T$ increases.  Thus, by choosing $\BFbeta$ that maximizes the infimum, one can accelerate the convergence of $\prob_{\BFpi_\BFm}(\Theta_{i^*(\BFpi_\BFm)})$. This optimization problem can be formulated as
\begin{oldequation}\label{opt:input_data_without_sim_error}
    \max_{\BFbeta}\inf_{\BFtheta \in \Theta^c_{i_0}}\sum\nolimits_{\ell = 1}^L \frac{\beta_\ell}{c_\ell} \dKL(\theta^\ell_0||\theta^\ell) \;\; \text{subject to} \;\; \BFbeta \geq \mathbf{0}_L, \mathbf{1}^\top_L\BFbeta = 1,
\end{oldequation}
where $\mathbf{0}_L$ and $\mathbf{1}_L$ are $L$-dimensional vectors with zeroes and ones, respectively, and $\BFbeta \geq \mathbf{0}_L$ represents elementwise inequalities. 

If $\BFtheta$ has a continuous support, then~\eqref{opt:input_data_without_sim_error} is difficult to solve in general even if we know $\Theta^c_{i_0}$ precisely without imposing a structural assumption (e.g., linearity or convexity in $\BFtheta$) 
on $\{\eta_i(\BFtheta)\}_{i=1}^k$. Instead, we first consider finite $\Theta$ as described in Assumption~\ref{asmp:supp_assumption}, and extend it to the continuous case in Section~\ref{sec:continuous.case.extension}.
\begin{assumption}\label{asmp:supp_assumption} 
We have $|\Theta| = B < \infty$.
\end{assumption}
Under Assumption~\ref{asmp:supp_assumption}, the posterior preference of  $i$ simplifies to
    $\prob_{\BFpi_\BFm}(\Theta_{i}) = \sum\nolimits_{\BFtheta_b \in \Theta}  \BFpi_{\BFm}(\BFtheta_b)1\{i = i^b\}$,
where $i^b  = \argmin_{1\leq h \leq k} \eta_h(\BFtheta_b)$ is the conditional optimum at $\BFtheta_b$. Additionally, Problem~\eqref{opt:input_data_without_sim_error} can be converted to the following LP:
\begin{oldequation} \label{opt:LP}
    \begin{aligned}
    &\max\nolimits_{D, \BFbeta} \quad D \\ &\text{ subject to } \quad D \leq \BFd_j^\top\BFbeta, 1\leq j\leq J, \;\;\mathbf{1}_L^\top\BFbeta = 1, \;\; \BFbeta \geq \mathbf{0}_L,
    \end{aligned}
\end{oldequation}
where $\Theta_{i_0}^c = \{\bm{\theta}_1, \bm{\theta}_2, \ldots, \bm{\theta}_J\}$ and
$\BFd_j:=(\dKL(\theta_0^\ell||\theta_j^\ell)/c_\ell)_{1\leq \ell \leq L}$ for each $\bm{\theta}_j \in \Theta_{i_0}^c$. 

Once $\BFtheta_0$ and $\Theta_{i_0}^c$ are given,~\eqref{opt:LP} can be solved easily by an LP solver. Clearly, both quantities are unknown in our problem.
A natural choice for a point estimator of $\BFtheta_0$ is $\hBFtheta = \argmax_{\BFtheta \in \Theta}\BFpi_\BFm(\BFtheta)$,  the MAP of $\BFtheta_0$. Estimating $\Theta_{i_0}^c$ is more challenging because we not only need to estimate $i_0$, but also identify all $\BFtheta\in\Theta$ at which $i_0$ is suboptimal. Although $i_0$ and $\Theta_{i_0}^c$  can be estimated by $i^*(\BFpi_\BFm)$ and $\Theta_{i^*(\BFpi_\BFm)}^c$, respectively, they  require  estimating $\eta_i(\BFtheta)$ at all $(i,\BFtheta)$ by running simulations.
In Section~\ref{sec:simulation_error}, we introduce simulation estimators for the unknown quantities and incorporate their estimation errors in the rate analysis.

\section{Optimal Learning of the Most Probable Best}\label{sec:simulation_error}

In this section, we propose a framework to optimize the allocation of  $T$ to both input and simulation sampling so that the estimated MPB converges to $i_0$ at the fastest rate considering both simulation error and posterior preference as $T$ increases. We impose Assumption~\ref{asmp:supp_assumption} throughout this section (finite $\Theta$).

Upon spending budget $T$, let $N_{i, T}(\BFtheta_b)$ denote the number of simulation replications allocated to $(i, \BFtheta_b)$ and $Y_{ir}(\BFtheta_b)$ be the $r$th simulation output. We estimate $\eta_i(\BFtheta_b)$ via its sample mean $\mu_{i, T}(\BFtheta_b) := \sum_{r=1}^{N_{i, T}(\BFtheta_b)} Y_{ir}(\BFtheta_b)/N_{i, T}(\BFtheta_b).$
To estimate the MPB, we first evaluate the sample posterior preference of each solution. 
 From the sample version of $i^b$, $i_T^b:= \argmin_{1\leq j\leq k}\mu_{j, T}(\BFtheta_b)$, $\Theta_j$ can be estimated by computing $\Theta_{j, T} := \{\BFtheta_b\in  \Theta| j = i_T^b\}$  for each $1\leq j \leq k$.  Then, the corresponding MPB estimator is $i_T^* := \argmax_{1\leq j\leq k}P_{\BFpi_\BFm}(\Theta_{j, T})$. Although suppressed for notational convenience, $i_T^*$ depends on $\BFpi_\BFm$.

In~\eqref{opt:input_data_without_sim_error} and~\eqref{opt:LP}, the objective is to allocate $\BFbeta$ so that the MPB's posterior preference converges to one as fast as possible. 
With  simulation error, we set the objective to maximizing the posterior probability assigned to the estimated favorable set of $i^*_T$ while correctly estimating $i^*_T$ to be $i^*(\BFpi_\BFm)$:
\begin{oldequation}\label{opt:input_data_collection}
    \begin{aligned}
    &\max\nolimits_{\mathbf{N}, \BFm}\; &\expec[\prob_{\BFpi_\BFm}(\Theta_{i_T^*, T})1\{i_T^* = i^*(\BFpi_\BFm)\}|\BFpi_\BFm] \;\; \\
    &\text{subject to} \;\; &\sum\nolimits_{i=1}^k \sum\nolimits_{b=1}^B N_{i, T}(\BFtheta_b) + \BFc^\top \BFm = T
    \end{aligned}
\end{oldequation}
The expectation in the objective function is with respect to the simulation error given $\BFpi_\BFm$ and $\mathbf{N} := \{N_{i,T}(\BFtheta_b)\}_{1\leq b\leq B, 1\leq i \leq k}$ while the indicator function accounts for the posterior preference of $i^*_T$, only if it correctly estimates $i^*(\BFpi_\BFm).$

Solving~\eqref{opt:input_data_collection} for finite $T$ is challenging as the analytical expression for the objective function is difficult to derive. Instead, we consider the asymptotic version of~\eqref{opt:input_data_collection} as $T$ increases under the following assumption.
\begin{assumption}\label{asmp:sim_size}
For each $(i, \BFtheta_b)$, $N_{i, T}(\BFtheta_b)\rightarrow \infty$ as $T\to\infty$. Moreover, $\lim_{T \rightarrow \infty} \frac{N_{i, T}(\BFtheta_b)}{T} = \alpha_i(\BFtheta_b)$ for some $\alpha_i(\BFtheta_b) \geq 0$ and $\sum_{i=1}^k\sum_{b=1}^B \alpha_i(\BFtheta_b)\leq 1$. 
\end{assumption}
Notice that $\alpha_i(\BFtheta_b)= 0$ is allowed under  Assumption~\ref{asmp:sim_size}, which means that we sample $(i, \theta_b)$ infinitely often but in a sublinear rate in $T$. As $T\to\infty$, the constraint of Program~\eqref{opt:input_data_collection} can be modified to $\sum_{i=1}^k\sum_{b=1}^B \alpha_i(\BFtheta_b) + \mathbf{1}_L^\top \BFbeta = 1$. Our goal is not only to show that the objective function of~\eqref{opt:input_data_collection} converges to~1 as $T\rightarrow \infty$, but also to maximize its convergence rate by optimizing $\{\alpha_i(\BFtheta_b)\}_{1\leq i \leq k, 1\leq b\leq B}$ and $\BFbeta$.
To set up the discussion, we define rate function $G_i(\BFtheta_b)$ for $i \neq i^b$ at each $\BFtheta_b$
\begin{oldequation}\label{eq:def_G}
    G_i(\BFtheta_b):=\lim_{T \rightarrow \infty}-\frac{1}{T}\log\prob\lt(\mu_{i, T}(\BFtheta_b) \leq \mu_{i^b, T}(\BFtheta_b)\rt),
\end{oldequation}
where the probability $\prob$ is taken with respect to the simulation error.
In words, $G_i(\BFtheta_b)$ is the convergence rate of the incorrect pairwise comparison between $i$ and $i^b$ and is well established in the literature since first discussed by~\cite{glynn2004large}. If $\alpha_i(\BFtheta_b) = 0$, we have $G_i(\BFtheta_b) = 0$, which means that $\prob\lt(\mu_{i, T}(\BFtheta_b) \leq \mu_{i^b, T}(\BFtheta_b)\rt)$ decays more slowly than at an exponential rate.
Theorem~\ref{thm:LDR_first} formally establishes strong consistency of $i^*_T$ and the rate at which the objective function of~\eqref{opt:input_data_collection} converges to one.

\begin{theorem}\label{thm:LDR_first}
    Under Assumptions~\ref{asmp:uniq.cond.opt}--~\ref{asmp:sim_size}, $i^*_T$ converges to $i_0$ almost surely. Moreover, we have 
\begin{oldequation}\label{eq:LDR_second}
    \begin{aligned}
        & \lim_{T \rightarrow \infty} -\frac{1}{T}\log(1 - \expec[\prob_{\BFpi_\BFm}(\Theta_{i_T^*, T})1\{i_T^* = i^*(\BFpi_\BFm)\}|\BFpi_\BFm]) \\
        & = \min_{\BFtheta_b \in \Theta} \left\{\sum\nolimits_{\ell=1}^L \frac{\beta_\ell}{c_\ell}\dKL(\theta_0^\ell||\theta_b^\ell)+ \min\nolimits_{i\neq i_0}G_i(\BFtheta_b) 1\{\BFtheta_b \in \Theta_{i_0}\}\right\} \;\; \text{almost surely.} 
    \end{aligned}
\end{oldequation}
\end{theorem}

To appreciate~\eqref{eq:LDR_second}, compare it with the rate function~\eqref{eq:ldr_mpb} that does not account for the simulation error.
First, notice that the outer minimum (infimum) is taken within $\Theta$, not just $\Theta_{i_0}^c$. For $\BFtheta_b\in\Theta_{i_0}^c$,~\eqref{eq:LDR_second} still compares the weighted sum of the KL divergences as $1\{\BFtheta_b \in \Theta_{i_0}\}=0$. For $\BFtheta_b\in\Theta_{i_0}$, however, the rate at which a false selection is made at $\BFtheta_b$, $\min\nolimits_{i\neq i_0}G_i(\BFtheta_b)$, is added to account for the simulation error. This is because as $\BFm$ increases, most of the probability mass assigned by $\BFpi_\BFm$ concentrates within $\Theta_{i_0}$ and making   correct selections at $\BFtheta_b\in\Theta_{i_0}$ is important to ensure fast convergence of 
$\prob_{\BFpi_\BFm}(\Theta_{i_T^*, T})$ to one, whereas  a correct selection at $\BFtheta_b\in\Theta_{i_0}^c$ is not as critical.

Thanks to Theorem~\ref{thm:LDR_first}, an asymptotic version of~\eqref{opt:input_data_collection} can be written as
\begin{oldequation}\label{opt:reformulation}
    \begin{aligned}
    &\max_{\{\alpha_i(\BFtheta_b)\}_{1\leq i\leq k, 1\leq b\leq B}, \BFbeta} \quad\quad \quad \min_{\BFtheta_b \in \Theta} \left\{\sum\nolimits_{\ell=1}^L \frac{\beta_\ell}{c_\ell} \dKL(\theta_0^\ell||\theta_b^\ell)+ \min_{i\neq i_0}G_i(\BFtheta_b) 1\{\BFtheta_b \in \Theta_{i_0}\}\right\} \\
    & \text{subject to } \quad \sum\nolimits_{i=1}^k\sum\nolimits_{b=1}^B \alpha_i(\BFtheta_b) + \mathbf{1}_L^\top \BFbeta = 1, \alpha_i(\BFtheta_b) \geq 0, \BFbeta \geq \mathbf{1}_L.
    \end{aligned}
\end{oldequation}

Let $\alpha(\BFtheta_b) = \sum_{i=1}^k \alpha_i(\BFtheta_b)$ and $G^*(\BFtheta_b) = \max_{\sum_{i=1}^k \alpha_i(\BFtheta_b) = 1} \min_{i \neq i^b} G_i(\BFtheta_b)$. The latter can be interpreted as the optimal convergence rate of the static allocation scheme of the classical R\&S problem for fixed $\BFtheta_b$~\citep{glynn2004large}. Notice that $G^*(\BFtheta_b)$ is obtained by assuming 100\% of the computation budget goes to find the best at $\BFtheta_b$. Observe the following:
\begin{oldequation}
    \begin{aligned}
    G_i(\BFtheta_b) = \lim_{T\rightarrow \infty}\frac{\sum_{i=1}^k N_{i, T}(\BFtheta_b)}{T}\frac{-1}{\sum_{i=1}^k N_{i, T}(\BFtheta_b)}\log\prob\lt(\mu_{i, T}(\BFtheta_b) \leq \mu_{i^b, T}(\BFtheta_b)\rt)
    = \alpha(\BFtheta_b) G^*_i(\BFtheta_b).
    \end{aligned}
 \end{oldequation}
Therefore, given some fixed $\alpha(\BFtheta_b)$, the maximum of $\min_{i\neq i^b} G_i(\BFtheta_b)$ can be computed as the product of $G_i^*(\BFtheta_b)$ and $\alpha(\BFtheta_b)$. Since~\eqref{opt:reformulation} is the maximization problem with respect to $\BFalpha$, $\min_{i\neq i_0} G_i(\BFtheta_b) = \alpha(\BFtheta_b)G^*(\BFtheta_b)$ holds at the optimal $\BFalpha$. Consequently, we can rewrite~\eqref{opt:reformulation} as
\begin{oldequation}\label{opt:reformulation2}
    \begin{aligned}
    &\max_{\BFalpha, \BFbeta} \quad\quad \quad \min_{\BFtheta_b \in \Theta} \left\{\sum\nolimits_{\ell=1}^L \frac{\beta_\ell}{c_\ell} \dKL(\theta_0^\ell||\theta_b^\ell) + \alpha(\BFtheta_b)G^*(\BFtheta_b) 1\{\BFtheta_b \in \Theta_{i_0}\}\right\}\\
    & \text{subject to } \quad \mathbf{1}_B^\top \BFalpha + \mathbf{1}_L^\top \BFbeta = 1, \BFalpha \geq \mathbf{0}_B, \BFbeta \geq \mathbf{0}_L,
    \end{aligned}
\end{oldequation}
where $\BFalpha = (\alpha(\BFtheta_b))_{1\leq b\leq B} \in \Real^B$. Once an optimal solution $(\BFalpha^*, \BFbeta^*)$ is determined by solving~\eqref{opt:reformulation2}, then \emph{any} rate-optimal classical R\&S algorithm can be applied to solve the subproblem at each $\BFtheta_b$ given $\BFalpha^*$. Furthermore,~\eqref{opt:reformulation2} is an LP in $\BFalpha$ and $\BFbeta$, which can be easily solved given its coefficients.

The objective function of~\eqref{opt:reformulation2} does not depend on $\alpha_i(\BFtheta_b)$ such that $\BFtheta_b \in \Theta_{i_0}^c$. Therefore, the optimal $\BFalpha^*$ should satisfy $\alpha_i^*(\BFtheta_b) = 0$ for all $1\leq i\leq k, \BFtheta_b \in \Theta_{i_0}^c$. Moreover, we have $\alpha^*(\BFtheta_0) > 0$. Observe that $\alpha(\BFtheta_0)G^*(\BFtheta_0)$ appears in the inner minimization term in the objective function of~\eqref{opt:reformulation2} when $\BFtheta_b = \BFtheta_0$. If $\alpha^*(\BFtheta_0) = 0$, then the objective value becomes zero. As a positive objective value can be attained by setting $\BFalpha = (B+L)^{-1}\BFone_B$ and $\BFbeta = (B+L)^{-1}\BFone_L$, optimal $\alpha^*(\BFtheta_0)$ must be positive. Corollary~\ref{cor:zero_ratio} below formalizes this discussion.
\begin{corollary}\label{cor:zero_ratio}
    Let $(\BFalpha^*, \BFbeta^*)$ be an optimal solution of~\eqref{opt:reformulation2}. Then, we have $\alpha^*(\BFtheta_b) = 0$ for all $\BFtheta_b \in \Theta_{i_0}^c$ and $\alpha^*(\BFtheta_b) > 0$ if $\BFtheta_b = \BFtheta_0$. 
\end{corollary}

The sequential learning algorithms introduced in Section~\ref{sec:seq.sampling} assign the computational budget by repeatedly solving the plug-in version of~\eqref{opt:reformulation2}. Therefore, if $\BFtheta_b$ is incorrectly classified to be in the estimated $\Theta_{i_0}^c$, then there is a nonzero probability that the empirical sampling fractions do not converge to $\BFalpha^*$ and $\BFbeta^*$ even if the budget increases to infinity. To prevent this, we consider the following $\epsilon$-optimal problem for small $\epsilon>0$:
\begin{oldequation}\label{opt:reformulation_eps}
    \begin{aligned}
    V(\epsilon) = &\max_{\BFalpha, \BFbeta} \min_{\BFtheta_b\in \Theta} \left\{\sum\nolimits_{\ell=1}^L \frac{\beta_\ell}{c_\ell} \dKL(\theta_0^\ell||\theta_b^\ell) + \alpha(\BFtheta_b)G^*(\BFtheta_b) 1\{\BFtheta_b \in \Theta_{i_0}\}\right\} \\
    & \text{subject to } \mathbf{1}_B^\top\BFalpha + \mathbf{1}_L^\top \BFbeta = 1, \BFalpha \geq \epsilon \mathbf{1}_B, \BFbeta \geq \epsilon \mathbf{1}_L
    \end{aligned}
\end{oldequation}
That is, all input and simulation sampling fractions are at least $\epsilon$. The optimal objective function value of~\eqref{opt:reformulation2} corresponds to $V(0)$. 
Although the practical implication of introducing $\epsilon$ can be made negligible by choosing small $\epsilon$, it is of theoretical interest to examine the loss in the convergence rate by requiring the minimal sampling fractions. To answer this question, 
let the solution to Program~\eqref{opt:reformulation_eps} be $(\BFalpha^*(\epsilon), \BFbeta^*(\epsilon))$. Proposition~\ref{prop:opt_gap} characterizes the optimality gap of $(\BFalpha^*(\epsilon), \BFbeta^*(\epsilon))$ for Program~\eqref{opt:reformulation2}.

\begin{proposition}\label{prop:opt_gap}
    For $\epsilon \in (0, \frac{1}{B+L})$, we have $0 \leq 1 - \frac{V(\epsilon)}{V(0)}\leq \epsilon\lt(B + L\rt)$.
\end{proposition}
Since $B$ and $L$ are known in advance, Proposition~\ref{prop:opt_gap} lets us control the relative optimality gap with arbitrary precision by choosing $\epsilon$.

\section{Sequential Sampling Budget Allocation Design}\label{sec:seq.sampling}

In this section, we propose the Optimal Sampling Allocation for R\&S (OSAR) that sequentially allocates the sampling budget based on the theoretical results in Section~\ref{sec:simulation_error}. Presented in Algorithm~\ref{alg:plain_OSAR}, OSAR repeatedly solves Program~\eqref{opt:plug_in_eps},  a plug-in version of~\eqref{opt:reformulation_eps}, to distribute the total budget, $T$, in batches of size $\Delta$. Here, the batch size is not a strict requirement as OSAR makes random budget assignments via sampling---$\Delta$ is the expected budget to be allocated at each iteration. Separating from the allocated budget counter, $t$, we introduce the batch iteration counter, $s$, for ease of exposition. The quantities labeled by $s$ are recomputed at each batch iteration.

\begin{algorithm}[!tb]
\caption{Optimal Sampling Allocation for R\&S (OSAR)}\label{alg:plain_OSAR}
\begin{algorithmic}[1]
    \State \textbf{Initialization}: $T:$ total sampling budget; $\Delta$: batch size; $(m_0, n_0)$: initial sample size for input data and simulation outputs; $s$:  batch allocation counter; $t$: expended budget counter; and select a rate-optimal classical R\&S algorithm to solve the subproblem. \label{step:init_start}
    \State Collect $m_0$ input data for each $\ell$-th input data sources and $n_0$ simulation outputs at each $(i, \BFtheta_b)$ for $1\leq i \leq k, 1\leq b\leq B$; $t \leftarrow m_0\sum_{\ell=1}^L c_\ell + Bkn_0$ and 
    $s\leftarrow 0$. \label{step:init_end}  
    \While{$t < T$}\label{step:stopping_rule}
        \State Update $\mu_{i, t}(\BFtheta_b)$ of $\eta_i(\BFtheta_b)$ for each $(i, \BFtheta_b)$ and $\BFpi_\BFm(\BFtheta_b)$ at all $\BFtheta_b\in\Theta$.\label{line:MPB_estimation}
        \State Construct $i_{t}^* = \argmax_{1\leq i\leq k} \sum_{b=1}^B \BFpi_\BFm(\BFtheta_b)1\{i = i_{t}^b\}$, its favorable set $\Theta_{i^*_{t}, t}$, and estimate the MAP $\hBFtheta = \argmin_{\Theta} \BFpi_\BFm(\BFtheta)$.
        \State For all $1\leq b \leq B$, compute $G^*_{t}(\BFtheta_b) = \min_{i \neq i^b_{t}}G_{i, t}(\BFtheta_b)$. Here $G_{i, t}(\BFtheta_b)$ is a plug-in version of $G_i(\BFtheta_b)$, which replaces unknown quantities with their estimates, respectively.\label{step:G_update}
        \State Solve the plug-in version of~\eqref{opt:reformulation_eps}:\label{step:start}
        \begin{oldequation}\label{opt:plug_in_eps}
        \begin{aligned}
            (\hat{\BFalpha}^s, \hat{\BFbeta}^s) = &\argmax_{\BFalpha, \BFbeta} \min_{\BFtheta_b\in \Theta} \left\{\sum\nolimits_{\ell=1}^L \frac{\beta_\ell}{c_\ell} \hat{\textsf{D}}^\text{KL}(\hat{\theta}^\ell|| \theta_b^\ell)
            + \alpha(\BFtheta_b)G_t^*(\BFtheta_b) 1\{\BFtheta_b \in \Theta_{i^*_{t}, {t}}\}\right\} \\
            & \text{subject to } \mathbf{1}_B^\top\BFalpha + \mathbf{1}_L^\top \BFbeta = 1, \BFalpha \geq \epsilon \mathbf{1}_B, \BFbeta \geq \epsilon \mathbf{1}_L.
        \end{aligned}
        \end{oldequation} 
        \State (Input Data Collection) Draw $(\hat{m}_{\ell,s})_{1\leq \ell \leq L} \sim \text{Bin}(\Delta, \hat{\beta}_{\ell, s}/c_\ell)$ and collect $\hat{m}_{\ell, s}$ observations from the $\ell$-th input data source.  \label{line:input_collection}
        \State (Simulation) Draw $\hat{n}_s(\BFtheta_b) \sim \text{Bin}(\Delta, \hat{\alpha}_{s}(\BFtheta_b))$ for each $1\leq b \leq B$. For each $\BFtheta_b$, allocate $\hat{n}_s(\BFtheta_b)$ to $\{(i, \BFtheta_b)\}_{1\leq i\leq k}$ based on the R\&S algorithm selected in Step~\ref{step:init_start}. Update $N_{i,t}(\BFtheta_b)$ for all $1\leq i \leq k$ and $1\leq b\leq B$.\label{line:simulation}
        \State Update $t \leftarrow t + \sum_{\ell=1}^L c_\ell\hat{m}_{\ell, s} + \sum_{b=1}^B \hat{n}_s(\BFtheta_b)$ and $s \leftarrow s+1$. \label{step:end}
    \EndWhile
    \State \textbf{return} $i^*_T$ as the optimal solution.
\end{algorithmic}
\end{algorithm}

In Step~\ref{step:G_update}, OSAR computes the plug-in estimate, $G_{i, t}(\BFtheta_b)$,  of $G_i(\BFtheta_b)$ by replacing the unknown parameters in the rate function with their respective estimates obtained up to $t\leq T$. For instance, if the simulation outputs are normally distributed, then $G_i(\BFtheta_b) = \frac{(\eta_i(\BFtheta_b) - \eta_{i^b}(\BFtheta_b))^2}{2(\lambda_i^2(\BFtheta_b)/\alpha_i(\BFtheta_b) + \lambda_{i^b}^2/\alpha_{i^b}(\BFtheta_b))}$ and $G_{i,t}(\BFtheta_b)$ can be computed by replacing $\eta_i(\BFtheta_b)$ and $\alpha_i(\BFtheta_b)$ with $\mu_{i, t}(\BFtheta_b)$ and $N_{i, t}(\BFtheta_b)/\sum_{j=1}^k N_{j, t}(\BFtheta_b)$, respectively. 

In Step~\ref{step:start}, Program~\eqref{opt:plug_in_eps} replaces $\BFtheta_0$ in~\eqref{opt:reformulation_eps} with its MAP $\hat\BFtheta$ and estimates $\dKL(\theta^\ell_0||\theta_b^\ell)$ with
\begin{oldequation}
    \hat{\textsf{D}}^\text{KL}(\hat{\theta}^\ell|| \theta_b^\ell) : = \frac{1}{m_{\ell}}\sum\nolimits_{j=1}^{m_\ell}\log\frac{f_{\hat{\theta}^\ell}(Z^\ell_{j})}{f_{\theta^\ell_b}(Z^\ell_{j})},
\end{oldequation}
for each $(\ell, \theta_b) \in \{1,2,\ldots, L\}\times \Theta$. For fixed  $\hat\BFtheta$, $\hat{\textsf{D}}^\text{KL}(\hat{\theta}^\ell|| \theta_b^\ell)$ is an unbiased estimator of $\expec_{Z \sim f^\ell_{\theta^\ell_0}}[\log ({f^\ell_{\hat\theta^\ell}(Z)}/{f^\ell_{\theta^\ell}(Z)})]$, which differs from $\dKL(\hat\theta^\ell||\theta_b^\ell)$ in that the expectation is taken with respect to the correct data-generating input distribution. In general, the analytical expression for $\dKL(\hat\theta^\ell||\theta_b^\ell)$ cannot be derived explicitly, although there are exceptions under some parametric models. Even when $\dKL(\hat\theta^\ell||\theta_b^\ell)$ has an explicit form, we have  observed that adopting $\hat{\textsf{D}}^\text{KL}(\hat{\theta}^\ell|| \theta_b^\ell)$ instead significantly improves the empirical performance of OSAR. 

The $s$-th batch iteration's solutions to~\eqref{opt:plug_in_eps}, $\hat\BFalpha^s$ and $\hat\BFbeta^s$, respectively decide how $\Delta$ is allocated to $L$ input data sources and the parameters in $\Theta$ in Steps~\ref{line:input_collection}--\ref{line:simulation} via binomial sampling with budget $\Delta$ and ``success probabilities'' stipulated by $\hat\BFalpha^s$ and $\hat\BFbeta^s$. This ensures that the number of input data or simulation replications to sample are integer-valued. 

We emphasize that in Step~\ref{line:simulation}, any classical R\&S algorithm whose asymptotic sampling ratios converge to the optimal allocation in~\cite{glynn2004large} can be applied. For instance, BOLD~\citep{chen2022BOLD}, mCEI~\citep{chen2019complete} and gCEI~\citep{Avic23:gECI} are suitable candidates, where the latter two are applicable when the simulation outputs are assumed Gaussian. In this paper, our default is set to the BOLD algorithm unless otherwise mentioned. Theorem~\ref{thm:convergence} below stipulates that  the cumulative sampling ratios of Algorithm~\ref{alg:plain_OSAR} converges to the optimal sampling ratios of~\eqref{opt:reformulation_eps},  $(\BFalpha^*(\epsilon), \BFbeta^*(\epsilon))$, almost surely.

\begin{theorem}\label{thm:convergence}
Suppose Assumptions~\ref{asmp:uniq.cond.opt}--\ref{asmp:GC_class} and~\ref{asmp:supp_assumption} hold. Further, assume that the algorithm used in Step~\ref{line:simulation} has strong consistency of the sampling ratios to the rate-optimal static ratios in \cite{glynn2004large}. Let $\BFalpha^s$ and $\BFbeta^s$ respectively represent the cumulative simulation and input data sampling ratios after the $s$-th batch allocation in OSAR. Then, we have $\lim_{s \rightarrow \infty} \BFalpha^s = \BFalpha^*(\epsilon)$ and $\lim_{s \rightarrow \infty} \BFbeta^s = \BFbeta^*(\epsilon)$ almost surely.
\end{theorem}

\subsection{Improving the Finite-Sample Performance with Kernel Ridge Regression}\label{subsec:krr}

 OSAR estimates $\eta_i(\BFtheta_b)$ with the sample average, $\mu_{i, t}(\BFtheta_b)$, of the simulation outputs drawn at $(i, \BFtheta_b)$. However, if $n_0$ is small, the initial estimate may be poor and lead to poor finite-sample performance. In this section, we discuss how to improve the finite-sample performance of OSAR by improving the estimation accuracy of $\eta_i(\BFtheta_b)$ by pooling the simulation outputs obtained at all parameters in $\Theta$ using KRR. 
 
 The theoretical results hereafter require the following additional assumption on the simulation output distribution.
\begin{assumption}\label{asmp:normality}
For each $1\leq i\leq k$ and $\BFtheta\in\Theta$,  $Y_i(\BFtheta)\sim N(\eta_i(\BFtheta), \lambda^2_i(\BFtheta))$ for known $\lambda^2_i(\BFtheta)$. 
\end{assumption}

Our convergence result in Theorem~\ref{thm:LDR_first} and OSAR only require knowing the distribution family so that the rate function $G$ can be derived.
However, we assume normality hereon since it allows us to derive a closed-form expression for the predictor, as described below.
The known variance assumption makes the large-deviation analysis simpler and lets us concentrate on the main message of the analysis. The same assumption has been adopted in~\cite{glynn2004large, chen2006OCBA, chen2019complete}.

 We introduce some notation to formally state the KRR scheme. Let $\mathcal{X}$ be the domain of $\BFtheta$. The kernel, $K$, is a symmetric real function defined on  $\mathcal{X}\times\mathcal{X}$ and called \emph{positive definite} (PD) if $\sum_{j=1}^J\sum_{j'=1}^J K(\BFtheta_j, \BFtheta_{j'})q_jq_{j'}\geq 0$ holds for any integer $J$ and any two real vectors $\left(q_j\right)_{1\leq j\leq J}$ and $\left(q_{j'}\right)_{1\leq j'\leq J}$. Given PD kernel $K$, we can generate a \emph{reproducing kernel Hilbert space} (RKHS), $\mathcal{K}$, defined as a closure of $\{h(\cdot)|h(\cdot) = \sum_{j=1}^J a_jK(\BFtheta_j, \cdot), \{\BFtheta_j\}_{1\leq j \leq J}\subseteq \mathcal{X}\}$ with the associated norm,  $||f||_\mathcal{K} = \sum_{j=1}^J\sum_{j'=1}^J a_ja_{j'}K(\BFtheta_j, \BFtheta_{j'})$, for any $f \in \mathcal{K}$. 
 
 In their MPB estimation algorithm, \cite{kimkimsong2022paper} replace the sample means at solution-parameter pairs with KRR estimators to improve the finite-sample performance of the algorithm. 
 We take the same approach. For each Solution $i$, we adopt an RKHS $\mathcal{K}_i$ associated with PD kernel $K_i$ and impose the model, $\eta_i(\BFtheta) = \gamma_i + g_i(\BFtheta)$, for some constant $\gamma_i$ and $g_i \in \mathcal{K}_i$. From Assumption~\ref{asmp:normality}, we choose the loss function to be  the Gaussian negative log-likelihood function and estimate $g(\BFtheta)$ by solving 
 \begin{oldequation}\label{eq:GNLLL}
    g^*_{i, t} = \argmin_{g \in \mathcal{K}_i}\left\{\sum\nolimits_{b=1}^B \frac{N_{i, t}(\BFtheta_b)}{2\lambda_i^2(\BFtheta_b)}(\mu_{i, t}(\BFtheta_b) - g(\BFtheta_b) - \gamma_i)^2 + \frac{\kappa}{2}||g||_{\mathcal{K}_i}^2\right\}.
\end{oldequation} 
Here, $\kappa$ controls the regularization term to avoid overfitting. Let $\mathbf{A} = \text{diag}(\mathbf{a})$ for some vector $\mathbf{a}$ be the diagonal matrix whose diagonal elements are given by $\mathbf{a}$.  Thanks to the properties of the RKHS, $\hat{\BFmu}_{i, t} = (\gamma_i + g^*_{i, t}(\BFtheta_b))_{1\leq b\leq B}$ can be derived analytically as: 
\begin{oldequation}\label{eq:KRR}
    \hat{\BFmu}_{i, t} := \gamma_i\mathbf{1}_B + \BFK_i^\top(\BFK_i + \kappa \Sigma_{i,t})^{-1}(\BFmu_{i, t} - \gamma_i \mathbf{1}_B),
\end{oldequation}
where $\BFmu_{i, t} = (\mu_{i, t}(\BFtheta_b))_{1\leq b\leq B}$, $\BFK_i  = (K_i(\BFtheta_b, \BFtheta_{b^\prime}))_{1\leq b, b^\prime\leq B}$ is the $B\times B$ Gram matrix of $K_i$ at $\BFtheta_1,\ldots,\BFtheta_B$, and $\Sigma_{i,t} = \text{diag}\left(\left(\frac{\lambda_i^2(\BFtheta_b)}{N_{i, t}(\BFtheta_b)}\right)_{1\leq b\leq B}\right)$.
For $\kappa = 1$,~\eqref{eq:KRR} is identical to the predictive mean of the GP evaluated at $\BFtheta_b, 1\leq b\leq B$, with prior mean $\gamma_i \mathbf{1}_B$ and covariance matrix $\BFK_i$.  We set $\kappa = 1$ in our empirical studies.

OSAR$^+$ presented in Algorithm~\ref{alg:OSAR+KRR} adopts~\eqref{eq:KRR} to estimate each $\eta_i(\BFtheta_b)$. Notice that we slightly abuse the notation in Step~4 and let $i^b_{t}$ denote the plug-in best at $\BFtheta_b$ when the true means are replaced by the KRR predictors, not sample means.

\begin{algorithm}[tbp!]
\caption{$\OSARkrr$}\label{alg:OSAR+KRR}
\begin{algorithmic}[1]
    \State Run Steps~\ref{step:init_start}--\ref{step:init_end} of Algorithm~\ref{alg:plain_OSAR}.
    \While{${t} < T$}
        \State Update $\mu_{i, {t}}(\BFtheta_b)$ for each $(i, \BFtheta_b)$ and $\BFpi_\BFm(\BFtheta_b)$ for each $\BFtheta_b$. Construct  $\hat{\BFmu}_{i, {t}}$ in~\eqref{eq:KRR}.\label{line:krr_start}
        \State Update $i^*_{t} = \argmax_{1\leq i\leq k} \sum_{b=1}^B \BFpi_\BFm(\BFtheta_b)1\{i_{t}^b = i\} $ and $\Theta_{i_{t}^*, t} = \{\BFtheta_b: i^*_{t} = i^b_{t}\}$ where $i^b_{t} = \argmin_{1\leq i\leq k} \hat{\mu}_{i, t}(\BFtheta_b)$ for each $\BFtheta_b$; update $\hBFtheta$. 
        \State For $1\leq b \leq B$, compute $G_{t}^*(\BFtheta_b) = \min_{i \neq i_{t}^b}G_{i, {t}}(\BFtheta_b)$, where $G_{i, {t}}(\BFtheta_b)$ is an estimator of $G_i(\BFtheta_b)$, which replaces $\eta_i(\BFtheta_b)$ and $\alpha^*_i(\BFtheta_b)$ with $\hat{\mu}_{i, {t}}(\BFtheta_b)$ and $N_{i, {t}}(\BFtheta_b)/\sum_{j=1}^k N_{j, {t}}(\BFtheta_b)$, respectively.\label{line:krr_end}
        \State Run Steps~\ref{step:start}--\ref{step:end} of Algorithm~\ref{alg:plain_OSAR}.
     \EndWhile
    \State \textbf{return} $i^*_T = \argmax_{1\leq i\leq k}   \sum_{b=1}^B \BFpi_\BFm(\BFtheta_b)1\{i_{T}^b = i\}$, where $i^b_{T} = \argmin_{1\leq i\leq k} \hat{\mu}_{i, T}(\BFtheta_b)$ for each $\BFtheta_b$, as the optimal solution.
\end{algorithmic}
\end{algorithm}

\cite{kimkimsong2022paper} show that under Assumption~\ref{asmp:normality}, $\BFmu_{i,t}$ and $\hat{\BFmu}_{i, t}$ have the same cumulant generating function (CGF) as $t$ increases. According to the G{\"a}rtner-Ellis theorem~\citep{dembo2009large}, the large-deviation rate (LDR) of a rare event of a random vector being included in a closed set is determined by its limiting CGF. Therefore, 
the convergence rates derived in Section~\ref{sec:simulation_error} are still valid without any modifications even if $\BFmu_{i, T}$ is replaced with $\hat{\BFmu}_{i, T}$ in OSAR$^+$. 
This lets us establish Theorem~\ref{thm:convergence_OSAR+} below, which verifies that  OSAR$^+$ achieves the same asymptotic convergence result as OSAR stated in Theorem~\ref{thm:convergence}. 
\begin{theorem}\label{thm:convergence_OSAR+}
    In addition to the assumptions made in Theorem~\ref{thm:convergence}, suppose Assumption~\ref{asmp:normality} holds. Let $\BFalpha^s$ and $\BFbeta^s$ respectively represent the cumulative simulation and input data sampling ratios after the $s$-th batch allocation in OSAR. Then, $(\BFalpha^s,\BFbeta^s)$ converges to 
    $(\BFalpha^*(\epsilon), \BFbeta^*(\epsilon))$ almost surely as $T\to\infty$. 
\end{theorem}

\begin{remark}\label{rmk:Sherman-Morrsion-OSAR+}
If $B$ is large, the computation in~\eqref{eq:KRR} can be costly due to the matrix inversion in~\eqref{eq:KRR}. 
However, this can be alleviated by applying the Sherman-Morrison-Woodbury formula. This approach requires inverting a $B\times B$ matrix only once at initialization and updates $\BFmu_{i,t}$ by recursion. 
Suppose $\hat{\BFmu}_{i,t}$ in~\eqref{eq:KRR} is given and let  $\BFC_{i, t}:= \BFK_i - \BFK_i(\BFK_i + \Sigma_{i, t})^{-1}\BFK_i$ for $1\leq i \leq k.$ 
When new observation $Y_i(\BFtheta_b)$ is obtained, we can update the KKR estimators at all $kB$ solution parameter pairs as 
\begin{oldequation}\label{eq:sherman_update_OSAR+}
    \begin{aligned}
    \hat{\BFmu}_{i,t+1} = \hat{\BFmu}_{i,t} + \frac{Y_i(\BFtheta_b) - \hat{\mu}_{i,t}(\BFtheta_b)}{\lambda_i^2(\BFtheta_b) + \BFC_{i, t}(b,b)}\BFC_{i, t} \BFe_b \;\; &\text{and}\;\; \hat{\BFmu}_{j, t+1} = \hat{\BFmu}_{j, t} \;\;\forall j\neq i,\\
    \BFC_{i, t+1} = \BFC_{i, t} - \frac{\BFC_{i,t}\BFe_b\BFe_b^\top\BFC_{i, t}}{\lambda_i^2(\BFtheta_b) + \BFC_{i, t}(b,b)}\;\; &\text{and}\;\; \BFC_{j, t+1} = \BFC_{j, t}, \;\; \forall j\neq i,
    \end{aligned}
\end{oldequation}
where  $\mathbf{A}(\ell_1, \ell_2)$ is the $(\ell_1, \ell_2)$-th element of matrix $\mathbf{A}$ and $\BFe_b$ is the  $b$-th basis vector. Since the bottleneck operation of~\eqref{eq:sherman_update_OSAR+} is the outer product of the $B$-dimensional vectors, it is much cheaper than computing~\eqref{eq:KRR} directly. We employ~\eqref{eq:sherman_update_OSAR+} for $\OSARkrr$. 
\end{remark}

\section{Extension to Continuous-valued Input Parameters}\label{sec:continuous.case.extension}

In Sections~\ref{sec:simulation_error} and~\ref{sec:seq.sampling}, $\Theta$ is assumed finite. In this section, we extend OSAR  to accommodate continuous $\Theta$ that includes infinitely many parameters. To avoid a pathological case, we focus on the problems where $\BFtheta_0$ is in the interior of $\Theta_{i_0}$. 
Indeed, this is guaranteed under Assumptions~\ref{asmp:uniq.cond.opt} and~\ref{asmp:continuity_y} when $\Theta$ is continuous; uniqueness of $i_0$ implies that $\eta_i(\BFtheta_0) - \eta_{i_0}(\BFtheta_0)>0$ for $i\neq i_0$, and continuity of $\eta_i(\cdot)$ implies that we can find a ball around $\BFtheta_0$ that is completely contained in $\Theta_{i_0}.$

For continuous $\Theta$, it is impossible to simulate $\{\eta_i(\BFtheta)\}_{1\leq i \leq k}$ at all $\BFtheta\in\Theta$ to estimate the MPB or its adversarial set. A naive way to tackle this issue is to approximate $\Theta$ with a finite subset, which we denote by $\Theta^+$, and apply OSAR replacing $\Theta$ with $\Theta^+$. In the subsequent discussion, we let $B=|\Theta^+|.$ Such $\Theta^+$ can be constructed from a space-filling design within $\Theta$, or sampling from prior probability density function (pdf) $\BFpi_0$. However, this naive version would generally perform poorly for the following reasons.

First, the MPB estimated by the modified OSAR described above may not converge to $i_0$ almost surely. The algorithm would estimate $\BFtheta_0$ with the posterior-maximizing parameter within $\Theta^+$,  $\hat{\BFtheta}^+ := \argmax_{\BFtheta_b \in \Theta^+} \BFpi_\BFm(\BFtheta_b)$, not the MAP, $\hBFtheta$, defined on  $\Theta$. Since $\BFtheta_0$ is unknown in advance, it is difficult to guarantee $\BFtheta_0 \in \Theta^+$. If $\BFtheta_0 \notin \Theta^+$, then $\hBFtheta^+$ does not converge to $\BFtheta_0$, but converges to the parameter in $\Theta^+$ with the smallest weighted KL divergence from $\BFtheta_0,$  $\BFtheta^+ := \argmin_{\BFtheta_b \in \Theta} \sum_{\ell=1}^L \frac{\beta_\ell}{c_\ell}\dKL(\theta_0^\ell||\theta_b^\ell)$ as $m_\ell \rightarrow \infty$ for all $1\leq \ell \leq L$ under Assumptions~\ref{asmp:cost} and~\ref{asmp:sim_size}~(Proposition~1 in \cite{KimSong:OptimizingMPB}). Because our primary goal is to identify $i_0$, we hope $\BFtheta^+ \in \Theta_{i_0}$, which unfortunately cannot be guaranteed in general. This implies that the MPB estimated with  $\Theta^+$ may not converge to $i_0$.

Second, even if $\BFtheta^+\in\Theta_{i_0}$, approximating $\Theta$ with finite $\Theta^+$ may cause a loss in the convergence rate because
the adversarial set of $i_0$ estimated within $\Theta^+$ does not match $\Theta_{i_0}^c$ even if $T\to\infty$. We later illustrate that this may cause the algorithm to allocate too much budget to simulation relative to input data sampling and slow down the convergence of the estimated MPB to $i_0.$

To tackle the first challenge,  $\OSARkrrPP$ presented in Algorithm~\ref{alg:OSAR++} defines a finite approximation of  
$\Theta$ that is dynamically updated throughout the algorithm. 
Let $\hat\BFtheta_{t}$ be the MAP of $\BFtheta_0$ computed after allocating $t$ sampling budget. We define $\Theta^+_t = \Theta^+ \cup \{\hat\BFtheta_{t}\}$, where $\Theta^+$ is determined when $\OSARkrrPP$ is initialized (Step~1). Thus, the first $B$ elements of $\Theta^+_{t}$ remain unchanged, while the last is updated each time new input data is obtained. 
For convenience, we label the elements of $\Theta_t^+$ as $\BFtheta_1,\ldots,\BFtheta_B,\BFtheta_{B+1}$, where $\BFtheta_{B+1} = \hat\BFtheta_t.$
 $\OSARkrrPP$ distributes the simulation portion of batch sampling budget to $k$ solutions paired with the parameters in  $\Theta^+_{t}$. We denote the fraction of sampling budget allocated to $\BFtheta_b\in\Theta^+_{t}$ by $\alpha(\BFtheta_b)$ as in Section~\ref{sec:seq.sampling},  and redefine $\BFalpha = (\alpha(\BFtheta_b))_{\BFtheta_b \in \Theta^+_t}$ to be a $(B+1)$-dimensional vector.
 
 Because $\hat\BFtheta_t$ converges to $\BFtheta_0$,  this modification makes $\BFpi_\BFm$ assign most of the posterior probability mass to $\hat\BFtheta_t$ relative to other parameters in $\Theta^+$ for sufficiently large $t$. Thus, when the estimated MPB within $\Theta^+_t$, $i_t^*$, is updated in Step~\ref{step:OSAR++_MPB_computation}, it would ultimately coincide with the estimated optimum at $\hat\BFtheta_t$. For sufficiently large $t$, we can ensure $\hat \BFtheta_t \in \Theta_{i_0}$, which resolves the first challenge mentioned above provided that a correct selection is made at $\hat\BFtheta_t.$

\begin{algorithm}[tbp!]
\caption{$\OSARkrrPP$}\label{alg:OSAR++}
\begin{algorithmic}[1]
    \State Run Steps~\ref{step:init_start}--\ref{step:init_end} of Algorithm~\ref{alg:plain_OSAR} and construct set $\Theta^+$ that approximates $\Theta$.\label{line:init_OSAR++}
    \While{${t} < T$}
        \State Find $\hBFtheta_{t} = \argmax_{\BFtheta \in \Theta}\BFpi_\BFm(\BFtheta)$ and construct $\Theta_{t}^+ = \Theta^+\cup\{\hBFtheta_{t}\}$. Update $\mu_{i, {t}}(\BFtheta_b)$ for each $(i,\BFtheta_b)\in \{1,\ldots,k\}\times \Theta_{t}^{+}$ and $\BFpi_\BFm(\BFtheta_b)$ at all $\BFtheta_b\in \Theta^+_{t}$. Compute  $\doublehat{\BFmu}_{i, {t}}$ in~\eqref{eq:krr_conti_ver}.\label{step:OSAR++_update_start}
        \State Construct $i^*_{t} = \argmax_{1\leq i\leq k} \sum_{b=1}^{B+1} \BFpi_\BFm(\BFtheta_b)1\{i_{t}^b = i\} $ and $\Theta_{i_{t}^*, t} = \{\BFtheta_b\in\Theta^+_{t}: i^*_{t} = i^b_{t}\}$ where $i^b_{t} = \argmin_{1\leq i\leq k} \doublehat{\mu}_{i, t}(\BFtheta_b)$ for each $\BFtheta_b\in\Theta^+_{t}$.\label{step:OSAR++_MPB_computation}
        \State For $1\leq b \leq B+1$, compute $G_{t}^*(\BFtheta_b) = \min_{i \neq i_{t}^b}G_{i, {t}}(\BFtheta_b)$, where $G_{i, {t}}(\BFtheta_b)$ is an estimator of $G_i(\BFtheta_b)$, which replaces $\eta_i(\BFtheta_b)$ and $\alpha^*_i(\BFtheta_b)$ with $\doublehat{\mu}_{i, {t}}(\BFtheta_b)$ and $N_{i, {t}}(\BFtheta_b)/\sum_{j=1}^k N_{j, {t}}(\BFtheta_b)$, respectively.\label{step:OSAR++_update_end}
        \State Solve the plug-in version of~\eqref{opt:reformulation_eps} replacing $\Theta$ with $\Theta^+_{t}$:\label{step:OSAR++_opt}
        \begin{oldequation}\label{opt:plug_in_eps_map}
        \begin{aligned}
            (\hat{\BFalpha}^s, \hat{\BFbeta}^s) = &\argmax_{\BFalpha, \BFbeta} \min_{\BFtheta_b\in \Theta^+_{t}} \left\{\sum\nolimits_{\ell=1}^L \frac{\beta_\ell}{c_\ell}\hat{\textsf{D}}^\text{KL}(\hat{\theta}_{t}^\ell||\theta_b^\ell) + \alpha(\BFtheta_b)G_{t}^*(\BFtheta_b) 1\{\BFtheta_b \in \Theta_{i^*_{t}, {t}}\}\right\} \\
            & \text{subject to } \mathbf{1}_{B+1}^\top\BFalpha + \mathbf{1}_L^\top \BFbeta = 1, \BFalpha \geq \epsilon \mathbf{1}_{B+1}, \BFbeta \geq \epsilon \mathbf{1}_L.
        \end{aligned}
        \end{oldequation} 
        \State Run Steps~\ref{line:input_collection}--\ref{step:end} of Algorithm~\ref{alg:plain_OSAR}.
     \EndWhile
    \State \textbf{return} $i^*_T = \argmax_{1\leq i\leq k} \sum_{b=1}^{B+1} \BFpi_\BFm(\BFtheta_b)1\{i_{T}^b = i\}$, where $i^b_{T} = \argmin_{1\leq i\leq k} \doublehat{\mu}_{i, T}(\BFtheta_b)$ for each $\BFtheta_b$, as the optimal solution.\label{line:OSAR++_MPB}
\end{algorithmic}
\end{algorithm}

When $\hat\BFtheta_t$ is newly updated in Step~\ref{step:OSAR++_update_start}, however, no simulations have been made at the $\hat\BFtheta_t$ yet. Nevertheless, we need the estimate of $\eta_i(\hat\BFtheta_t)$ at each $i$ so that the batch sampling budget allocation can be made.
$\OSARkrrPP$ addresses this by applying KRR to predict $\eta_i(\hat\BFtheta_t)$. Furthermore, $\OSARkrrPP$ pools the simulations made at the past MAPs of $\BFtheta_0$ to improve the KRR prediction model for each solution instead of discarding them. While this is clearly beneficial in terms of prediction precision, it comes at the cost of computation since the updated MAPs are distinct with nonzero probability in general, which makes the size of the Gram matrix for the KRR prediction model increase. To avoid this, we adopt the \emph{Nystr{\"o}m method}~\citep{williams2000using}, which aims to achieve dimension reduction by replacing $\mathcal{K}_i$ with its subspace $\bar{\mathcal{K}}_i$ spanned by the kernel functions evaluated at the points in $\Theta_T^+$ only. Namely, $\bar{\mathcal{K}}_i :=\{f\in \mathcal{K}_i: f(\cdot) = \sum_{\BFtheta_b \in \Theta_T^+} c_b K_i(\BFtheta_b, \cdot), \{c_b\}_{1\leq b\leq B+1} \in \mathbb{R}^{B+1}\}$. 
Then,~\eqref{eq:GNLLL} is modified to
\small 
\begin{oldequation}\label{eq:GNLLL_conti}
    g^*_{i, t} = \argmin_{g \in \bar{\mathcal{K}}_i}\left\{\sum_{b=1}^B \frac{N_{i, t}(\BFtheta_b)}{2\lambda_i^2(\BFtheta_b)}(\mu_{i, T}(\BFtheta_b) - g_i(\BFtheta_b) - \gamma_i)^2 + \sum_{r=1}^{N_{i, t}(\hBFtheta)} \frac{(Y_i(\hBFtheta_{i, r}) - g(\hBFtheta_{i,r}) - \gamma_i)^2}{2\lambda_i^2(\hBFtheta_{i, r})} +  \frac{\kappa}{2}||g||_{\mathcal{K}_i}^2\right\},
\end{oldequation} \normalsize
where  $N_{i, t}(\hBFtheta)$ is the number of times  $i$ has been simulated in combination with an MAP of $\BFtheta_0$ until $t$ and $\hBFtheta_{i,r}$ is the MAP at which the $r$-th simulation run is made. The value of $\hBFtheta_{i,r}$ may remain unchanged for several $r$s if no additional input data are collected during those iterations. 

In Step~\ref{step:OSAR++_update_start}, $\OSARkrrPP$ computes $\doublehat{\BFmu}_{i, t} := (\gamma_i + g^*_{i,t}(\BFtheta_{b}))_{1\leq b\leq B+1}$, the solution to~\eqref{eq:GNLLL_conti} evaluated at $\Theta_t^+$.
Proposition~\ref{prop:revised_KRR} below presents a closed-form expression for $\doublehat{\BFmu}_{i, t}$. 
We first state some required definitions; let $\BFK_{i, B+1}$ be the Gram matrix of $K_i$ constructed from the parameters in $\Theta^+\cup\{\theta_0\}$ and $\bar{\BFK}_{i, t} := (K_i(\BFtheta_b, \BFtheta_{b^\prime}))_{\BFtheta_b, \BFtheta_{b^\prime} \in \Theta_t^+}$. The following matrices and vector increase in their sizes as new MAPs of $\BFtheta_0$ are computed:  $\widetilde{\BFK}_{i, t} := (K_i(\BFtheta_b, \BFtheta_{b^\prime}))_{\BFtheta_{b}\in \Theta_t^+, \BFtheta_{b^\prime} \in \Theta_b\cup \{\hBFtheta_{i, r}\}_{1\leq r\leq N_{i, t}(\BFtheta)}}$, $\BFY_i := (Y_i(\hBFtheta_{i,r}))_{1\leq r\leq N_{i, t}(\hBFtheta)}$, and $\widetilde{\Sigma}_{i, t} := \text{diag}\left(\left(\frac{\lambda_i^2(\BFtheta_b)}{N_{i, t}(\BFtheta_b)}\right)_{1\leq b \leq B} \cup \left(\lambda_i^2(\hBFtheta_{i, r})\right)_{1\leq r \leq N_{i, t}(\hBFtheta)}\right)$. 
Furthermore, $A^\dagger$ denotes the Moore-Penrose inverse of $A$.

    \begin{proposition}\label{prop:revised_KRR}
    Suppose $\BFK_{i, B+1}$ is invertible. Then,
   \begin{oldequation}\label{eq:krr_conti_ver}
        \begin{aligned}
        &\doublehat{\BFmu}_{i, t} =  \gamma_i \BFone_{B + N_{i, t}(\hBFtheta)} + \bar{\BFK}_{i, t}(\widetilde{\BFK}_{i, t} \widetilde{\Sigma}_{i, t}^{-1}\widetilde{\BFK}^\top_{i,t} + \kappa \bar{\BFK}_{i, t})^\dagger\widetilde{\BFK}_{i, T}\widetilde{\Sigma}_{i, t}^{-1}\left([\BFmu_{i, t}^\top \;\; \BFY_i^\top]^\top - \gamma_i \BFone_{B + N_{i, t}(\hBFtheta)}\right).
        \end{aligned}
    \end{oldequation} 
\end{proposition}

If $\BFtheta_0 \in \Theta^+$, then  $\BFK_{i, B+1}$ has duplicate columns and thus, is non-invertible. However, we argue that this is unlikely to occur. If $\Theta^+$ is generated by sampling from continuous prior $\pi$, then 
this event is negligible because the event, $\BFtheta_0 \in \Theta^+$, occurs with zero probability. Provided that $\BFK_{i, B+1}$ has unique rows and columns, its invertibility can be guaranteed by selecting the kernel function from a subclass of PD kernel called \emph{strictly positive definite} (SPD) kernel, whose Gram matrix evaluated at distinct parameters is always PD. For instance, the squared-exponential  (or Gaussian) kernel is an SPD kernel~\citep{hoffman:08Kernel}, which we employ in the numerical section.

Steps 4 and 5 of $\OSARkrrPP$ adopt $\doublehat{\BFmu}_{i, t}$ to estimate $\{\eta_i(\BFtheta_b)\}_{\BFtheta_b \in \Theta_{t}^{+}}$. We abuse the notation, $i^b_{t}$, to denote the plug-in best at each $\BFtheta_b \in \Theta^+_{t}$ computed from $\doublehat{\BFmu}_{i, t}$. $\OSARkrrPP$ estimates $\prob_{\BFpi_\BFm}(\Theta_{i})$ by $\sum_{b=1}^{B+1}\BFpi_\BFm(\BFtheta_b)\mathbf{1}\{i_{t}^b = i\}$ and selects $i^*_{t}$ based on the estimates. Because $\BFtheta$ is continuous-valued,  the estimated posterior preferences of all $k$ solutions do not add to one even under Assumption~\ref{asmp:uniq.cond.opt}. Nevertheless, the estimates correctly order the posterior preferences as $T$ increases as shown below in Theorem~\ref{thm:KRR_conti_consistency}.

\begin{theorem}\label{thm:KRR_conti_consistency}
    Suppose $\BFK_{i, B+1}$ is invertible, $\lambda_i(\BFtheta)$ is continuous in $\BFtheta\in\Theta$ and $\BFK_i(\BFtheta_1, \BFtheta_2)$ is continuous in $\BFtheta_1,  \BFtheta_2\in\Theta$ for all $1\leq i\leq k$. 
    Under Assumptions~\ref{asmp:uniq.cond.opt}--\ref{asmp:continuity} and~\ref{asmp:normality}, 
 $i_T^*$ returned by $\OSARkrrPP$ converges to $i_0$ as $T \rightarrow \infty$, almost surely. 
\end{theorem}

\begin{figure}[tbp!]
\centering
\includegraphics[width = \textwidth]{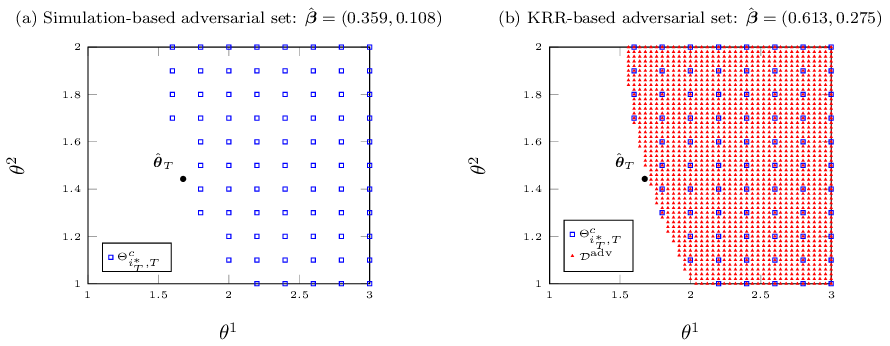}
    \caption{Panels (a) and (b) visualize the simulated and predicted adversarial sets, respectively. We indicate the optimal $\BFbeta$ by solving Programs~\eqref{opt:plug_in_eps} without and with $\mathcal{D}$. }
    \label{fig:dense_grid}
\end{figure}

Although the strong consistency of $i^*_T$ is assuring, it says little about how fast $i^*_T$ converges. Indeed, as pointed out earlier as the second challenge, $\OSARkrrPP$ suffers from an  ``over-simulation'' issue stemming from that $\Theta_{i^*_T}^c$ is estimated with a discrete set. 
To illustrate the point,  Figure~\ref{fig:dense_grid}(a) presents an example of   $\Theta_{i^*_T, T}^c$ estimated by $\OSARkrrPP$ for the numerical example in Section~\ref{subsec:synthetic}. In this problem, $\Theta = [1, 3]\times[1,2]$ is approximated by a two-dimensional grid $\Theta^+$ with $B = 121$ points equally spaced in each dimension.  For $\OSARkrrPP$ to balance the input and simulation sampling, 
the parameter in $\Theta_{i^*_T, T}^c$ that minimizes the weighted sum of the KL divergences in~\eqref{opt:plug_in_eps_map} plays a critical role at increasing the input sampling ratios relative to the simulation sampling ratios. 
However, the discrete approximation shown in Figure~\ref{fig:dense_grid}(a) tends to overestimate the minimal weighted sum of KL divergences within the estimated adversarial set as the boundary of the set is not approximated accurately. Consequently, $\OSARkrrPP$ spends too much of $T$ for simulation and not enough for input data collection, and slows down the convergence of $i^*_T.$

To fix this issue, we select large finite subset $\mathcal{D}$  of $\Theta$ such that $|\mathcal D| \gg B$.  The points in  $\mathcal{D}$ are not considered for simulation, but incorporated to construct a much denser estimate of the adversarial set than  $\Theta_{i^*_t, t}^c$ from the  KRR prediction model. 
From~\eqref{eq:GNLLL_conti}, we can predict the values of $\eta_i(\BFtheta)$ at all $\BFtheta\in\mathcal{D}$ by
\begin{oldequation}\label{eq:krr_conti_pred}
    (g_{i, t}^*(\BFtheta))_{\BFtheta \in \mathcal{D}} = \gamma_i \BFone_{|\mathcal{D}|} + {\BFK}^\mathcal{D}_i (\widetilde{\BFK}_{i, t} \widetilde{\Sigma}_{i, t}^{-1}\widetilde{\BFK}^\top_{i, t} + \kappa \bar{\BFK}_{i, t})^\dagger\widetilde{\BFK}_{i, t}\widetilde{\Sigma}_{i, t}^{-1}\left([\BFmu_{i, t}^\top \;\; \BFY_i^\top]^\top - \gamma_i \BFone_{B + N_{i, t}(\hBFtheta)}\right),
\end{oldequation}
where ${\BFK}^\mathcal{D}_i = (K(\BFtheta_b, \BFtheta_{b^\prime}))_{\BFtheta_b \in \mathcal{D}, \BFtheta_{b^\prime} \in \Theta^+_t}$ is a matrix of size $|\mathcal{D}| \times (B+1)$. We estimate $\Theta_{i^*_t}^c$ with 
\begin{oldequation}\label{eq:predicted_adv_set}
    \mathcal{D}^\text{adv} = \left\{\BFtheta \in \mathcal{D}: i^*_{t} \neq \argmin\nolimits_{1\leq i\leq k} g_{i, t}^*(\BFtheta)\right\}.
\end{oldequation} 
Figure~\ref{fig:dense_grid}(b) overlays an example of $\mathcal{D}^\text{adv}$ constructed from $|\mathcal{D}|=2601$ equally-spaced points over  $\Theta_{i^*_t,t}^c$. For the same $\BFbeta$, it is clear that the minimal weighted sum of the KL divergences between $\hat\BFtheta$ and the parameters within $\mathcal{D}^\text{adv}$ is smaller than that within $\Theta_{i^*_t,t}^c.$

We define  $\OSARkrrFD$ that adopts $\mathcal{D}^\text{adv}$ to approximate the adversarial set of the estimated MPB, where `FD' refers to fixed dense grid.

\begin{definition}
    \textbf{$\OSARkrrFD$} is identical to $\OSARkrrPP$ except that: 1) in Step 1, $\mathcal{D}$ is chosen to approximate $\Theta$; and 2) in Step~6, $\mathcal{D}^\text{adv}$ is estimated from~\eqref{eq:krr_conti_pred} and the solution space, $\Theta^+_{t}$, of the inner minimization problem of~\eqref{opt:plug_in_eps_map} is replaced with $\Theta^+_{t} \cup \mathcal{D}^\text{adv}$. 
\end{definition}

For high-dimensional $\Theta$, however, filling $\Theta$ densely with a uniform grid as in Figure~\ref{fig:dense_grid}(b) suffers from the curse of dimensionality. For instance, suppose $\mathcal{D}$ for $\Theta = [0, 2]^{10}$ is defined by splitting each coordinate into ten equally spaced points. Then, $|\mathcal{D}| = 10^{10}$;   the memory requirement and computational cost for prediction are prohibitively large for such $\mathcal{D}.$ On the other hand, a large portion of $\mathcal{D}$ does not contribute much for improving the performance of $\OSARkrrPP$. Given $\BFbeta$, the parameter that minimizes the weighted sum of KL divergences from $\hat\BFtheta_t$ would lie at the boundary of the adversarial set near  $\hat\BFtheta_t$.
From this observation, we propose $\OSARkrrPS$ that dynamically updates $\mathcal{D}$ whenever a new batch of input data is collected by sampling $|\mathcal{D}|$ points from $\BFpi_\BFm(\BFtheta)$, where `PS' stands for posterior sampling. The resulting $\mathcal{D}$ tends to be more clustered near $\hat\BFtheta_t$.

\begin{definition}\label{def:OSAR+PS}
    \textbf{$\OSARkrrPS$} is identical to {$\OSARkrrFD$} except that at each $t,$ $\mathcal{D}$ is updated with a new size-$|\mathcal{D}|$ sample $\BFtheta_b \stackrel{\text{i.i.d.}}{\sim} \BFpi_\BFm, 1\leq b\leq |\mathcal{D}|$, before estimating $\mathcal{D}^\text{adv}$.
\end{definition}

In Section~\ref{sec:num.exp}, we empirically compare the performances of $\OSARkrrFD$ and $\OSARkrrPS$.

\begin{remark}\label{rmk:Sherman-Morrsion-OSAR_conti}
Notice that $\BFc_{i, t} = (\widetilde{\BFK}_{i, t} \widetilde{\Sigma}_{i, t}^{-1}\widetilde{\BFK}^\top_{i, t} + \kappa \bar{\BFK}_{i, t})^\dagger\widetilde{\BFK}_{i, t}\widetilde{\Sigma}_{i, t}^{-1}\left([\BFmu_{i, t}^\top \;\; \BFY_i^\top]^\top - \gamma_i \BFone_{B + N_{i, t}(\hBFtheta)}\right)$ appears in both~\eqref{eq:krr_conti_ver} and~\eqref{eq:krr_conti_pred}. Section~\ref{ec:recursive} in the Electronic Companion introduces a computationally efficient method for updating $\BFc_{i, t}$ to $\BFc_{i, t+1}$. Once we update $\BFc_{i, t+1}$,~\eqref{eq:krr_conti_ver} and~\eqref{eq:krr_conti_pred} can be updated easily via matrix multiplications.
\end{remark}
		
\section{Numerical Experiments}\label{sec:num.exp}

In this section, we demonstrate the empirical performances of OSAR and its variants by applying them to synthetic examples and a simplified simulation study on a food supply chain network to control contamination. The former examines robustness of our procedures under various experiment settings. The latter demonstrates how our framework can optimize input and simulation data collection in applications.  

We benchmark our algorithms with the Bayesian Information Collection and Optimization (BICO)  proposed by~\cite{ungredda2022}. See Section~\ref{ec:ungredda} for implementation details on BICO. Unless otherwise mentioned, all OSAR variants adopt BOLD~\citep{chen2022BOLD} as a subroutine to allocate $\hat{n}_s(\BFtheta_b)$.

\subsection{Synthetic Examples}\label{subsec:synthetic}

In this section, we construct some synthetic examples with known  $\eta_i(\BFtheta)$ to investigate the performances of the algorithms in comparison under different settings. 
The examples have $k = 10$ solutions that share the two exponential input distributions with respective mean parameters $\theta^1$ and $\theta^2$. 
We examine both discrete and continuous supports of $\BFtheta$; we adopt $\Theta = [1,3]\times[1,2]$ for the latter and take its discrete subset for the former.
For the discrete case, we set $\theta^1_{0} = 1.6$ and $\theta^2_0 = 1.4$. For the continuous case, $\theta^1_0 = \pi/2$ and $\theta^2_0 = \sqrt{2}$. 

\begin{figure}[tbp!]
\centering
\includegraphics[width=0.9\textwidth]{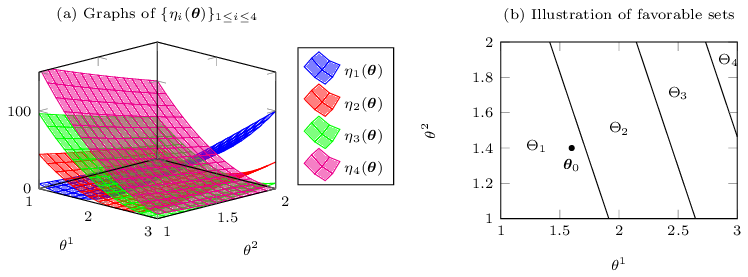}
    \caption{Panels (a) and (b) visualize the mean functions $\{\eta_i(\BFtheta)\}_{1\leq i\leq 4}$ and each solution's favorable set, respectively.}
    \label{fig:mean_ft}
\end{figure}

Given any $(i,\BFtheta)$, we have 
$\eta_i(\BFtheta) = (\BFa^\top\BFtheta - 10\sqrt{i})^2, $
where $\BFa = [5\;\; 5/2]^\top$, 
and $Y_i(\BFtheta)$ is normally distributed with variance $\lambda_i^2(\BFtheta)>0$ that depends on the scenario settings as described later in this section. For both discrete and continuous cases, Solution~1 is optimal at $\BFtheta_0$, i.e., $i_0 = 1,$
and  Solutions~1--4 are conditional optima at some $\BFtheta\in \Theta$ given $\eta_i(\BFtheta)$ while all others are suboptimal. Figure~\ref{fig:mean_ft}(a) displays $\{\eta_i(\BFtheta)\}_{1\leq i\leq 4}$ in $[1,3]\times[1,2]$. Figure~\ref{fig:mean_ft}(b) partitions $[1,3]\times[1,2]$ into the favorable sets of Solutions~1--4, $\{\Theta_i\}_{1\leq i\leq 4}$.
The discrete problem's $\BFtheta_0$ is marked in Figure~\ref{fig:mean_ft}(b), which belongs to $\Theta_1$.

In addition to $\Theta,$ we vary several factors to test the algorithms under different scenarios. The \textit{Baseline} case has $\Delta = 50$ and  $\epsilon = 10^{-4}$ for OSAR and its variants, and $c_1 = c_2 = 1$ and $\lambda_i(\BFtheta) = 8$ for all $1\leq i\leq k, \BFtheta \in \Theta$. Additional scenarios are constructed by modifying Baseline as follows: 
 \begin{itemize}
     \item Scenario 1.  $\lambda_i(\BFtheta) = 6- 2\max(|\theta^1-\theta_0^1|, |\theta^2-\theta_0^2|)$
     \item Scenario 2.  $\lambda_i(\BFtheta) = 2 + 2\max(|\theta^1-\theta_0^1|, |\theta^2-\theta_0^2|)$
     \item Scenario 3.  $(c_1, c_2) = (1, 2)$
     \item Scenario 4.  $\BFtheta_0 = (1.4, 1.2)$ 
     \item Scenario 5.  $\epsilon = 10^{-3}$
 \end{itemize}
Scenarios~1 and 2 test the effect of the simulation output variances; the former makes $\BFtheta$ closer to $\BFtheta_0$ have larger variances, and the latter does the opposite. In Scenario~3, collecting the second input data costs double the first. In Scenario~4, $\BFtheta_0$ is farther away from the boundary of the favorable set, $\Theta_{i_0}$, which makes the posterior preference of $i_0$ converges to $1$ faster when $\BFm$ increases.  Thus, an efficient sampling algorithm would spend less effort on input data collection and more on simulation than in Baseline. Lastly, Scenario~5 empirically assesses the loss in the optimal convergence rate of OSAR  theoretically analyzed in Proposition~\ref{prop:opt_gap}. For all algorithms including BICO, we set  $(m_0, n_0) = (50, 1)$ for initial sampling.

In Sections~\ref{subsubsec:discrete}--\ref{subsubsec:continuous}, we provide detailed problem settings and experiment results for discrete and continuous support cases, respectively. 

\subsubsection{Discrete Support}
\label{subsubsec:discrete}

We adopt $\Theta=\{\BFtheta|\theta_1 \in \lt\{1 + 2j/10\rt\}_{0\leq j \leq 10}, \theta_2 \in \lt\{1 + j/10\rt\}_{0\leq j \leq 10}\}$ as the support of the input parameter vector, which makes $B=121$. We impose a non-informative prior in that $\BFpi_0$ assigns probability mass $1/B$ to each $\BFtheta\in\Theta.$

\begin{figure}[tbp!]
\centering
\includegraphics[width=\textwidth]{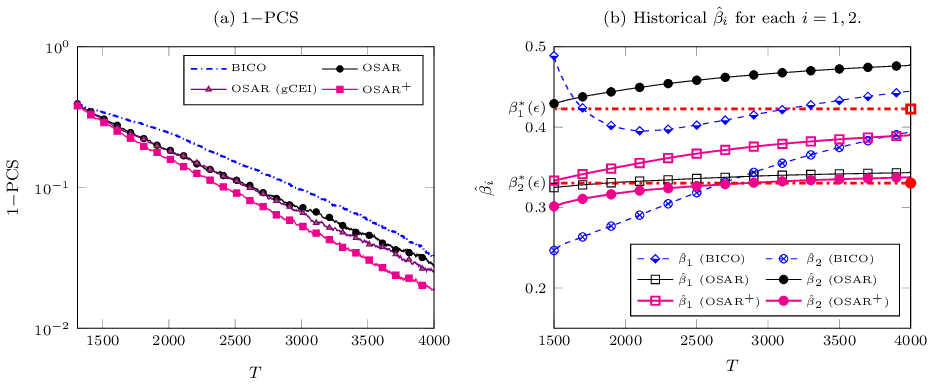}
 \caption{The left panel displays $1-$PCS in log scale. The right panel presents input data sampling ratios $\hat\beta_{i,T}$ for $i=1,2$. The results are obtained by averaging over 10,000 macroruns. }
    \label{fig:PCS_m_ratio}
\end{figure}

We test four algorithms: OSAR, OSAR$^+$, OSAR (gCEI), and BICO. OSAR (gCEI) adopts gCEI~\citep{Avic23:gECI} instead of BOLD as the R\&S subroutine. 
For the Baseline case, the left panel of Figure~\ref{fig:PCS_m_ratio} displays the logarithm of $1-$PCS for each algorithm as $T$ increases.  $\OSARkrr$ shows the fastest convergence in PCS followed by  OSAR and OSAR (gCEI). The latter two algorithms behave very similarly, which indicates robustness of the OSAR framework with respect to the choice of the employed R\&S subroutine. All three OSAR variants outperform BICO. This is indeed notable since it has been reported in the literature that the plug-in type R\&S algorithms derived from the optimal convergence rate analyses tend to underperform dynamic programming-based algorithms that maximize the value of information, especially for smaller $T$. 

The right panel of Figure~\ref{fig:PCS_m_ratio} displays the 
input data sampling ratios $\hat\BFbeta_T$ each algorithm spends; OSAR (gCEI) is dropped from comparison as its behavior is similar to OSAR. The $\epsilon$-optimal input data sampling ratios, $\beta_1^*(\epsilon)$ and $\beta_2^*(\epsilon)$, obtained by solving~\eqref{opt:reformulation_eps} with  $\epsilon = 10^{-4}$ are marked on the $y$-axis, which stipulate sampling more frequently from the first input distribution. 
The empirical sampling ratios of $\OSARkrr$ appear to converge to $(\beta_1^*(\epsilon),\beta_2^*(\epsilon))$ as $T$ increases, while those of OSAR have their orders reversed (i.e., it collects more from the second input distribution).  
The performance difference between OSAR and OSAR$^+$ implies that the KRR effectively improves the prediction of the simulation output means at the solution-parameter pairs. BICO's $\hat\beta_{1,T}$ is arguably closest to $\beta_1^*(\epsilon)$ in the range of $T$ displayed here, however, it over-samples the second input data. Moreover,  neither $\hat\beta_{1,T}$ nor $\hat\beta_{2,T}$ seems to converge to its respective $\epsilon$-optimal ratio.

\begin{table}[tbp!]
\centering
\caption{Comparisons of BICO, OSAR, and OSAR$^+$ under the six defined scenarios. We report total input sample sizes ($m_1$ and $m_2$) and the number of simulation runs ($n$), and the empirical PCS at $T=4{,}000$. 
All statistics are computed from $1{,}000$ macroruns.}
\label{tab:numerics.with.warmstart}
\resizebox{\textwidth}{!}{
\begin{tabular}{cccc|ccc|ccc}
\multirow{2}{*}{Scenario} & \multicolumn{3}{c|}{BICO} & \multicolumn{3}{c|}{OSAR}      & \multicolumn{3}{c}{$\OSARkrr$}   \\ \cline{2-10} 
                          & $(m_1, m_2)$ & $n$     & PCS   & $(m_1, m_2)$ & $n$     & PCS   & $(m_1, m_2)$ & $n$     & PCS\\ \hline
Baseline                  & (1246, 1110)  & 1.64E+3 & 0.967 & (982, 1330)  & 1.70E+3 & 0.972 & (1099, 967)  & 1.95E+3 & {\bf 0.981} \\ \hline
Scenario 1                & (1353, 1208) & 1.44E+3 & 0.983 & (1089, 1313) & 1.61E+3 & 0.976 & (1257, 971)  & 1.78E+3 & {\bf 0.987} \\ \hline
Scenario 2                & (1416, 1278) & 1.31E+3 & 0.983 & (1442, 1139) & 1.43E+3 & 0.991 & (1530, 983)  & 1.50E+3 & {\bf 0.998} \\ \hline
Scenario 3                & (893, 736) & 1.64E+3 & 0.935 & (750, 810) & 1.64E+3 & 0.959 & (867, 616)  & 1.91E+3 & {\bf 0.974} \\ \hline
Scenario 4                & (1394, 1382) & 1.22E+3 & 1.000 & (1275, 1135) & 1.60E+3 & 1.000 & (1087, 903)  & 2.02E+3 & {\bf 1.000} \\ \hline
Scenario 5                & (1244, 1103)  & 1.65E+3 & 0.969 & (896, 1162)  & 1.95E+3 & 0.966 & (981, 847)   & 2.18E+3 & {\bf 0.972} \\ \hline
\end{tabular}
}
\end{table}

Next, we compare the performances of OSAR, OSAR$^+$, and BICO under all six scenarios. Table~\ref{tab:numerics.with.warmstart} displays the average input sample sizes $(m_1,m_2)$, simulation sample size ($n$), and  empirical PCS at $T=4{,}000$ computed from $10{,}000$ macroruns of each algorithm.    OSAR$^+$ shows higher PCS than OSAR for all scenarios (equal in Scenario~4), which is consistent to the observation made in Figure~\ref{fig:PCS_m_ratio}.

All algorithms run more simulations in Scenario~1  than Scenario~2 because the R\&S problem at $\BFtheta_0$ is more difficult in Scneario~1  than in Scenario~2 as the simulation error variances of solutions are higher near $\BFtheta_0$ in the former. In Scenario~3, we collect fewer input and simulation data compared to other scenarios because $c_2$ is higher, which also reduces the sample sizes from other sources since $T$ must be split among them. Consequently, Scenario~3 has the lowest PCS among all scenarios.  The opposite can be observed for  Scenario~4 since  $\BFtheta_0$ is farther from the boundary of $\Theta_{i_0}$ rendering it an ``easier'' problem. Lastly, in Scenario~5, both OSAR and OSAR$^+$ have PCS values close to those in Baseline, which shows that the effect of increasing $\epsilon$ from $10^{-4}$ to $10^{-3}$ is negligible in terms of practical performances.

\subsubsection{Continuous Support}
\label{subsubsec:continuous}
We impose the noninformative priors, $\BFpi^1_0(\theta^1)\sim U(1,3)$ and $\BFpi^2_0(\theta^2)\sim U(1, 2),$ on the continuous support.  
The performances of $\text{OSAR}^\text{++}$, $\OSARkrrFD$,  $\OSARkrrPS$, and BICO are compared in this section. 
The first three algorithms adopt $\Theta^+ = \{\BFtheta|\theta^1 \in \lt\{1 + 2j/10\rt\}_{0\leq j \leq 10}, \theta^2 \in \lt\{1 + j/10\rt\}_{0\leq j \leq 10}\}$ as the initial discrete approximation of $\Theta$ and $|\Theta^+| = 121.$ This choice makes $\BFtheta_0 \notin \Theta^+$. 
 
For  $\OSARkrrFD$, $\mathcal{D}$ is constructed as $\{\BFtheta|\theta^1 \in \lt\{1 + 2j/50\rt\}_{0\leq j \leq 50}, \theta^2 \in \lt\{1 + j/50\rt\}_{0\leq j \leq 50}\}$, which results in $|\mathcal{D}|=2601$. $\OSARkrrPS$ adopts the same size, $|\mathcal{D}|=2601$, but resamples $\mathcal{D}$ each iteration from $\BFpi_\BFm(\BFtheta)$.

\begin{figure}[tbp!]
\centering
\includegraphics[width=\textwidth]{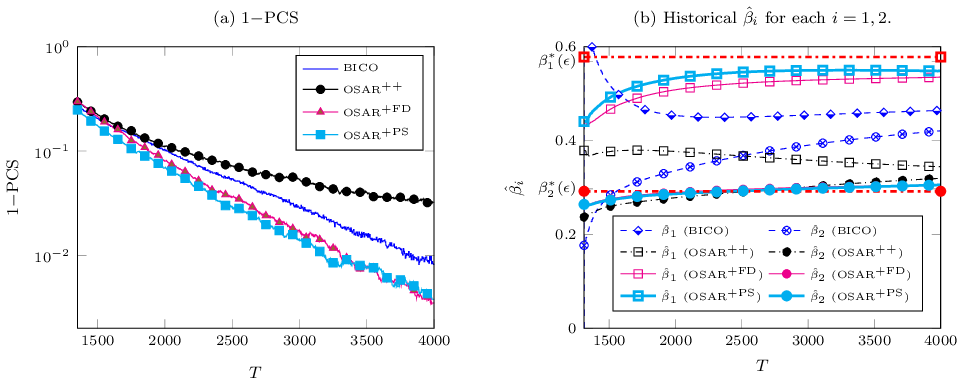}
 \caption{The left panel displays $1-$PCS in log scale. The right panel presents input data sampling ratios $\hat\beta_{i,T}$ for $i=1,2$. The results are obtained by averaging over 10{,}000 macroruns.}
    \label{fig:PCS_m_ratio_conti}
\end{figure}

Figure~\ref{fig:PCS_m_ratio_conti} summarizes all four algorithms'  performances applied to the Baseline scenario. The PCSs of all OSAR variants converge to $1$ even if $\BFtheta_0\neq \Theta^+$ demonstrating the effectiveness of including the historical MAPs of $\BFtheta_0$ in simulation sampling decisions and KRR predictions. 
$\text{OSAR}^\text{++}$ shows the worst performance in the logarithmic $1-$PCS (left panel). The empirical convergence rate degrades in $T$, indicating that $\text{OSAR}^\text{++}$  does not achieve the exponential convergence rate of PCS. 
On the other hand, $\OSARkrrFD$ and $\OSARkrrPS$ outperform BICO. This demonstrates that adopting $\mathcal{D}$ to improve the estimator of $\Theta_{i_0}^c$ affects the convergence rate of the PCS. Although 
$\OSARkrrPS$ slightly outperforms $\OSARkrrFD$ initially, their empirical PCS are indistinguishable for larger $T$. 

The right panel of Figure~\ref{fig:PCS_m_ratio_conti} visualizes the input data sampling ratios, $\hat{\BFbeta}_T$, each algorithm spends. The $\epsilon$-optimal ratios, $\BFbeta^*(\epsilon) = (\beta_1^*(\epsilon), \beta_1^*(\epsilon))$, marked on the y-axis are obtained by approximately solving the following program:
\begin{oldequation}\label{opt:true_opt_beta}
    \begin{aligned}
    &\max\nolimits_{\BFalpha, \BFbeta} \min\nolimits_{\BFtheta_b\in \Theta^+ \cup \Theta^\text{adv}} && \left\{\sum\nolimits_{\ell=1}^L \frac{\beta_\ell}{c_\ell} \dKL(\theta_0^\ell||\theta_b^\ell) + \alpha(\BFtheta_b)G^*(\BFtheta_b) 1\{\BFtheta_b \in \Theta^+, i^b = i_0\}\right\} \\
    & \text{subject to } &&\mathbf{1}_{B+1}^\top\BFalpha + \mathbf{1}_L^\top \BFbeta = 1, \BFalpha \geq \epsilon \mathbf{1}_{B+1}, \BFbeta \geq \epsilon \mathbf{1}_L,
    \end{aligned}
\end{oldequation}
where $\Theta^\text{adv} = \{\BFtheta: {i_0} \neq \argmin_{1 \leq i \leq k} \eta_i(\BFtheta)\}$. Because $\Theta^\text{adv}$ is a continuous set,  we approximate it with an extremely dense grid of size $2001^2 \approx 4\times 10^6$ and solve~\eqref{opt:true_opt_beta} with the discrete approximation. 
Observe that $\hat{\beta}_{1, T}$ of $\OSARkrr$ converges to the wrong limit because it estimates $\Theta_{i_0}^c$ to be a subset of $\Theta^+$. On the other hand, the sampling ratios of $\OSARkrrFD$ and $\OSARkrrPS$ approach the targets as $T$ increases. 
The sampling ratios of BICO are quite different from $\beta_1^*(\epsilon)$ and $\beta_2^*(\epsilon)$ although we do not expect them to match since the set of parameters within which BICO makes the simulation sampling decisions differs from the others.

\subsubsection{Computational Overhead Comparison}

\begin{table}[tbp!]
\centering
\caption{Average computation time for a single run with $T = 4,000$ of each algorithm. The experiment settings are identical to Baseline for the discrete $\Theta$ and the setup for continuous $\Theta$, respectively. We averaged over 1,000 statistics for all algorithms.}\label{tab:numerics.time}
\begin{tabular}{ccccc|cccc}
\hline
                                      & \multicolumn{4}{c|}{Discrete } & \multicolumn{4}{c}{Continuous }              \\ \hline
Algorithms                            & OSAR   & OSAR (gCEI)  & $\OSARkrr$  & BICO   & $\text{OSAR}^\text{++}$ & $\OSARkrrPS$ & $\OSARkrrFD$ &  BICO   \\ \hline
Avg time/single run & 0.43s  & 0.37s      & 1.01s       & 24.48s & 1.28s & 5.38s               & 1.48s             & 50.72s \\ \hline
\end{tabular}
\end{table}
Table~\ref{tab:numerics.time} reports the wall-clock time for each algorithm for discrete and continuous $\Theta$ cases averaged over 1{,}000 macroruns until each exhausts $T=4{,}000$ simulation budget applied to the Baseline scenario.
Since the simulation run times are negligible for this problem, the reported times measure the computational overhead of the algorithms.
All algorithms are written in \textsc{Matlab} 2024a and run on a Macbook Pro with an Apple M1 Pro chip and 16GB memory. 

For the discrete support case,  OSAR and OSAR (gCEI) exhibit similar computing times, whereas the BICO takes about 66 times longer than OSAR.  $\OSARkrr$ requires additional calculations to fit the KRR model, but the results show the increase in the computational time is minor compared to BICO. 

For the continuous case,  $\OSARkrrFD$ and $\text{OSAR}^\text{++}$  complete the job at a similar speed to the discrete case. However, $\OSARkrrPS$ is slower since it samples a new $\mathcal{D}$  from the posterior distribution of $\BFtheta$ each iteration and KRR predictions are made at the new points in $\mathcal{D}$. Nonetheless, $\OSARkrrPS$ is much cheaper than BICO, which also requires sampling from the posterior of $\BFtheta$ to make the input data sampling decisions. 

BICO's computational overhead can be attributed to the fact that its sampling decisions are based on values of information of input or simulation data sampling, which do not have analytical expressions and must be estimated via Monte Carlo (MC) simulations. 
The computational costs of OSAR variants that adopt KRR increase in the dimension of $\BFtheta$, however, BICO suffers from the same curse of dimensionality from the GP regression.

\subsection{Contamination Control in the Food Supply Chain}\label{subsec:contamination}

\citet{foodSCM} present a simulation study on contamination control of a multi-stage food supply chain network. In this section, we present an example inspired by their work. 

\begin{figure}[tbp!]
\centering
\scalebox{0.8}{
\begin{tikzpicture}[
  > = stealth',
  auto,
  prob/.style = {sloped, near start, anchor=south, inner sep=3pt,font=\scriptsize},
  node distance = 1cm and 2.5cm
  ]
  \node[state, rectangle, minimum width = 1.5cm, minimum height = 8cm, align=center, font = \scriptsize] (a) {Supplier };
  \node[state, align=center, font = \scriptsize] (b) [right=of a] {{Processing} \\ {center~2}{($Q_2$)}};
  \node[state, align=center, font = \scriptsize] (c) [above=of b] {{Processing} \\ {center~1}{($Q_1$)}};
  \node[state] (d) [below=of b, align=center, font = \scriptsize] {{Processing} \\ {center~3}{($Q_3$)}};
  \node[state, align=center, font = \scriptsize] (e) [right=of c] {{Processing} \\ {center~4}{($Q_4$)}};
  \node[state, align=center, font = \scriptsize] (f) [right=of b] {{Processing} \\ {center~5}{($Q_5$)}};
  \node[state, align=center, font = \scriptsize] (g) [right=of d] {{Processing} \\ {center~6}{($Q_6$)}};
  \node[state, rectangle, minimum width = 1.5cm, minimum height = 8cm, align=center, font = \scriptsize] (h) [right=of f] {Retailer};

  \draw[->] ($(b.west)-(2.5,0)$) -- ($(b.west)$) node [pos=.5, above, sloped, near start, font = \scriptsize] (TextNode1) {$1/3$};
  \draw[->] ($(c.west)-(2.5,0)$) -- ($(c.west)$) node [pos=.5, above, sloped, near start, font = \scriptsize] (TextNode2) {$1/3$};
  \draw[->] ($(d.west)-(2.5,0)$) -- ($(d.west)$) node [pos=.5, above, sloped, near start, font = \scriptsize] (TextNode3) {$1/3$};

  \draw[->] ($(e.east)$) -- ($(e.east) + (2.5, 0)$) node [pos=.5, above, sloped, near start, font = \scriptsize] (TextNode4) {$\BFq_{4, 7}$};
  \draw[->] ($(f.east)$) -- ($(f.east) + (2.5, 0)$) node [pos=.5, above, sloped, near start, font = \scriptsize] (TextNode5) {$\BFq_{5, 7}$};
  \draw[->] ($(g.east)$) -- ($(g.east) + (2.5, 0)$) node [pos=.5, above, sloped, near start, font = \scriptsize] (TextNode6) {$\BFq_{6, 7}$};

  \path[->]
            (b) edge node[prob]{$\BFq_{2, 5}$}  (f)
                edge node[prob]{$\BFq_{2, 4}$}  (e)
                edge node[prob]{$\BFq_{2, 6}$}  (g)

            (c) edge node[prob]{$\BFq_{1, 5}$}  (f)
                edge node[prob]{$\BFq_{1, 4}$}  (e)

            (d) edge node[prob]{$\BFq_{3, 5}$}  (f)
                edge node[prob]{$\BFq_{3, 6}$}  (g);

\end{tikzpicture}
}
\caption{An illustration of the two-stage food supply chain example. Each $\BFq_{i, j} = (c_{i, j}, p_{i, j})$ consists of two quantities $c_{i,j}$ and $p_{i, j}$, which represent the contamination rate and routing probability when a product transits from Processing center $i$ to $j$, respectively.}
\label{fig:food_chain}
\end{figure}
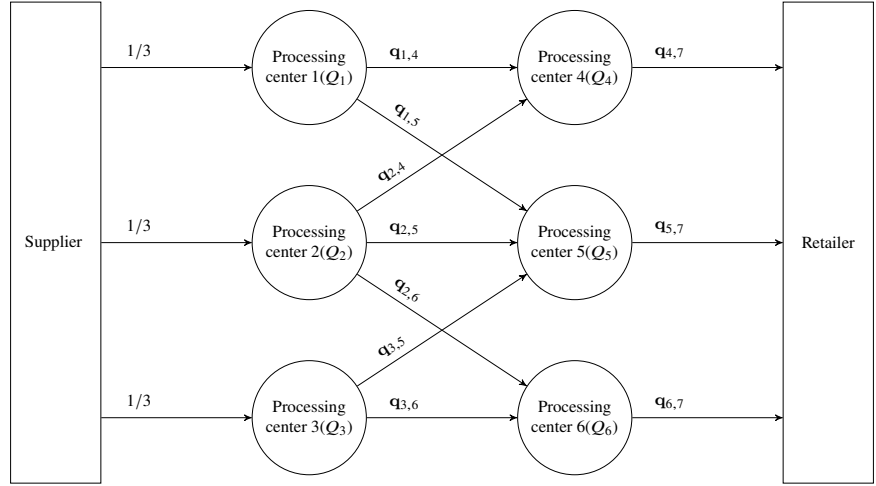

Figure~\ref{fig:food_chain} illustrates a supply chain network, where Processing Centers~1--3 receive the raw material from the suppliers to complete the first-stage process. The intermediate products are then routed to Processing Centers~4--6 for the second-stage process before being delivered to the retailers. 
When the products are in transit, they are exposed to the risk of contamination and contaminated food must be discarded. To reduce the profit loss due to the disposal, we consider installing post-processing facilities to slow down the contamination. Since the installation and operation of the facility is costly, the goal is to carefully choose one processing center expected to reduce the number of disposed products the most from a simulation study.

We formulate this problem as R\&S with six solutions (processing centers), where the objective is to maximize the expected throughput of the food products. A stochastic simulator is built to evaluate each solution.
The total number of units of raw materials of the food products delivered by the suppliers at each epoch is distributed uniformly between  $475$ and $525$, i.e., $U[475, 525]$. These are then routed randomly to Processing Centers~1--3 according to the routing probabilities as depicted in Figure~\ref{fig:food_chain}. 
For $1\leq i \leq 6$, $Q_i$ denotes the number of units of uncontaminated food products at the $i$th processing center. We assume that the raw materials are not contaminated until they reach the first-stage processing centers. Thus, $(Q_1, Q_2, Q_3)|Q_{\text{tot}} \sim \text{MN}(Q^\text{tot}, [1/3, 1/3, 1/3])$, where $Q^\text{tot} \sim U[475, 525]$ is a realization of the total number of units of raw materials and MN refers to the multinomial distribution. 

Once the first-stage process of the raw materials is complete, the products are routed to the second stage. Let $p_{i,j}$ be the routing probability from Processing Center $i$ to $j$ and $N_{i,j}$ be the number of units routed to Processing Center $j$ out of $Q_i$ for each $i$. Therefore, $\sum_{j\neq i} p_{i,j} =~1$ and  $\sum_{j\neq i} N_{i,j} = Q_i$ for each $i$. Given $Q_i$, $\{N_{i,j}\}_{j\neq i}$ follow MN$(Q_i, \{p_{i,j}\}_{j\neq i})$. For instance, $(N_{2, 4}, N_{2, 5}, N_{2, 6})|Q_2 \sim \text{MN}(Q_2, [p_{2, 4}, p_{2, 5}, p_{2, 6}])$. However, some products routed to the next stage become contaminated and are discarded. We denote the probability of each product being contaminated when it is on the route from $i$ to $j$ by $c_{i,j}$. Therefore, the number of uncontaminated products delivered to second-stage Processing Center $j$ is a sum of binomial random variables: $Q_j = \sum_{i: p_{i,j}>0} \text{Bin}(N_{i,j}, 1-c_{i,j})$ for $j=4,5,6$.

We assume that once the contamination control facility is installed at a processing center, there is no further contamination at the center. For instance, installing the facility at Processing Center $3$ makes $c_{3,5}=c_{3,6}=0.$

In this simulation model, the routing probabilities $\{p_{i, j}\}$ and contamination rates $\{c_{i, j}\}$ determine the randomness. We suppose that $\{p_{i, j}\}$ are subject to the operational decision, and thus their values are known. On the other hand, each $c_{i, j}$ needs to be learned by inspecting randomly sampled products. Consequently, we view $\BFtheta = \{c_{i, j}\}\in\mathbb{R}^{10}$ as the input model parameter. 
We impose the truncated Beta conjugate prior for $\BFtheta$ by setting the prior density $\BFpi_0^\ell(\theta^\ell) \propto (\theta^\ell(1 - \theta^\ell))^{0.5}1\{0.1 \leq \theta^\ell \leq 0.3\}$.

Let $Y_i(\BFtheta)$ be the simulated amount of food product that the retailer receives if the contamination control facility is installed at Processing Center $i$ and the contamination rate is set to  $\BFtheta$. The true rates, $\BFtheta_0$, are given in Table~\ref{tab:true_par_contamination}. To evaluate the algorithms' performances, $\expec[Y_i(\BFtheta_0)]$ for all $1\leq i\leq k$ are estimated via $10^6$ MC simulation runs from which $i_0 = 2$ is found.

\begin{table}[tbp!]
\centering
\caption{The true contamination probabilities for the food contamination control example.}
\label{tab:true_par_contamination}
\resizebox{\textwidth}{!}{\begin{tabular}{|c|c|c|c|c|c|c|c|c|c|c|}
\hline
$\BFtheta_0$ & $\theta_0^1 = c_{1,4}$ & $\theta_0^2 = c_{1,5}$ & $\theta_0^3 = c_{2,4}$ & $\theta_0^4 = c_{2,5}$ & $\theta_0^5 = c_{2,6}$ & $\theta_0^6 = c_{3,5}$ & $\theta_0^7 = c_{3,6}$ & $\theta_0^8 = c_{4,7}$ & $\theta_0^9 = c_{5,7}$ & $\theta_0^{10} = c_{6, 7}$ \\ \hline
Value        & $0.1516$               & $0.2384$               & $0.2707$               & $0.2986$               & $0.2987$               & $0.2121$               & $0.1556$               & $0.1485$               & $0.1110$               & $0.1478$                   \\ \hline
\end{tabular}}
\end{table}

As $\BFtheta$ in this example has a continuous support, we compare OSAR$^{++}$, $\OSARkrrFD$, and $\OSARkrrPS$ with BICO. For all algorithms we adopt $(m_0, n_0) = (5, 5)$. For OSAR variants, $\Theta^{+}$ is constructed by sampling 100 points from the prior, $\BFpi_0(\BFtheta)$. For $\OSARkrrFD$, we construct $\mathcal{D}$ by generating 10{,}000 ten-dimensional low-discrepancy sequence---Halton sequence~\citep{kocis1997computational}---over $\Theta$. For all algorithms, we run a batch of 10 replications and take their average as the simulation output for simulation sampling; this makes the simulation output distribution to be better approximated by a normal distribution, which is the assumption all algorithms in comparison are based upon (including BICO). Since the simulation variance at each $(i,\BFtheta)$, $\lambda^2_i(\BFtheta),$ is unknown in this example, we use a plug-in estimate.
If $\BFtheta \in \Theta^+$, the sample variance of simulation outputs is employed. Otherwise (i.e., $\BFtheta = \hBFtheta_t$), we calculate the sample variance of the recent half of the replications made at the MAPs of $\BFtheta_0$ thus far, $\{Y_i(\hBFtheta_{i, r})\}_{1\leq r\leq N_{i,T}(\hBFtheta)}$. Note that these replications are not i.i.d.\ as the MAPs may have been updated over the iterations, however, their sample variance converges to $\lambda^2(\BFtheta_0)$ almost surely as $T\to\infty$ and discarding the first half of replications mitigates its finite-sample bias. 

\begin{figure}[tbp!]
\centering
\includegraphics[width = \textwidth]{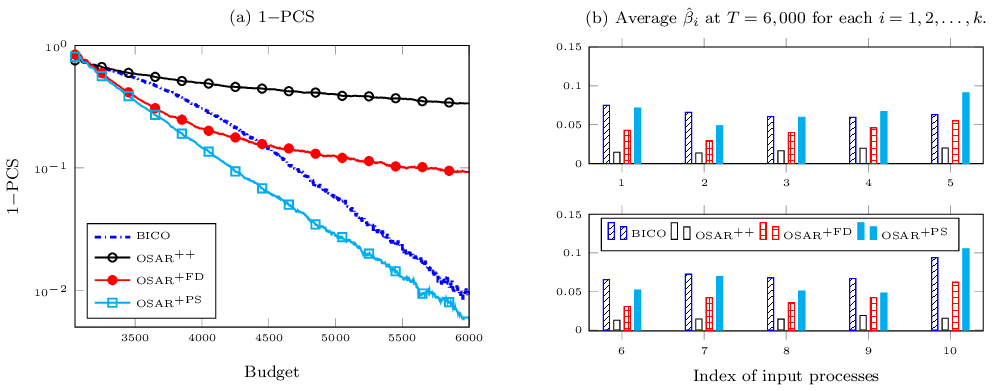}
 \caption{For the food contamination example, (a) displays $1-$PCS  in log scale and (b) presents input data sampling ratios $\hat\beta_{i,T}$ for $1\leq i\leq 10$ at $T=6{,}000$. The results are obtained by averaging over 10{,}000 macroruns. }
    \label{fig:Contamination_results}
\end{figure}

Figure~\ref{fig:Contamination_results} summarizes the empirical results from 10{,}000 macroruns of all four algorithms. Panel~(a) of Figure~\ref{fig:Contamination_results} shows the convergence rate of the PCS for each algorithm. Similar to the results in Section~\ref{subsubsec:continuous}, 
OSAR$^{++}$ performs the worst. Meanwhile, there is a clearer performance difference between $\OSARkrrFD$ and $\OSARkrrPS$ in this example, which indicates the advantage of posterior sampling over using fixed $\mathcal{D}$ when the input parameter dimension is high. BICO is outperformed by all OSAR variants in the beginning, then gradually catches up with OSAR$^{++}$ and $\OSARkrrFD$, but is consistently outperformed by $\OSARkrrPS$.

On Panel (b), we present the empirical input data sampling ratios, $\{\hat{\beta}_i\}_{1\leq i\leq 10}$, at $T=6{,}000$ averaged over the macroruns. OSAR$^{++}$ collects less input data from all 10 sources implying that it tends to simulate more. Consequently, the posterior probability does not concentrate at $\BFtheta_0$ quickly enough compared to other algorithms, 
which explains the convergence rate degradation of OSAR$^{++}$ in Panel (a). 
The ratios of BICO, $\OSARkrrFD$ and $\OSARkrrPS$ across all $10$ inputs show no clear dominance. All three algorithms agree that $\BFtheta_{10}$ is the most critical input parameter to learn to determine $i_0$.

 \section{Conclusion}

This paper investigates the optimal way to allocate data collection budget to input and simulation sampling to reduce the uncertainty in the R\&S problem. Under Bayesian input modeling, we adopt the MPB as an estimator of the true optimal solution. We demonstrate the MPB approaches the true optimum at the true input parameter and derive the exponential convergence rate of its posterior preference in the presence of simulation error. Building upon these results, we propose a sequential budget allocation algorithm named OSAR, which assigns simulation and input data budget in a batch. Given that the input parameter space is discrete, the historical sampling ratios guided by OSAR are shown to converge to the optimal ratios. 

We further extend OSAR to a continuous input parameter space.  Employing the KRR as an inferential tool, we provide several recipes to ensure that the algorithm returns the optimum asymptotically while achieving good empirical convergence rate. Compared to a state-of-the-art algorithm based on Bayesian optimization framework, OSAR and its variants exhibit superior empirical performances.  

Several research questions remain. First, an extension to the nonparametric input model is worth investigating. Extension of OSAR  to such a problem is not straightforward because the nonparametric input model space is infinite-dimensional. 
Second, for fixed input parameters, applying the common random numbers is expected to enhance the overall performance by rapidly identifying each conditional optimum. Investigating this requires revisiting the LDR of the posterior preference of the MPB. Finally, considering estimators other than the MPB to establish OSAR  is another interesting research direction. For example, one can take a risk-neutral approach to estimate the true optimal as done in~\cite{ungredda2022}.

\bibliographystyle{informs2014} 
\bibliography{reference.bib} 

\ECSwitch

\noindent {\bf\large Electronic companion to ``Optimizing Input Data Collection for Ranking and Selection"}

\section{Proofs and Discussions on the Theory in Section~\ref{sec:MPB_asymp}}

\subsection{Discussion on Assumption~\ref{asmp:continuity} and Proof of Proposition~\ref{prop:exp.fam} }

We first show that the strong consistency of the MAP to $\BFtheta_0$ can be guaranteed under Assumption~\ref{asmp:continuity} as long as $m_\ell \rightarrow \infty$ for all $1\leq \ell\leq L$, which is weaker than Assumption~\ref{asmp:data_size} since it does not require the convergence of ratio $c_\ell m_\ell/T$. 
\begin{lemma}\label{lem:consistency.MAP}
    If Assumption~\ref{asmp:continuity} holds and $\lim_{T\rightarrow \infty} m_\ell = \infty$ for all $1\leq \ell \leq L$, then $\hBFtheta_T \xrightarrow{a.s.}  \BFtheta_0$. 
\end{lemma}
\proof{Proof.}
    Observe that $\theta_0^\ell = \argmin_{\theta^\ell}\dKL(\theta^\ell_0||\theta^\ell) = \argmin_{\theta^\ell} \expec_{Z^\ell \sim f_{\theta_0^\ell}}\left[-\log \left(\frac{f_{\theta^\ell}(Z^\ell)}{f_{\theta_0^\ell}(Z^\ell)}\right) \right]$ and 
        $\hat{\theta}_T^\ell = \argmin\nolimits_{\theta^\ell} \frac{1}{m_\ell}\sum\nolimits_{i=1}^{m_\ell} -\log f_{\theta^\ell}(Z_i^\ell) = \argmin\nolimits_{\theta^\ell} \frac{1}{m_\ell}\sum\nolimits_{i=1}^{m_\ell} -\log \left({f_{\theta^\ell}(Z_i^\ell)}/{f_{\theta_0^\ell}(Z_i^\ell)}\right).$
    Considering the two optimization problems' objective functions,  $\lim_{m_\ell \rightarrow \infty}\frac{1}{m_\ell}\sum\nolimits_{i=1}^{m_\ell} -\log \left(\frac{f_{\theta^\ell}(Z_i^\ell)}{f_{\theta_0^\ell}(Z_i^\ell)}\right) = \expec_{Z^\ell \sim f_{\theta_0^\ell}}\left[-\log \left(\frac{f_{\theta^\ell}(Z^\ell)}{f_{\theta_0^\ell}(Z^\ell)}\right) \right]$ almost surely uniformly over $\theta^\ell \in \Theta^\ell$ from Assumption~\ref{asmp:continuity}(c). 
    By Proposition~3 in~\cite{shapiro2003monte_ec}, the convergence of the optimal solution then follows, i.e., $\hat{\theta}_T^\ell \xrightarrow{a.s.} \theta_0$. \qed
\endproof

\vspace{3mm}

Some sufficient conditions for GC class have been studied in the empirical process literature. We briefly mention three conditions. One is provided in Theorem 19.4 in \cite{van2000asymptotic_ec} by exploiting a bracketing number of a function class. From Examples 19.7 and 19.8 in \cite{van2000asymptotic_ec}, we can derive sufficient conditions for Assumption~\ref{asmp:GC_class}(c) by assuming Lipschitz continuity and existence of an integrable envelope function, respectively. The detailed theory is beyond the scope of this paper; see \cite{van2000asymptotic_ec} and \cite{van1996weak_ec} for further reading.

Proposition~\ref{prop:exp.fam}  states a sufficient condition that implies Assumption~\ref{asmp:continuity}(c). Below, we proof it for the case of $L=1$, which can be extended to when $L>1$ easily.

\proof{\underline{Proof of Proposition~\ref{prop:exp.fam} ($L = 1$).}}
From the properties of exponential family: $\expec[S(\BFZ)] = \nabla A(\theta_0)$, $\mathrm{Cov}[S(\BFZ)] = \nabla^2 A(\theta_0)$, and the KL divergence is $\dKL(\theta_0||\theta) = A(\theta) - A(\theta_0) - \inner{\nabla A(\theta_0), \theta - \theta_0}$. 
Therefore,
\begin{oldequation}
    \begin{aligned}
        \sup_{\theta \in \Theta}\lt|\frac{1}{m}\sum\nolimits_{i=1}^m \lt(-\log\frac{f_{\theta}}{f_{\theta_0}}\rt)(\BFZ_i) - \dKL(\theta_0||\theta)\rt| &= \sup_{\theta \in \Theta}\lt|\inner{\theta_0 - \theta, \frac{1}{m}\sum\nolimits_{i=1}^m (S(\BFZ_i) - \nabla A(\theta_0))}\rt| \\
        &\leq 2C \frac{1}{m}\sum_{i=1}^m \norm{(S(\BFZ_i) - \nabla A(\theta_0))}_{1}  =: F
    \end{aligned}
\end{oldequation}
where $C = \max_{\theta \in \Theta} \norm{\theta}_{\infty}$, which is bounded from compactness of $\Theta$, and the inequality follows from H\"{o}lder's inequality. From Example 19.8 in~\cite{van2000asymptotic_ec}, if $F$ is integrable, then $\{\log(f_\theta/f_{\theta_0})\}_{\theta \in \Theta}$ is $\prob_{\theta_0}$-Glivenko-Cantelli. Hence, it suffices to show that $F$ is integrable. By the second-order differentiability of $A(\theta)$, $F$ is indeed integrable since
\begin{oldequation}
    \expec[F]  = 2C \sum\nolimits_{j=1}^{\dim(S)}\expec\lt[\norm{S(\BFZ) - \nabla A(\theta_0)}_1\rt] \leq 2C \sum\nolimits_{j=1}^{\dim(S)} \sqrt{\mathrm{Var}(S_j(\BFZ))} = \sum\nolimits_{j=1}^{\dim(S)} \sqrt{\frac{\partial^2}{\partial \theta_j^2} A(\theta_0)} < \infty,
\end{oldequation}
where $S_j$ represents the $j$-th element of $S$. For the first inequality we use $\expec[X]^2 \leq \expec[X^2]$. \qed
\endproof

\subsection{Proof of Theorem~\ref{thm:LDR_input}}
Let us first show Part (b). Because $\prob_{\BFpi_\BFm}(\Theta) = 1$, observe that
\begin{oldequation}\label{eq:post.prob}
    \begin{aligned}
    -\dfrac{1}{T}\log \prob_{\BFpi_\BFm}(\Theta^c_{i_0}) &= -\dfrac{1}{T}\log \prob_{\BFpi_\BFm}(\Theta^c_{i_0}) + \frac{1}{T}\log \prob_{\BFpi_\BFm}(\Theta)\\
    &= -\dfrac{1}{T}\log\int_{\Theta^c_{i_0}} \dfrac{\BFpi_\BFm(\BFtheta)}{\BFpi_\BFm(\BFtheta_0)}d\BFtheta + \dfrac{1}{T}\log\int_{\Theta} \dfrac{\BFpi_\BFm(\BFtheta)}{\BFpi_\BFm(\BFtheta_0)}d\BFtheta.
    \end{aligned}
\end{oldequation}

To bound the integral, we first show the following uniform convergence of the log-likelihood ratio, $\log\lt(\BFpi_\BFm(\BFtheta)/\BFpi_\BFm(\BFtheta_0)\rt)$:
\begin{oldequation}\label{eq:unif.epsilon}
    \epsilon_T : = \sup_{\BFtheta \in \Theta}\lt|\frac{1}{T}\log\lt(\dfrac{\BFpi_\BFm(\BFtheta)}{\BFpi_\BFm(\BFtheta_0)}\rt) + \sum\nolimits_{\ell = 1}^L \frac{\beta_\ell}{c_\ell}\dKL(\theta^\ell_0||\theta^\ell)\rt| \rightarrow 0 \;\; \prob_{\BFtheta_0}\text{-a.s.}
\end{oldequation}
as $m_\ell \rightarrow \infty$ for all $1\leq \ell \leq L$. From the definition of $\BFpi_\BFm$, we have
\begin{oldequation}\label{eq:split}
    \begin{aligned}
         &\frac{1}{T}\log\lt(\frac{\BFpi_\BFm(\BFtheta)}{\BFpi_\BFm(\BFtheta_0)}\rt) + \sum_{\ell = 1}^L \frac{\beta_\ell}{c_\ell}\dKL(\theta^\ell_0||\theta^\ell)
         = \frac{1}{T}\sum_{\ell = 1}^L \log\lt(\frac{\BFpi^\ell_0(\theta^\ell)}{\BFpi^\ell_0(\theta^\ell_0)}\rt) + \sum_{\ell = 1}^L \frac{m_\ell}{T}\lt(\frac{1}{m_\ell}\sum_{i=1}^{m_\ell} \log\frac{f^\ell_{\theta^\ell}(Z^\ell_i)}{f^\ell_{\theta^\ell_0}(Z^\ell_i)} + \dKL(\theta^\ell_0||\theta^\ell)\rt). 
    \end{aligned}
\end{oldequation}
For each $\ell$, combining Assumption~\ref{asmp:GC_class} (c) with $\dKL(\theta^\ell_0||\theta^\ell) := -\expec_{Z^\ell \sim f^\ell_{\theta^\ell_0}}\lt[\log \lt({f^\ell_{\theta^\ell}(Z^\ell)}/f^\ell_{\theta^\ell_0}(Z^\ell)\rt)\rt]$, each term in the second summation converges to zero uniformly over $\theta^\ell \in \Theta^\ell$ almost surely. Further, the first term also converges to zero uniformly by the boundedness of $\BFpi^\ell_0$.
Thus, $\epsilon_T \rightarrow 0$ almost surely. For simplicity, we denote $\mathcal{I}(\BFtheta): = \sum_{\ell = 1}^L\frac{\beta_\ell}{c_\ell}\dKL(\theta^\ell_0||\theta^\ell)$ for $\BFtheta = (\theta^1, \ldots, \theta^L)$ from now on.

From~\eqref{eq:unif.epsilon}, we have the following two-sided bound
\begin{oldequation}
    \exp\lt(-T\mathcal{I}(\BFtheta) - T\epsilon_T\rt) \leq \frac{\BFpi_\BFm(\BFtheta)}{\BFpi_\BFm(\BFtheta_0)} \leq \exp\lt(-T\mathcal{I}(\BFtheta) + T\epsilon_T\rt).
\end{oldequation}
By integrating all sides over $\BFtheta \in \Theta^c_{i_0}$, we get
\begin{oldequation}
    \exp\lt(-T\epsilon_T\rt)\int_{\Theta^c_{i_0}}\exp\lt(-T\mathcal{I}(\BFtheta))\rt)d\BFtheta \leq \int_{\Theta^c_{i_0}}\dfrac{\BFpi_\BFm(\BFtheta)}{\BFpi_\BFm(\BFtheta_0)}d\BFtheta \leq \exp\lt(T\epsilon_T\rt)\int_{\Theta^c_{i_0}}\exp\lt(-T\mathcal{I}(\BFtheta)\rt)d\BFtheta.
\end{oldequation}
After taking logarithms on all sides, dividing by $T$, and letting $T \rightarrow \infty$, we obtain
\begin{oldequation}\label{eq:post_to_laplace}
    \lim_{T \rightarrow \infty} \lt\{\dfrac{1}{T}\log \int_{\Theta^c_{i_0}}\dfrac{\BFpi_\BFm(\BFtheta)}{\BFpi_\BFm(\BFtheta_0)}d\BFtheta - \frac{1}{T} \log \int_{\Theta^c_{i_0}}\exp\lt(-T\mathcal{I}(\BFtheta)\rt)d\BFtheta\rt\} = 0, \;\; \prob_{\BFtheta_0}\text{-a.s.}
\end{oldequation}
thanks to~\eqref{eq:unif.epsilon}. Our next goal is to approximate the second term in~\eqref{eq:post_to_laplace}. Let $\BFtheta_1$ be a minimizer of $\mathcal{I}(\BFtheta)$ in the closure of $\Theta^c_{i_0}$. Then, we have
\begin{oldequation}
    \begin{aligned}
    \frac{1}{T} \log \int_{\Theta^c_{i_0}}\exp\lt(-T\mathcal{I}(\BFtheta)\rt)d\theta + \mathcal{I}(\BFtheta_1) \leq \frac{1}{T}\log \lt(\exp(-T\mathcal{I}(\BFtheta_1))\int_{\Theta^c_{i_0}}d\theta\rt) + \mathcal{I}(\BFtheta_1) = \frac{1}{T}\log \text{Vol}(\Theta^c_{i_0})
    \end{aligned}
\end{oldequation}
where the volume $\text{Vol}(\tilde{\Theta}) : = \int_{\tilde{\Theta}}d\BFtheta$. Taking the limsup with respect to $T$ on both sides of the inequality, we have
\begin{oldequation}
    \limsup_{T \rightarrow \infty} \frac{1}{T} \log \int_{\Theta^c_{i_0}}\exp\lt(-T\mathcal{I}(\BFtheta)\rt)d\BFtheta + \mathcal{I}(\BFtheta_1) \leq 0.
\end{oldequation}
Equivalently, 
\begin{oldequation}\label{eq:limsup}
    \liminf_{T \rightarrow \infty} -\dfrac{1}{T} \log \int_{\Theta^c_{i_0}}\exp\lt(-T\mathcal{I}(\BFtheta)\rt)d\BFtheta \geq \inf_{\BFtheta \in \Theta^c_{i_0}} \mathcal{I}(\BFtheta) \;\; \prob_{\BFtheta_0}\text{-a.s.}
\end{oldequation}

Next, we aim to show 
\begin{oldequation}\label{eq:liminf}
    \limsup_{T \rightarrow \infty} -\frac{1}{T} \log\int_{\Theta^c_{i_0}}\exp\lt(-T\mathcal{I}(\BFtheta)\rt)d\BFtheta \leq \inf_{\BFtheta \in \text{int}(\Theta^c_{i_0})} \mathcal{I}(\BFtheta)\;\; \prob_{\BFtheta_0}\text{-a.s.}
\end{oldequation}
When $\text{int}(\tilde{\Theta}) = \emptyset$,~\eqref{eq:liminf} is straightforward since the infimum over the empty set is $\infty$ by convention. Suppose $\text{int}(\tilde{\Theta}) \neq \emptyset$. Since a Cartesian product of compact sets is compact, Assumption~\ref{asmp:input_LDR}(a) implies that $\Theta = \prod_{\ell = 1}^L \Theta^\ell$ is also compact. Assumptions~\ref{asmp:input_LDR}(a) and (d) imply uniform continuity of $\mathcal{I}(\BFtheta)$ for $\BFtheta \in \Theta$. In other words, for any $\epsilon > 0$, there exists $\delta > 0$ such that $\norm{\BFtheta - \bm{\nu}}_{\infty}\leq \delta$ implies $|\mathcal{I}(\BFtheta) - \mathcal{I}(\bm{\nu})| \leq \epsilon$. For such $\delta > 0$, compactness of $\Theta$ implies that there exists finite open cover $\lt\{o_1, o_2,\cdots, o_M\rt\}$, where $o_j$ is a $\delta$-radius open ball in $\Theta$ such that $\Theta \subset \cup_{j=1}^M o_j$. Let us remove all balls that do not intersect with $\text{int}(\Theta^c_{i_0})$. Abusing the notation, let us say $\text{int}(\Theta^c_{i_0}) \subset \cup_{j=1}^M o_j$. Denote $\BFtheta_2$ be a minimizer of $\inf_{\BFtheta \in \text{int}(\Theta^c_{i_0})}\mathcal{I}(\BFtheta)$. Then, $\BFtheta_2 \in \text{cl}(\text{int}(\Theta^c_{i_0}))$ and $\BFtheta_2$ must be contained in at least one of $\lt\{\text{cl}(o_j), 1\leq j \leq M\rt\}$. Without loss of generality, suppose $\BFtheta_2 \in \text{cl}(o_1)$. Furthermore, $\text{Vol}(o_1 \cap \text{int}(\Theta^c_{i_0})) > 0$ since $o_1 \cap \text{int}(\Theta^c_{i_0})$ is a nonempty open set. Consequently,
\begin{oldequation}
    \begin{aligned}
    \frac{1}{T} \log \int_{\Theta^c_{i_0}}\exp\lt(-T\mathcal{I}(\BFtheta)\rt)d\BFtheta +\mathcal{I}(\BFtheta_2)& \geq \frac{1}{T} \log \int_{\text{int}(\Theta^c_{i_0})\cap o_1}\exp\lt(-T\mathcal{I}(\BFtheta)\rt)d\BFtheta + \mathcal{I}(\BFtheta_2) \\
    &\geq \frac{1}{T} \log \int_{\text{int}(\Theta^c_{i_0})\cap o_1}\exp\lt(-T\mathcal{I}(\BFtheta) - T\epsilon\rt)d\BFtheta + \mathcal{I}(\BFtheta_2) \\
    & = \frac{1}{T}\log\text{Vol}(\text{int}(\Theta^c_{i_0})\cap o_1) - \epsilon
    \end{aligned}
\end{oldequation}
where the second inequality holds by the uniform continuity.
Hence, we can see that $$\liminf_{T \rightarrow \infty} \lt\{\frac{1}{T} \log \int_{\Theta^c_{i_0}}\exp\lt(-T\mathcal{I}(\BFtheta)\rt)d\theta + \mathcal{I}(\BFtheta_2))\rt\} \geq -\epsilon,$$ a.s., for any $\epsilon >0$, which implies~\eqref{eq:liminf}.

By Assumption~\ref{asmp:continuity_y}, we have $\BFtheta_0 \in \text{int}(\Theta)$. It implies that $\inf_{\BFtheta \in \text{int}(\Theta)} \mathcal{I}(\BFtheta) = \inf_{\theta \in \Theta} \mathcal{I}(\BFtheta) = 0$. Therefore, we obtain $\lim_{T \rightarrow \infty} -\frac{1}{T} \log \int_{\Theta}\exp\lt(-T\mathcal{I}(\BFtheta)\rt)d\BFtheta = 0$, $\prob_{\BFtheta_0}$-a.s. Finally, we have
\begin{oldequation}
    \begin{aligned}
    \liminf_{T \rightarrow \infty} -\frac{1}{T}\log \prob_{\BFpi_\BFm}(\Theta^c_{i_0}) &\underset{\eqref{eq:post.prob}}{=} \liminf_{T \rightarrow \infty}\lt\{-\frac{1}{T}\log\int_{\Theta^c_{i_0}} \frac{\BFpi_\BFm(\BFtheta)}{\BFpi_\BFm(\BFtheta_0)}d\BFtheta + \frac{1}{T}\log\int_{\Theta} \frac{\BFpi_\BFm(\BFtheta)}{\BFpi_\BFm(\BFtheta_0)}d\BFtheta\rt\}\\
    &\underset{\eqref{eq:post_to_laplace}}{=} \liminf_{T \rightarrow \infty}\lt\{-\frac{1}{T}\log\int_{\Theta^c_{i_0}} \exp(-T \mathcal{I}(\BFtheta))d\BFtheta + \frac{1}{T}\log\int_{\Theta} \exp(-T \mathcal{I}(\BFtheta))d\BFtheta\rt\}\\
    &\underset{\eqref{eq:limsup}}{\geq} \inf_{\BFtheta \in \Theta^c_{i_0}}\mathcal{I}(\BFtheta)
    \end{aligned}
\end{oldequation}
as desired. To show $\limsup_{T \rightarrow \infty} -\frac{1}{T}\log \prob_{\BFpi_\BFm}(\Theta^c_{i_0}) \leq \inf_{\BFtheta \in \textup{int}(\Theta^c_{i_0})} \mathcal{I}(\BFtheta)$, a similar argument can be applied to~\eqref{eq:liminf}. Consequently,  $\lim_{T\rightarrow \infty} -\dfrac{1}{T}\log \prob_{\BFpi_\BFm}(\Theta^c_{i_0}) = \mathcal{I}(\BFtheta)$ for some $\mathcal{I}(\BFtheta) > 0$ from Assumption~\ref{asmp:data_size}.

We proceed to show Part (a). From Part (b), $\prob_{\BFpi_\BFm}(\Theta_{i_0}) \rightarrow 1$ and $\prob_{\BFpi_\BFm}(\Theta_{j}) \rightarrow 0$ for $j \neq i_0$. Let us define $\mathcal{E} \subset {\Omega}^{\mathbb{N}}$ such that $\lim_{T \rightarrow \infty} \prob_{\BFpi_\BFm}(\Theta_i) = 1\{i = i_0\}$ for all sample paths of input data that belong in $\mathcal{E}$. Then, for each sample path in $\mathcal{E}$, there exists $T_1$ such that for all $T\geq T_1$, $\prob_{\BFpi_\BFm}(\Theta_{i_0}) > 1/2$ and $\prob_{\BFpi_\BFm}(\Theta_{i}) < 1/2$. This implies $i^*(\BFpi_\BFm) = i_0$ for all $T\geq T_1$. Since $\prob_{\BFtheta_0}^{\mathbb{N}}(\mathcal{E}) = 1$, the proof is complete. \qed
\endproof

\vspace{3mm} 

\begin{remark}
For the case when the input model belongs to an exponential family, Proposition 5 of \cite{russo2020simple_ec} also derives a posterior large deviation rate. This result coincides with our Theorem~\ref{thm:LDR_input} when $L=1$, however, Theorem~\ref{thm:LDR_input} is more general as it does not restrict the distribution family.
\end{remark}

\section{Proof of Statements in Section~\ref{sec:simulation_error}}

\subsection{Proof of Theorem~\ref{thm:LDR_first}}

{\bf Strong consistency of $i_T^*$}. As shown in Theorem~\ref{thm:LDR_input}, we have $\lim_{T\rightarrow \infty} \BFpi_\BFm(\BFtheta_0) = 1$ almost surely. Hence, $i_T^*$  converges to the estimated conditional optimum at $\BFtheta_0$ almost surely. Thanks to $N_{i, T}(\BFtheta_b) \rightarrow \infty$ for all $(i, \BFtheta_b)$ in Assumption~\ref{asmp:sim_size}, we can apply the strong law of large numbers, and this yields the estimated conditional optimum at $\BFtheta_0$ converges to $i_0 = \argmin_{1\leq i\leq k} \eta_i(\BFtheta_0)$ almost surely.

\vspace{3mm}
\noindent {\bf The convergence rate of $\expec[\prob_{\BFpi_\BFm}(\Theta_{i_T^*, T})1\{i_T^* = i^*(\BFpi_\BFm)\}|\BFpi_\BFm]$}. 
For each $\BFtheta_b\in\Theta$,  define
\begin{oldequation}
    \widetilde{G}_i(\BFtheta_b):=\lim_{T \rightarrow \infty}-\frac{1}{T}\log\prob\lt(i_T^b = i\rt).
\end{oldequation}
Namely, $\widetilde{G}_i(\BFtheta_b)$ is the exponential decay rate of the probability such that $i (\neq i^b)$ is falsely selected as $i^b$. The expression for $\widetilde{G}_i(\BFtheta_b)$ is derived under Assumption~\ref{asmp:normality} by~\cite{kimkimsong2022paper_ec} to analyze the convergence rate of $\expec[1\{i^*_T=i^*(\BFpi_\BFm)\}|\BFpi_\BFm]$ for fixed $\BFpi_\BFm$, i.e., no additional input data sampling is allowed.
However, in our problem, as $\BFpi_\BFm$ becomes more concentrated with additional input data, $\widetilde{G}_i(\BFtheta_b)$ can eventually be eliminated in the rate analysis. Hence, we do not present an analytical expression of $\widetilde{G}_i(\BFtheta_b)$.

Let us first establish the following: \small
\begin{oldequation}\label{eq:LDR_first}
    \begin{aligned}
    &\lim_{T \rightarrow \infty} -\frac{1}{T}\log(1 - \expec[\prob_{\BFpi_\BFm}(\Theta_{i_T^*, T})1\{i_T^* = i^*(\BFpi_\BFm)\}|\BFpi_\BFm]) \\
    & = \min\left\{\min_{i \neq i_0} G_i(\BFtheta_0), \min_{\wTheta \in \mathcal{A}} \bigg(\min_{\BFtheta \in \wTheta^c}\sum\nolimits_{\ell=1}^L \frac{\beta_\ell}{c_\ell}\dKL(\theta_0^\ell||\theta^\ell) + \sum\nolimits_{\BFtheta_b \in \Theta_{i_0}\setminus \wTheta} \min_{i \neq i_0} G_i(\BFtheta_b) + \sum\nolimits_{\BFtheta_b \in \wTheta\setminus \Theta_{i_0}} \widetilde{G}_{i_0}(\BFtheta_b)\bigg)\right\},
    \end{aligned}
\end{oldequation} \normalsize
where $\mathcal{A} := \{\wTheta \subseteq \Theta| \BFtheta_0 \in \wTheta\}$ is the collection of all subsets of $\Theta$ that include $\BFtheta_0$.
The proof will proceed by first showing some intermediate results in the following steps that will be needed to complete the proof.
\vspace{3mm}

\noindent\textbf{Step~1}. For sufficiently large $T$ such that $i^*(\BFpi_\BFm) = i_0$ and $\BFpi_\BFm(\BFtheta_0) > 1/2$, we have $\expec[\prob_{\BFpi_\BFm}(\Theta^c_{i_0, T})1\{i_T^* = i_0\}|\BFpi_\BFm] = \sum_{\wTheta \in \mathcal{A}}\prob_{\BFpi_{\BFm}}(\wTheta^c)\prob(\Theta_{i_0, T} = \wTheta)$.
\proof{\underline{Proof of Step~1.}}
    Suppose $T$ is sufficiently large as stated. Let $\mathcal{B}$ be the power set of $\{\BFtheta_1, \ldots, \BFtheta_B\}$. Observe that 
    \begin{oldequation}
    \begin{aligned}
        \expec[\prob_{\BFpi_\BFm}(\Theta^c_{i_0, T})1\{i_T^* = i_0\}|\BFpi_\BFm] &= \sum\nolimits_{\wTheta \in \mathcal{B}}\expec[\prob_{\BFpi_\BFm}(\wTheta^c)1\{i_T^* = i_0, \Theta_{i_0, T} = \wTheta\}|\BFpi_\BFm] \\
        &= \sum\nolimits_{\wTheta \in \mathcal{B}}\prob_{\BFpi_{\BFm}}(\wTheta^c)\prob(i_T^* = i_0, \Theta_{i_0, T} = \wTheta).
    \end{aligned}
    \end{oldequation}
    When $\BFpi_{\BFm}(\BFtheta_0)>1/2$, $\BFtheta_0 \in \Theta_{i_0, T}$  if and only if $i_T^* = i_0$. Thus,
    if $\BFtheta_0 \notin \wTheta$, then  $\prob\{i_T^* = i_0, \Theta_{i_0, T} = \wTheta\} = 0$. Moreover, if $\BFtheta_0 \in \wTheta$, then  $1\{i_T^* = i_0, \Theta_{i_0, T} = \wTheta\} = 1\{\Theta_{i_0, T} = \wTheta\}$. Consequently,
    {\small$$\sum\nolimits_{\wTheta \in \mathcal{B}}\prob_{\BFpi_{\BFm}}(\wTheta^c)\prob(i_T^* = i_0, \Theta_{i_0, T} = \wTheta) = \sum\nolimits_{\wTheta \in \mathcal{A}}\prob_{\BFpi_{\BFm}}(\wTheta^c)\prob(i_T^* = i_0, \Theta_{i_0, T} = \wTheta) = \sum\nolimits_{\wTheta \in \mathcal{A}}\prob_{\BFpi_{\BFm}}(\wTheta^c)\prob(\Theta_{i_0, T} = \wTheta)$$}
    as desired. \halmos
\endproof
\vspace{3mm}
\noindent\textbf{Step~2.} For each $\wTheta \in \mathcal{A}$, $\lim_{T \rightarrow \infty} -\frac{1}{T}\log \prob(\Theta_{i_0, T} = \wTheta) = \sum_{\BFtheta_b \in \Theta_{i_0}\setminus \wTheta} \min_{i\neq i_0}G_i(\BFtheta_b) + \sum_{\BFtheta_b \in \wTheta \setminus \Theta_{i_0}}\widetilde{G}_{i_0}(\BFtheta_b)$.
\proof{\underline{Proof of Step~2.}}
Notice that event $\{\wTheta = \Theta_{i_0, T}\}$ can be written as
\begin{oldequation}\label{eq:set_decomp}
    \begin{aligned}
    \{\Theta_{i_0, T} = \wTheta\} = \left(\cap_{\BFtheta_b \in \Theta_{i_0} \setminus \wTheta } \{i_T^b \neq i_0\}\right) \cap \left(\cap_{\BFtheta_b \in \wTheta \setminus\Theta_{i_0} } \{i_T^b = i_0\}\right) \cap \left(\cap_{\BFtheta_b \in \wTheta \cap \Theta_{i_0}} \{i_T^b = i_0\}\right).
    \end{aligned}
\end{oldequation}
By independence across $b$, we have
\begin{oldequation}
    \lim_{T \rightarrow \infty} -\frac{1}{T}\log \prob(\Theta_{i_0, T} = \wTheta) = \lim_{T \rightarrow \infty}\left\{\sum\nolimits_{\BFtheta_b \in \Theta_{i_0} \setminus \wTheta}-\frac{1}{T}\log \prob(i_T^b \neq i_0) + \sum\nolimits_{\BFtheta_b \in \wTheta \setminus \Theta_{i_0}}-\frac{1}{T}\log \prob(i_T^b = i_0)\right\},
\end{oldequation}
where the convergence rate of $\cap_{\BFtheta_b \in \wTheta \cap \Theta_{i_0}} \{i_T^b = i_0\}$ in~\eqref{eq:set_decomp} is negligible since $\lim_{T\rightarrow \infty} \prob(i_T^b = i_0) = 1$, if $\BFtheta_b \in \wTheta \cap \Theta_{i_0}$. Combining this with $\lim_{T \rightarrow \infty} -\frac{1}{T}\log\prob(i_T^b \neq i_0) = \min_{i\neq i_0}G_i(\BFtheta_b)$ and $\lim_{T \rightarrow \infty} -\frac{1}{T}\log\prob(i_T^b = i_0) = \widetilde{G}_{i_0}(\BFtheta_b)$ completes the proof. \halmos
\endproof
\vspace{3mm}
Thanks to Step~1, for all sufficiently large $T$ such that $i^*(\BFpi_\BFm) = i_0$ and $\BFpi_\BFm(\BFtheta_0)>1/2$, we have
\begin{oldequation}
    \begin{aligned}
        1 - \expec[\prob_{\BFpi_\BFm}(\Theta_{i_T^*, T})1\{i_T^* = i^*(\BFpi_\BFm)\}|\BFpi_\BFm]
        & = \prob(i_T^* \neq i_0) + \expec[\prob_{\BFpi_\BFm}(\Theta^c_{i_0, T})1\{i_T^* = i_0\}|\BFpi_\BFm]\\
        & = \prob(i_T^* \neq i_0) + \sum_{\wTheta \in \mathcal{A}}\expec[\prob_{\BFpi_\BFm}(\wTheta^c)1\{i_T^* = i_0, \Theta_{i_0, T} = \wTheta\}|\BFpi_\BFm]\\
        & = \prob(i_T^* \neq i_0) + \sum_{\wTheta \in \mathcal{A}}\prob_{\BFpi_{\BFm}}(\wTheta^c)\prob(\Theta_{i_0, T} = \wTheta).
    \end{aligned}
\end{oldequation}
Then, from Lemma~\ref{lem:min.property}, 
\begin{oldequation}
    \begin{aligned}
    &\lim_{T \rightarrow \infty} -\frac{1}{T}\log(1 - \expec[\prob_{\BFpi_\BFm}(\Theta_{i_T^*, T})1\{i_T^* = i^*(\BFpi_\BFm)\}|\BFpi_\BFm]) \\
    &= \min\left\{\lim_{T \rightarrow \infty} - \frac{1}{T}\log \prob(i_T^* \neq i_0), \min_{\wTheta \in \mathcal{A}}\left(\lim_{T\rightarrow \infty} - \frac{1}{T}\log \prob_{\BFpi_\BFm}(\wTheta^c)  + \lim_{T\rightarrow \infty} - \frac{1}{T}\log\prob(\Theta_{i_0, T} = \wTheta)\right)\right\}.
    \end{aligned}
\end{oldequation}
Because $\BFpi_\BFm(\BFtheta_0)>1/2$, $\{i_T^*\neq i_0\}$ is equivalent to $\{i_T^b \neq i_0\}$ at $\BFtheta_b = \BFtheta_0$. Therefore, $\lim_{T \rightarrow \infty} - \frac{1}{T}\log \prob(i_T^* \neq i_0) =\min_{i \neq i_0} G_i(\BFtheta_0)$, where $\min_{i \neq i_0} G_i(\BFtheta_0)$ is the exponential convergence rate of the PFS at $\BFtheta_0$. Combining Step~2 with Theorem~\ref{thm:LDR_input}, we can further observe
\begin{oldequation}\label{eq:LDR_first2}
    \begin{aligned}
    &\lim_{T\rightarrow \infty} - \frac{1}{T}\log \prob_{\BFpi_\BFm}(\wTheta^c)  + \lim_{T\rightarrow \infty} - \frac{1}{T}\log\prob(\Theta_{i_0, T} = \wTheta)\\
    & = \min_{\BFtheta \in \wTheta^c}\sum\nolimits_{\ell=1}^L \frac{\beta_\ell}{c_\ell}\dKL(\theta_0^\ell||\theta^\ell) + \sum\nolimits_{\BFtheta_b \in \Theta_{i_0}\setminus \wTheta} \min_{i \neq i_0} G_i(\BFtheta_b) + \sum\nolimits_{\BFtheta_b \in \wTheta\setminus \Theta_{i_0}} \widetilde{G}_{i_0}(\BFtheta_b).
    \end{aligned}
\end{oldequation}
Putting all results together~\eqref{eq:LDR_first} follows.

To show the main result, we further simplify~\eqref{eq:LDR_first2} next.  Pick $\wTheta \in \mathcal{A}$ such that $\wTheta \setminus \Theta_{i_0} \neq \emptyset$. Define $\bar{\Theta} = \wTheta \cap \Theta_{i_0}$. Since $\BFtheta_0 \in \bar{\Theta}$, we have $\bar{\Theta} \in \mathcal{A}$, $\bar{\Theta}\setminus \Theta_{i_0} = \emptyset$, $\Theta_{i_0}\setminus \wTheta = \Theta_{i_0}\setminus \bar{\Theta}$, and $\wTheta^c \subseteq \bar{\Theta}^c$. Hence, 
\begin{oldequation}\label{eq:minimal_theta}
    \begin{aligned}
        &\min_{\BFtheta \in \bar{\Theta}^c}\sum\nolimits_{\ell=1}^L \frac{\beta_\ell}{c_\ell}\dKL(\theta_0^\ell||\theta^\ell) + \sum\nolimits_{\BFtheta_b \in \Theta_{i_0}\setminus \bar{\Theta}} \min_{i \neq i^s} G_i(\BFtheta_b) + \sum\nolimits_{\BFtheta_b \in \bar{\Theta}\setminus \Theta_{i_0}} \widetilde{G}_{i_0}(\BFtheta_b) \\
        & \leq \min_{\BFtheta \in \wTheta^c}\sum\nolimits_{\ell=1}^L \frac{\beta_\ell}{c_\ell}\dKL(\theta_0^\ell||\theta^\ell) + \sum\nolimits_{\BFtheta_b \in \Theta_{i_0}\setminus \wTheta} \min_{i \neq i^s} G_i(\BFtheta_b) + \sum\nolimits_{\BFtheta_b \in \Theta\setminus \Theta_{i_0}} \widetilde{G}_{i_0}(\BFtheta_b).
    \end{aligned}
\end{oldequation}
Define $\mathcal{A}_1 = \{\wTheta \in \mathcal{A}| \wTheta \subseteq \Theta_{i_0}\}$ as a subcollection of $\mathcal{A}$. From~\eqref{eq:minimal_theta}, it follows that the minimum function over $\wTheta \in \mathcal{A}$ in~\eqref{eq:LDR_first} is attained at $\wTheta$ such that $\wTheta \subseteq \Theta_{i_0}$, i.e., $\wTheta \in \mathcal{A}_1$. Namely,
\begin{oldequation}
    \begin{aligned}
    &\min_{\wTheta \in \mathcal{A}} \left(\min_{\BFtheta \in \wTheta^c}\sum\nolimits_{\ell=1}^L \frac{\beta_\ell}{c_\ell}\dKL(\theta_0^\ell||\theta^\ell) + \sum\nolimits_{\BFtheta_b \in \Theta_{i_0}\setminus \wTheta} \min_{i \neq i_0} G_i(\BFtheta_b) + \sum\nolimits_{\BFtheta_b \in \wTheta\setminus \Theta_{i_0}} \widetilde{G}_{i_0}(\BFtheta_b)\right)\\
    &=\min_{\wTheta \in \mathcal{A}_1} \left(\min_{\BFtheta \in \wTheta^c}\sum\nolimits_{\ell=1}^L \frac{\beta_\ell}{c_\ell}\dKL(\theta_0^\ell||\theta^\ell) + \sum\nolimits_{\BFtheta_b \in \Theta_{i_0}\setminus \wTheta} \min_{i \neq i_0} G_i(\BFtheta_b)\right)\\
    \end{aligned}
\end{oldequation}
In the following, we simplify~\eqref{eq:LDR_first} by showing that we can substitute $\mathcal{A}_1$ with $\widetilde{\mathcal{A}} := \{\wTheta \in \mathcal{A}_1| |\Theta_{i_0}\setminus \wTheta| \leq 1\}$. The proof is divided into Steps~3 and~4.

\vspace{3mm}

\noindent\textbf{Step~3.} We have 
    \begin{oldequation}\label{eq:ldr_temp}
        \begin{aligned}
            & \min_{\wTheta \in \mathcal{A}_1} \left(\min_{\BFtheta \in \wTheta^c}\sum\nolimits_{\ell=1}^L \frac{\beta_\ell}{c_\ell}\dKL(\theta_0^\ell||\theta^\ell) + \sum\nolimits_{\BFtheta_b \in \Theta_{i_0}\setminus \wTheta} \min_{i \neq i_0} G_i(\BFtheta_b)\right)\\           
            & =\min\left\{\min_{\BFtheta \in \Theta_{i_0}^c}\sum_{\ell=1}^L \frac{\beta_\ell}{c_\ell}\dKL(\theta_0^\ell||\theta^\ell), \min_{\BFtheta_b \in \Theta_{i_0}\setminus \{\BFtheta_0\}} \left(\min_{\BFtheta \in \left(\Theta_{i_0}\setminus\{\BFtheta_b\}\right)^c}\sum_{\ell=1}^L \frac{\beta_\ell}{c_\ell}\dKL(\theta_0^\ell||\theta^\ell) + \min_{i \neq i_0} G_i(\BFtheta_b)\right) \right\}.
        \end{aligned}
    \end{oldequation}

\proof{\underline{Proof of Step 3.}} We will first show that
\begin{oldequation}\label{eq:target}
    \min_{\wTheta \in {\mathcal{A}}_1} \left(\min_{\BFtheta \in \wTheta^c}\sum_{\ell=1}^L \frac{\beta_\ell}{c_\ell}\dKL(\theta_0^\ell||\theta^\ell) + \sum_{\BFtheta_b \in \Theta_{i_0}\setminus \wTheta} \min_{i \neq i_0} G_i(\BFtheta_b)\right) = \min_{\wTheta \in \widetilde{\mathcal{A}}} \left(\min_{\BFtheta \in \wTheta^c}\sum_{\ell=1}^L \frac{\beta_\ell}{c_\ell}\dKL(\theta_0^\ell||\theta^\ell) + \sum_{\BFtheta_b \in \Theta_{i_0}\setminus \wTheta} \min_{i \neq i_0} G_i(\BFtheta_b)\right).
\end{oldequation}

Let us pick $\wTheta \in \mathcal{A}_1 \setminus \widetilde{\mathcal{A}}$, i.e., $|\Theta_{i_0} \setminus \wTheta| \geq 2$. For fixed $\BFbeta$, $\bar{\BFtheta}:=\argmin_{\BFtheta \in \Theta^c} \sum_{\ell=1}^L \frac{\beta_\ell}{c_\ell}\dKL(\theta_0^\ell||\theta^\ell)$ belongs to either (a) $\Theta_{i_0}^c$ or (b) $\Theta_{i_0}\setminus \wTheta$. If Case (a) holds, then pick arbitrary $\widetilde{\BFtheta} \in \Theta_{i_0}\setminus \wTheta$, and define $\bar{\Theta} = \Theta_{i_0}\setminus\{\widetilde{\BFtheta}\}$. If Case (b) is satisfied, then let $\bar{\Theta} = \Theta_{i_0}\setminus\{\bar{\BFtheta}\}$. Then, for both cases, we have $\bar{\Theta} \subseteq \wTheta$, which implies
\begin{oldequation} \label{eq:input_temp}
\begin{aligned}
    \min_{\BFtheta \in \wTheta^c}\sum\nolimits_{\ell=1}^L \frac{\beta_\ell}{c_\ell}\dKL(\theta_0^\ell||\theta^\ell) &\leq \min_{\BFtheta \in \bar{\Theta}^c}\sum\nolimits_{\ell=1}^L \frac{\beta_\ell}{c_\ell}\dKL(\theta_0^\ell||\theta^\ell),\\
     \sum\nolimits_{\BFtheta_b \in \Theta_{i_0}\setminus \bar{\Theta}} \min_{i \neq i_0} G_i(\BFtheta_b) &\leq \sum\nolimits_{\BFtheta_b \in \Theta_{i_0}\setminus {\wTheta}} \min_{i \neq i_0} G_i(\BFtheta_b).
\end{aligned}
\end{oldequation}
From the definition of $\bar{\Theta}$, we have $\bar{\BFtheta} \in \bar{\Theta} \subseteq \mathcal{A}_1$. Hence, the equality holds for~\eqref{eq:input_temp}. This leads us to the following inequality.
\begin{oldequation}\label{eq:temp2}
    \min_{\BFtheta \in \bar{\Theta}^c}\sum_{\ell=1}^L \frac{\beta_\ell}{c_\ell}\dKL(\theta_0^\ell||\theta^\ell) + \sum_{\BFtheta_b \in \Theta_{i_0}\setminus \bar{\Theta}} \min_{i \neq i_0} G_i(\BFtheta_b)\leq \min_{\BFtheta \in \wTheta^c}\sum_{\ell=1}^L \frac{\beta_\ell}{c_\ell}\dKL(\theta_0^\ell||\theta^\ell) + \sum_{\BFtheta_b \in \Theta_{i_0}\setminus \wTheta} \min_{i \neq i_0} G_i(\BFtheta_b).
\end{oldequation}
Combining \eqref{eq:temp2} with $\bar{\Theta} \in \mathcal{A}_1$ yields that the minimum over $\widetilde{\mathcal{A}}$ in~\eqref{eq:target} is always attained at some $\bar{\Theta} \in \mathcal{A}_1$, which proves~\eqref{eq:target}. 

 Now it suffices to show that
 \begin{oldequation}\label{eq:inner_tot}
    \begin{aligned}
     &\min_{\wTheta \in \widetilde{\mathcal{A}}} \left(\min_{\BFtheta \in \wTheta^c}\sum_{\ell=1}^L \frac{\beta_\ell}{c_\ell}\dKL(\theta_0^\ell||\theta^\ell) + \sum_{\BFtheta_b \in \Theta_{i_0}\setminus \wTheta} \min_{i \neq i_0} G_i(\BFtheta_b)\right) \\
     & = \min\left\{\min_{\BFtheta \in \Theta_{i_0}^c}\sum_{\ell=1}^L \frac{\beta_\ell}{c_\ell}\dKL(\theta_0^\ell||\theta^\ell), \min_{\BFtheta_b \in \Theta_{i_0}\setminus \{\BFtheta_0\}} \left(\min_{\BFtheta \in \left(\Theta_{i_0}\setminus\{\BFtheta_b\}\right)^c}\sum_{\ell=1}^L \frac{\beta_\ell}{c_\ell}\dKL(\theta_0^\ell||\theta^\ell) + \min_{i \neq i_0} G_i(\BFtheta_b)\right) \right\}.
    \end{aligned}
 \end{oldequation}
Let $\mathcal{B}_i := \{\wTheta \in \widetilde{\mathcal{A}}| |\wTheta_{i_0} \setminus \Theta| = i\}$ for $i = 0, 1$. Then, we have $\widetilde{\mathcal{A}} = \mathcal{B}_0 \cup \mathcal{B}_1$. Because $\mathcal{B}_0 = \{\Theta_{i_0}\}$, we have 
\begin{oldequation}\label{eq:inner_1}
    \min_{\wTheta \in \mathcal{B}_0} \left(\min_{\BFtheta \in \wTheta^c}\sum_{\ell=1}^L \frac{\beta_\ell}{c_\ell}\dKL(\theta_0^\ell||\theta^\ell) + \sum_{\BFtheta_b \in \Theta_{i_0}\setminus \Theta} \min_{i \neq i_0} G_i(\BFtheta_b)\right) = \min_{\BFtheta \in \Theta_{i_0}^c}\sum_{\ell=1}^L \frac{\beta_\ell}{c_\ell}\dKL(\theta_0^\ell||\theta^\ell).
\end{oldequation}
Since each $\wTheta \in \mathcal{B}$ can be written as $\wTheta = \Theta_{i_0}\setminus \{\BFtheta_b\}$ for each $\BFtheta_b \in \Theta_{i_0} \setminus\{\BFtheta_0\}$, 
\begin{oldequation}\label{eq:inner_2}
    \begin{aligned}
    &\min\nolimits_{\wTheta \in \mathcal{B}_1} \left(\min\nolimits_{\BFtheta \in \wTheta^c}\sum\nolimits_{\ell=1}^L \frac{\beta_\ell}{c_\ell}\dKL(\theta_0^\ell||\theta^\ell) + \sum\nolimits_{\BFtheta_b \in \Theta_{i_0}\setminus \Theta} \min_{i \neq i_0} G_i(\BFtheta_b)\right)\\
    &= \min\nolimits_{\BFtheta_b \in \Theta_{i_0}\setminus\{\BFtheta_0\}}\left(\min\nolimits_{\BFtheta \in \left(\Theta_{i_0}\setminus\{\BFtheta_b\}\right)^c}\sum\nolimits_{\ell=1}^L \frac{\beta_\ell}{c_\ell}\dKL(\theta_0^\ell||\theta^\ell) + \min\nolimits_{i \neq i_0} G_i(\BFtheta_b)\right).
    \end{aligned}
\end{oldequation}
Combining~\eqref{eq:inner_1} with~\eqref{eq:inner_2} yields~\eqref{eq:inner_tot}, which completes the proof. \halmos

\vspace{3mm}
The following step further simplifies~\eqref{eq:ldr_temp}.

\noindent{\textbf{Step~4.}} We have 
    \begin{oldequation}\label{eq:ldr_temp2}
        \begin{aligned}          
            & \min\left\{\min_{\BFtheta \in \Theta_{i_0}^c}\sum_{\ell=1}^L \frac{\beta_\ell}{c_\ell}\dKL(\theta_0^\ell||\theta^\ell), \min_{\BFtheta_b \in \Theta_{i_0}\setminus \{\BFtheta_0\}} \left(\min_{\BFtheta \in \left(\Theta_{i_0}\setminus\{\BFtheta_b\}\right)^c}\sum_{\ell=1}^L \frac{\beta_\ell}{c_\ell}\dKL(\theta_0^\ell||\theta^\ell) + \min_{i \neq i_0} G_i(\BFtheta_b)\right) \right\}\\
            & = \min\left\{\min_{\BFtheta \in \Theta_{i_0}^c}\sum_{\ell=1}^L \frac{\beta_\ell}{c_\ell}\dKL(\theta_0^\ell||\theta^\ell),\min_{\BFtheta_b \in \Theta_{i_0}\setminus \{\BFtheta_0\}} \left(\sum_{\ell=1}^L \frac{\beta_\ell}{c_\ell}\dKL(\theta_0^\ell||\theta^\ell_s) + \min_{i \neq i_0} G_i(\BFtheta_b)\right)\right\}.
        \end{aligned}
    \end{oldequation}

\proof{\underline{Proof of Step~4.}}
    As the first inner minimization function does not depend on $\BFtheta_b$, we can switch the order of the minimization functions as {\small
    \begin{oldequation}\label{eq:LHS}
        \begin{aligned}
            &\min\left\{\min_{\BFtheta \in \Theta_{i_0}^c}\sum_{\ell=1}^L \frac{\beta_\ell}{c_\ell}\dKL(\theta_0^\ell||\theta^\ell), \min_{\BFtheta_b \in \Theta_{i_0}\setminus \{\BFtheta_0\}} \left(\min_{\BFtheta \in \left(\Theta_{i_0}\setminus\{\BFtheta_b\}\right)^c}\sum_{\ell=1}^L \frac{\beta_\ell}{c_\ell}\dKL(\theta_0^\ell||\theta^\ell) + \min_{i \neq i_0} G_i(\BFtheta_b)\right) \right\}\\
            &= \min_{\BFtheta_b \in \Theta_{i_0}\setminus \{\BFtheta_0\}}\left\{\min\left\{\min_{\BFtheta \in \Theta_{i_0}^c}\sum_{\ell=1}^L \frac{\beta_\ell}{c_\ell}\dKL(\theta_0^\ell||\theta^\ell), \min_{\BFtheta \in \left(\Theta_{i_0}\setminus\{\BFtheta_b\}\right)^c}\sum_{\ell=1}^L \frac{\beta_\ell}{c_\ell}\dKL(\theta_0^\ell||\theta^\ell) + \min_{i \neq i_0} G_i(\BFtheta_b)\right\}\right\}.
        \end{aligned}
    \end{oldequation}}
    Similarly, we have
    \begin{oldequation}\label{eq:RHS}
        \begin{aligned}
            &\min\left\{\min_{\BFtheta \in \Theta_{i_0}^c}\sum\nolimits_{\ell=1}^L \frac{\beta_\ell}{c_\ell}\dKL(\theta_0^\ell||\theta^\ell),\min_{\BFtheta_b \in \Theta_{i_0}\setminus \{\BFtheta_0\}} \left(\sum\nolimits_{\ell=1}^L \frac{\beta_\ell}{c_\ell}\dKL(\theta_0^\ell||\theta^\ell_s) + \min_{i \neq i_0} G_i(\BFtheta_b)\right)\right\}\\
            &= \min_{\BFtheta_b \in \Theta_{i_0}\setminus \{\BFtheta_0\}}\left\{\min\left\{\min_{\BFtheta \in \Theta_{i_0}^c}\sum\nolimits_{\ell=1}^L \frac{\beta_\ell}{c_\ell}\dKL(\theta_0^\ell||\theta^\ell), \sum\nolimits_{\ell=1}^L \frac{\beta_\ell}{c_\ell}\dKL(\theta_0^\ell||\theta^\ell_s) + \min_{i \neq i_0} G_i(\BFtheta_b)\right\}\right\}.
        \end{aligned}
    \end{oldequation}
    Combining~\eqref{eq:LHS} with~\eqref{eq:RHS}, it suffices to show that for each $\BFtheta_b \in \Theta_{i_0}\setminus \{\BFtheta_0\}$,
    \begin{oldequation}\label{eq:temp3}
        \begin{aligned}
            &\min\left\{\min_{\BFtheta \in \Theta_{i_0}^c}\sum\nolimits_{\ell=1}^L \frac{\beta_\ell}{c_\ell}\dKL(\theta_0^\ell||\theta^\ell), \min_{\BFtheta \in \left(\Theta_{i_0}\setminus\{\BFtheta_b\}\right)^c}\sum\nolimits_{\ell=1}^L \frac{\beta_\ell}{c_\ell}\dKL(\theta_0^\ell||\theta^\ell) + \min_{i \neq i_0} G_i(\BFtheta_b)\right\}\\
            & = \min\left\{\min_{\BFtheta \in \Theta_{i_0}^c}\sum\nolimits_{\ell=1}^L \frac{\beta_\ell}{c_\ell}\dKL(\theta_0^\ell||\theta^\ell), \sum\nolimits_{\ell=1}^L \frac{\beta_\ell}{c_\ell}\dKL(\theta_0^\ell||\theta^\ell_s) + \min_{i \neq i_0} G_i(\BFtheta_b)\right\}.
        \end{aligned}
    \end{oldequation}
    Let $\widetilde{\BFtheta} = \argmin_{\BFtheta \in \left(\Theta_{i_0}\setminus\{\BFtheta_b\}\right)^c}\sum_{\ell=1}^L \frac{\beta_\ell}{c_\ell}\dKL(\theta_0^\ell||\theta^\ell)$. There are two possibilities a) $\widetilde{\BFtheta} \in \Theta_{i_0}^c$ or b) $\widetilde{\BFtheta} = \BFtheta_b$. In Case a), we have $\sum_{\ell=1}^L \frac{\beta_\ell}{c_\ell}\dKL(\theta_0^\ell||\theta^\ell_s) > \min_{\BFtheta \in \Theta_{i_0}^c}\sum_{\ell=1}^L \frac{\beta_\ell}{c_\ell}\dKL(\theta_0^\ell||\theta^\ell)$. Therefore, both sides of~\eqref{eq:temp3} attain the minimum, $\min_{\BFtheta \in \Theta_{i_0}^c}\sum_{\ell=1}^L \frac{\beta_\ell}{c_\ell}\dKL(\theta_0^\ell||\theta^\ell)$, and the equality holds. For Case b),~\eqref{eq:temp3} holds straightforwardly. This concludes the proof.    
\endproof

To summarize, the LDR of the posterior preference can be characterized by
\begin{oldequation}
    \begin{aligned}
        &\lim_{T \rightarrow \infty} -\frac{1}{T}\log(1 - \expec[\prob_{\BFpi_m}(\Theta_{i_{T}^*, T})1\{i_{T}^* = i^*(\BFpi_\BFm)\}|\BFpi_\BFm]) \\
        & = \min\left\{\min_{i \neq i_0} G_i(\BFtheta_0), \min_{\BFtheta \in \Theta_{i_0}^c}\sum\nolimits_{\ell=1}^L \frac{\beta_\ell}{c_\ell}\dKL(\theta_0^\ell||\theta^\ell), \min_{\BFtheta_b \in \Theta_{i_0}\setminus \{\BFtheta_0\}} \left(\sum\nolimits_{\ell=1}^L \frac{\beta_\ell}{c_\ell}\dKL(\theta_0^\ell||\theta^\ell_s) + \min_{i \neq i_0} G_i(\BFtheta_b)\right)\right\}\\
        & = \min\left\{\min_{\BFtheta \in \Theta_{i_0}^c}\sum\nolimits_{\ell=1}^L \frac{\beta_\ell}{c_\ell}\dKL(\theta_0^\ell||\theta^\ell), \min_{\BFtheta_b \in \Theta_{i_0}} \left(\sum\nolimits_{\ell=1}^L \frac{\beta_\ell}{c_\ell}\dKL(\theta_0^\ell||\theta^\ell_s) + \min_{i \neq i_0} G_i(\BFtheta_b)\right)\right\}.
    \end{aligned}
\end{oldequation}
The last equality holds from noticing $\dKL(\theta_0^\ell||\theta_b^\ell) = 0$ when $\BFtheta_b = \BFtheta_0$.  \qed
\endproof

\section{Proof of Statements in Section~\ref{sec:simulation_error}}

\subsection{Proof of Proposition~\ref{prop:opt_gap}}
     Program~\eqref{opt:reformulation_eps} can be rewritten as 
\begin{oldequation}\label{opt:V_eps}
    \begin{aligned}
       V(\epsilon) = \max \quad & C \\
       \textrm{subject to} \quad &  C \leq \alpha(\BFtheta_b)G^*(\BFtheta_b) + \sum\nolimits_{\ell=1}^L \frac{\beta_\ell}{c_\ell} \dKL(\theta_0^\ell||\theta^\ell_s), \forall \theta_b \in \Theta_{i_0},\\
        & C \leq \sum\nolimits_{\ell=1}^L \frac{\beta_\ell}{c_\ell} \dKL(\theta_0^\ell||\theta^\ell_s), \forall \theta_b \in \Theta^c_{i_0},\\
       &  \mathbf{1}_B^\top\BFalpha + \mathbf{1}_L^\top \BFbeta = 1,\;\; \alpha(\BFtheta_b) \geq \epsilon, \beta_\ell \geq \epsilon, \;\; \forall 1\leq b \leq B, 1\leq \ell\leq L. 
    \end{aligned}
\end{oldequation}

    Since~\eqref{opt:V_eps} has a smaller constraint on $\BFalpha$ and $\BFbeta$, we have $V(0)\geq V(\epsilon)$, which proves the first inequality. For the second inequality, let us define $\alpha^\epsilon(\BFtheta_b)$ and $\BFbeta^\epsilon = (\beta_\ell^\epsilon)_{1\leq \ell\leq L}$ as
    \begin{oldequation}
        \alpha^\epsilon(\BFtheta_b) = \frac{\alpha(\BFtheta_b) - \epsilon}{1 - \epsilon(B + L)} \;\; \text{and} \;\;  \beta^\epsilon_\ell = \frac{\beta_\ell - \epsilon}{1 - \epsilon(B + L)}.
    \end{oldequation}
    Then, after some algebras, Program~\eqref{opt:V_eps} can be written as
    \begin{oldequation}\label{opt:V_eps_temp}
    \begin{aligned}
       V(\epsilon) = \max \quad & C \\
       \textrm{s.t.} \quad &  \frac{C - \epsilon\left(G^*(\BFtheta_b) + \sum_{\ell=1}^L \dKL(\theta_0^\ell||\theta_b^\ell)/c_\ell\right)}{{1 - \epsilon(B + L)}} \leq \alpha^\epsilon(\BFtheta_b)G^*(\BFtheta_b) + \sum\nolimits_{\ell=1}^L \frac{\beta^\epsilon_\ell}{c_\ell} \dKL(\theta_0^\ell||\theta^\ell_b), \forall \theta_b \in \Theta_{i_0},\\
        &\frac{C - \epsilon\left(\sum\nolimits_{\ell=1}^L \dKL(\theta_0^\ell||\theta_b^\ell)\right)/c_\ell}{{1 - \epsilon(B + L)}} \leq \sum\nolimits_{\ell=1}^L \frac{\beta^\epsilon_\ell}{c_\ell} \dKL(\theta_0^\ell||\theta^\ell_b), \forall \theta_b \in \Theta^c_{i_0},\\
       &  \mathbf{1}_B^\top \BFalpha^\epsilon(\BFtheta_b) + \mathbf{1}_L^\top \BFbeta^\epsilon = 1, \;\;\alpha^\epsilon(\BFtheta_b) \geq 0, \beta^\epsilon_\ell \geq 0, \;\; \forall 1\leq b \leq B, 1\leq \ell\leq L. 
    \end{aligned}
    \end{oldequation}
Let $C_0 = \min\left\{\min_{\BFtheta_b \in \Theta_{i_0}}\left(G^*(\BFtheta_b) + \sum_{\ell=1}^L \dKL(\theta_0^\ell||\theta_b^\ell)/c_\ell\right), \min_{\BFtheta_b \in \Theta_{i_0}^c} \sum_{\ell=1}^L \dKL(\theta_0^\ell||\theta_b^\ell)/c_\ell\right\}$. Then, one can see that $V(\epsilon)$ is bounded from below by $\bar{V}(\epsilon)$, which is defined as

    \begin{oldequation}\label{opt:V_eps_var}
    \begin{aligned}
       \bar{V}(\epsilon) = \max \quad & C \\
       \textrm{s.t.} \quad &  \frac{C - \epsilon C_0}{{1 - \epsilon(B + L)}} \leq \alpha^\epsilon(\BFtheta_b)G^*(\BFtheta_b) + \sum\nolimits_{\ell=1}^L \frac{\beta^\epsilon_\ell}{c_\ell} \dKL(\theta_0^\ell||\theta^\ell_b), \forall \theta_b \in \Theta_{i_0},\\
        &\frac{C - \epsilon C_0}{{1 - \epsilon(B + L)}} \leq \sum\nolimits_{\ell=1}^L \frac{\beta^\epsilon_\ell}{c_\ell}\dKL(\theta_0^\ell||\theta^\ell_b), \forall \theta_b \in \Theta^c_{i_0},\\
       &  \mathbf{1}_B^\top \BFalpha^\epsilon(\BFtheta_b) + \mathbf{1}_L^\top \BFbeta^\epsilon = 1, \;\; \alpha^\epsilon(\BFtheta_b) \geq 0, \beta^\epsilon_\ell \geq 0, \;\; \forall 1\leq b \leq B, 1\leq \ell\leq L. 
    \end{aligned}
    \end{oldequation}
From the definition of $V(0)$, we have $\bar{V}(\epsilon) = (1 - \epsilon(B + L))V(0) + \epsilon C_0$. Consequently, we obtain that
\begin{oldequation}
    V(0) - V(\epsilon) \leq V(0) - \bar{V}(\epsilon) = \epsilon\left((B+L)V(0) - C_0\right) \leq \epsilon(B+L)V(0).
\end{oldequation}
Dividing $V(0)$ on both sides the second inequality follows. \qed
\endproof

\subsection{Proof of Theorem~\ref{thm:convergence}}
We first define $N^s_{i}(\BFtheta_b)$  
and $m_{\ell,s}$ to be the cumulative numbers of replications made at $(i, \BFtheta_b)$ and data collected from the $\ell$-th input, respectively, after the
$s$-th batch allocation. As $T\to\infty,$ we have $s\to\infty.$

    From Steps~\ref{line:input_collection}--\ref{line:simulation} of Algorithm~\ref{alg:plain_OSAR}, we have $\hat{n}_{s_1}(\BFtheta_b) \sim \text{Bin}(\Delta, \hat{\alpha}_{s_1}(\BFtheta_b))$ and $\hat{m}_{\ell, s_1} \sim \text{Bin}(\Delta, \hat{\beta}_{\ell, s_1}/c_\ell)$ for each $1\leq s_1\leq s$. We will proceed by the following steps.
    
    \noindent\textbf{Step 1}. With probability one, $\lim_{s \rightarrow \infty} \sum_{i=1}^k N_{i}^s(\BFtheta_b) = \infty$ and $\lim_{s \rightarrow \infty} m_{\ell, s} = \infty$ for all $1\leq b\leq B$ and $1\leq \ell \leq L$. Moreover,  $\lim_{T \rightarrow \infty} N_{i, T}(\BFtheta_b) = \infty$ for all $1\leq i\leq k$, $1\leq b\leq B$.

     \proof{\underline{Proof of Step 1}.} Since $\hat{\alpha}_{s_1}(\BFtheta_b) \geq \epsilon$ and $\hat{\beta}_{\ell, s_1} \geq \epsilon$ for all $1\leq b \leq B$ and $1\leq \ell\leq L$, we have $\lim_{t \rightarrow \infty}\sum_{s_1=1}^s \hat{\alpha}_{s_1}(\BFtheta_b) = \infty$ and $\lim_{s \rightarrow \infty}\sum_{s_1=1}^s \hat{\beta}_{\ell, s_1} = \infty$. From Lemma~\ref{lem:second_Borel}, one can show that $\lim_{s \rightarrow \infty} \sum_{i=1}^k N_{i}^s(\BFtheta_b) = \lim_{s\rightarrow \infty} \sum_{i=1}^k \sum_{s_1=1}^s \hat{n}_{s_1}(\BFtheta_b) = \infty$ and $\lim_{s \rightarrow \infty} m_{\ell, s} = \lim_{s\rightarrow \infty} \sum_{s_1=1}^s \hat{m}_{\ell, s_1} = \infty$, which means that we run infinitely many simulation replications at all $\BFtheta_b\in\Theta$. Consequently, $\lim_{T \rightarrow \infty} N_{i, T}(\BFtheta_b) = \infty$ because the R\&S subroutine of Algorithm~\ref{alg:plain_OSAR} simulates all solutions at each $\BFtheta_b$ infinitely often w.~p.~1. \halmos
    \endproof
    \vspace{3mm}

    Combining Step~1 with that the subroutine algorithm achieves the rate-optimal static allocation yields that $\lim_{T \rightarrow \infty} G_{i, T}(\BFtheta_b) = G^*(\BFtheta_b)$ holds almost surely for all $1\leq b\leq B$.

    \vspace{3mm}
    
    \noindent\textbf{Step 2}. $\lim_{s \rightarrow \infty} \hat{\textsf{D}}^\text{KL}(\hat{\theta}^\ell|| \theta_b^\ell) = \dKL(\theta_0^\ell||\theta_b^\ell)$ almost surely.

    \proof{\underline{Proof of Step 2}.} Since $\lim_{s \rightarrow \infty} m_{\ell, s} = \infty$ holds almost surely for all $1\leq \ell \leq L$,  $\lim_{s\rightarrow \infty} \hat{\theta}^\ell = \theta_0^\ell$ a.s.~from Lemma~\ref{lem:consistency.MAP}. Since $\Theta$ is finite, for all sufficiently large $s$, $\hat{\theta}^\ell = \theta_0^\ell$. Hence, $\hat{\textsf{D}}^\text{KL}(\hat{\theta}^\ell|| \theta_b^\ell) = \hat{\textsf{D}}^\text{KL}({\theta}^\ell_0|| \theta_b^\ell)$ holds for all sufficiently large $s$. Combining this with Glivenko-Cantelli property in Assumption~\ref{asmp:GC_class}(c), we have $\lim_{s \rightarrow \infty}\hat{\textsf{D}}^\text{KL}(\hat{\theta}^\ell|| \theta_b^\ell) = \dKL(\theta_0^\ell||\theta_b^\ell)$ a.s., as desired. \halmos
    
    \endproof

    \vspace{3mm}
    \noindent\textbf{Step 3}. For $\hat{\BFalpha}^s$ and $\hat{\BFbeta}^s$ in~\eqref{opt:plug_in_eps}, we have $\lim_{s \rightarrow \infty} \hat{\BFalpha}^s = \BFalpha^*$ and $\lim_{s \rightarrow \infty} \hat{\BFbeta}^s = \BFbeta^*$ almost surely.

    \proof{\underline{Proof of Step 3}.} Recall that Algorithm~1 solves the plug-in Program~\eqref{opt:plug_in_eps} with coefficients $G_{i, T}(\BFtheta_b)$ and $\hat{\textsf{D}}^\text{KL}(\hat{\theta}^\ell|| \theta_b^\ell)$ at each $s$ to obtain $\hat\BFalpha^s$ and $\hat\BFbeta^s$. Combining Step~2 with the strong consistency of $G_{i, T}(\BFtheta_b)$ shows that each plug-in coefficient  in~\eqref{opt:plug_in_eps} converges to the corresponding quantity in~\eqref{opt:reformulation_eps} almost surely. Furthermore, $i^*_{T} = i_0$ and $\Theta_{i^*_{T}, T} = \Theta_{i_0}$  for all sufficiently large $T$ w.~p.~1. 
    
    Next, it remains to show that the optimal solution of~\eqref{opt:plug_in_eps} converges to that of~\eqref{opt:reformulation_eps}. The objective function of~\eqref{opt:plug_in_eps} is continuous and concave. Thus, from Proposition 24 and Theorem 25 in~\cite{shapiro2003monte_ec}, the convergence of the optimal solution follows.
    \halmos
    \endproof

    \vspace{3mm}

    \noindent\textbf{Step 4}. We  have $\lim_{s\rightarrow \infty} \BFalpha^s = \BFalpha^*$ and $\lim_{s \rightarrow \infty} \BFbeta^s = \BFbeta^*$. 

    \proof{\underline{Proof of Step 4}.} For ease of exposition, we assume $m_0 = 0$ and $n_0 = 0$; the extension to nonzero $m_0$ and $n_0$ case is straightforward.
    Then, for all $1\leq b \leq B$ and $1 \leq \ell \leq L$,
    \begin{oldequation}
       \alpha_s(\BFtheta_b) = (s\Delta)^{-1}\sum\nolimits_{s_1 = 1}^s \hat{n}_{s_1}(\BFtheta_b) \;\; \text{and} \;\; \beta_{\ell, s} = (s\Delta)^{-1}\sum\nolimits_{s_1 = 1}^s \hat{m}_{\ell, s_1}.
    \end{oldequation}
    From Lemma~\ref{lem:second_Borel} in Section~\ref{ec:aux}, we can show that
    \begin{oldequation}\label{eq:borel.cantelli}
        \frac{\sum_{s_1=1}^s  \hat{n}_{s_1}(\BFtheta_b)}{\Delta \sum_{s_1=1}^s \hat{\alpha}_{s_1}(\BFtheta_b)} \rightarrow 1 \;\; \text{and} \;\; \frac{\sum_{s_1=1}^s  \hat{m}_{\ell, s_1}}{\Delta\sum_{s_1=1}^s \hat{\beta}_{\ell, s_1}}\rightarrow 1.
    \end{oldequation}
    From the convergence of the Cesaro's mean, we can verify that for all $1\leq b\leq B$ and $1\leq \ell \leq L$, 
    \begin{oldequation}\label{eq:cesaro}
        s^{-1} \sum\nolimits_{s_1=1}^s \hat{\alpha}_{s_1}(\BFtheta_b)\rightarrow \alpha^*(\BFtheta_b) \;\; \text{and} \;\; s^{-1}{\sum\nolimits_{s_1=1}^s \hat{\beta}_{\ell, s_1}} \rightarrow \beta^*_{\ell}.
    \end{oldequation}
    Combining~\eqref{eq:borel.cantelli} with~\eqref{eq:cesaro} completes the proof. \qed
    \endproof

\section{Proof of Statements in Sections~\ref{sec:seq.sampling} and~\ref{sec:continuous.case.extension}}

\subsection{Proof of Theorem~\ref{thm:convergence_OSAR+}}

This proof proceeds in a similar manner as that of Theorem~\ref{thm:convergence}. Steps 1,~2, and 4 can be proved without modification. To see Step 3, it suffices to show the convergence of $G_{i, T}(\BFtheta_b)$ to $G^*(\BFtheta_b)$ in Algorithm~\ref{alg:OSAR+KRR}. From the closed-form expression of $\hat{\BFmu}_{i, T}(\theta_b)$ in~\eqref{eq:KRR}, one can confirm $\lim_{T \rightarrow \infty} \hat{\BFmu}_{i, T} = (\eta_i(\BFtheta_b))_{1\leq b\leq B}$ a.s. The rest follows from Step~3 in the proof of Theorem~\ref{thm:convergence}.
\qed

\subsection{Proof of Proposition~\ref{prop:revised_KRR}}
The optimal solution to~\eqref{eq:GNLLL_conti} can be written as $g_{i, T}(x) - \gamma_i = \sum_{b=1}^{B+1} c_b K_i(\BFtheta_b, x)$. Then, we have 
    \begin{oldequation}
        [(g_{i,T}(\BFtheta_b))_{1\leq b\leq B}^\top \;\; (g_{i,T}(\hBFtheta_{i, r}))_{1\leq r\leq N_{i, T}(\BFtheta)}^\top ]^\top - \gamma_i\BFone_{B+N_{i, T}(\hBFtheta)} = \widetilde{\BFK}_{i, T}^\top \BFc_{i, T},
    \end{oldequation}
    where $\BFc_{i, T} = (c_b)_{1\leq b\leq B+1}$ is a $(B+1)$-dimensional column vector. Then,~\eqref{eq:GNLLL_conti} can be rewritten as
    \begin{oldequation}\label{eq:opt_cit}
        \BFc_{i, T} = \argmin\nolimits_{\BFc}\left\{\frac{1}{2}(\BFx - \widetilde{\BFK}_{i, T}^\top \BFc)^\top \widetilde{\Sigma}^{-1}_{i, T}(\BFx - \widetilde{\BFK}_{i, T}^\top \BFc) + \frac{\kappa}{2}\BFc^\top \bar{\BFK}_{i, T} \BFc\right\}.
    \end{oldequation}
    Here $\BFx = [\BFmu_{i, T}^\top \;\; \BFY_i^\top]^\top - \gamma_i \BFone_{B + N_{i, T}(\hBFtheta)}$. After some algebra,~\eqref{eq:opt_cit} is reformulated as
    \begin{oldequation}\label{eq:opt_cit2}
        \BFc_{i, T} = \argmin\nolimits_{\BFc}\left\{\frac{1}{2}\BFc^\top(\widetilde{\BFK}_i\widetilde{\Sigma}_{i, T}^{-1}\widetilde{\BFK}_{i, T}^\top + \kappa\bar{\BFK}_{i, T})\BFc - (\widetilde{\BFK}_{i, T}\widetilde{\Sigma}_{i, T}^{-1}\BFx)^\top \BFc\right\}.
    \end{oldequation}
    Let $\BFA_{i, T} = \widetilde{\BFK}_{i, T}\widetilde{\Sigma}_{i, T}^{-1}\widetilde{\BFK}_{i, T}^\top + \kappa\bar{\BFK}_{i, T}$. Since $\BFA_{i,T}$ is symmetric and positive semidefinite (psd), its eigendecomposition can be written as
    \begin{oldequation}\label{eq:eig_decomp}
        \BFA_{i, T} = \BFV \BFD \BFV^\top
    \end{oldequation}
    for orthogonal matrix $\BFV$ and $\BFD = \text{diag}(\{\sigma_b\}_{1\leq b\leq B+1})$, where $\{\sigma_b\}_{1\leq b\leq B+1}$ is the set of the eigenvalues of $\BFA_{i, T}$. From the positive semi-definiteness of $\BFA_{i, T}$, we have $\sigma_b \geq 0$ for all $1 \leq b\leq B+1$. Without loss of generality, assume $\sigma_b > 0$ for all $1\leq b\leq R$ and $\sigma_b = 0$ for all $R+1\leq b\leq B+1$ for some nonnegative integer $R$. In this case,~\eqref{eq:eig_decomp} can be regarded as the singular value decomposition of $\BFA_{i, T}$, and the pseudoinverse of $\BFA_{i, T}$, $\BFA_{i, T}^\dagger$, is given as $\BFA_{i, T}^\dagger = \BFV\BFD^\dagger\BFV^\top$, where $\BFD^\dagger = \text{diag}(\{\sigma_1^{-1}, \ldots, \sigma_R^{-1}, 0, \ldots, 0\})$

    Letting $\BFd = \BFV^\top \BFc$,~\eqref{eq:opt_cit2} can be written as
    \begin{oldequation}\label{eq:opt_cit3}
        \BFc_{i, T} = \BFV\BFd^* \;\;\text{where}\;\; \BFd^* = \argmin\nolimits_{\BFd}\left\{\frac{1}{2}\BFd^\top \BFD\BFd - (\BFV^\top\widetilde{\BFK}_{i, T}\widetilde{\Sigma}_{i, T}^{-1}\BFx)^\top \BFd\right\}.
    \end{oldequation}
    Let the $b$th elements of  $\BFd$ and $\BFV^\top\widetilde{\BFK}_{i, T}\widetilde{\Sigma}_{i, T}^{-1}\BFx$ be $d_b$ and $z_b$, respectively, for $b=1\leq b\leq B+1$. We show that $z_b = 0$  for all $b \geq R+1$ for which it suffices to have $\widetilde{\BFK}_{i, T}^\top \BFV \BFe_b = \BFzero_{(B+1)\times 1}$ for all $b \geq R+1$. To see this, observe that $\BFe_b^\top \BFV^\top\BFA_{i, T}\BFV\BFe_b = \BFe_b^\top \BFD\BFe_b = 0$ holds for all $b \geq R+1$. Thus, 
    $0 = \BFe_b^\top \BFV^\top\BFA_{i, T}\BFV\BFe_b  = (\BFV\BFe_b)^\top (\widetilde{\BFK}_{i, T}\widetilde{\Sigma}_{i, T}^{-1}\widetilde{\BFK}_{i, T}^\top + \kappa\bar{\BFK}_{i, T})\BFV\BFe_b\geq (\widetilde{\BFK}_{i, T}\BFV\BFe_b)^\top \widetilde{\Sigma}^{-1}_{i, T}(\widetilde{\BFK}_{i, T}\BFV\BFe_b) \geq 0$, where the first inequality follows because $\bar{\BFK}_{i, T}$ is a psd matrix. Therefore,  $\widetilde{\BFK}_{i, T}\BFV\BFe_b = \BFzero_{(B+1)\times 1}$ 
    as desired, which implies
    \begin{oldequation}
        \BFd^* = \argmin\nolimits_{d_b, 1\leq b\leq R} \left\{\sum\nolimits_{b=1}^R\left(\frac{\sigma_b}{2}d_b^2 - z_bd_b\right)\right\}
    \end{oldequation}
    From the first-order condition, the optimal solution is attained at $d_b = z_b/\sigma_b$. Consequently, $\BFd^* = \BFD^\dagger \BFV^\top\widetilde{\BFK}_{i, T}\widetilde{\Sigma}_{i, T}^{-1}\BFx$ and $\BFc_{i,T} = \BFV\BFD^\dagger \BFV^\top\widetilde{\BFK}_{i, T}\widetilde{\Sigma}_{i, T}^{-1}\BFx = \BFA_{i, T}^\dagger \widetilde{\BFK}_{i, T}\widetilde{\Sigma}_{i, T}^{-1}\BFx$, as desired. \qed 
\endproof

\vspace{3mm}

\subsection{Proof of Theorem~\ref{thm:KRR_conti_consistency}}
In this proof, we introduce notation $\Theta_{\text{a.s.}}$ to compare two positive random sequences $\{a_T\}$ and $\{b_T\}$. Specifically, $b_T = \Tas(a_T)$ means that $0 < \liminf_{T \rightarrow\infty} {b_T}/{a_T} \leq \limsup_{T \rightarrow\infty} {b_T}/{a_T} < \infty$ holds almost surely. 

We proceed with the following steps.    
\vspace{3mm}
    
    \noindent{\bf Step 1}. We have $N_{i, T}(\BFtheta_b) \xrightarrow{a.s.} \infty$ almost surely and $N_{i, T}(\BFtheta_b) = \Tas(T)$, for all $1\leq i\leq k$ and $1\leq b\leq B$. 

    \proof{\underline{Proof of Step~1}}. Proceeding similarly as in Step 1 of the proof of Theorem 3, we have $\lim_{T \rightarrow \infty} N_{i, T}(\BFtheta_b) = \infty$ for all $1\leq i\leq k, 1\leq b\leq B+1$. 
    
    To show $N_{i, T}(\BFtheta_b) = \Tas(T)$, first note that $N_{i, T}(\BFtheta_b)\leq T$ as the number of simulation replications at $(i,\BFtheta_b)$ cannot exceed the budget, $T$, which implies that  $\limsup_{T\rightarrow \infty}\frac{N_{i, T}(\BFtheta_b)}{T} <\infty$ for all $1 \leq i\leq k$.  Since $\alpha(\BFtheta_b) \geq \epsilon$ for all $1\leq b\leq B$, we have $\epsilon \leq \liminf_{T \rightarrow \infty}\frac{\sum\nolimits_{i=1}^k N_{i, T}(\BFtheta_b)}{T}$. Therefore, there exists $1\leq j_1 \leq k$ such that $\liminf_{T\rightarrow \infty} \frac{N_{j_1, T}(\BFtheta_b)}{T} > 0$ at $\BFtheta_b$. As the R\&S subroutine in Algorithm~\ref{alg:OSAR++} achieves the asymptotically optimal allocation ratios in~\citep{glynn2004large_ec}, the following  must be satisfied at each $\BFtheta_b$:
    \begin{oldequation}\label{eq:two.balance.condition}
        \begin{aligned}
            &\lim_{T\rightarrow \infty} T^{-2}\left\{\frac{N^2_{i^b, T}(\BFtheta_b)}{\lambda_{i^b}^2(\BFtheta_b)} - \sum\nolimits_{i\neq i^b}\frac{N^2_{i, T}(\BFtheta_b)}{\lambda_i^2(\BFtheta_b)}\right\} = 0,\\
            &\lim_{T\rightarrow \infty}\frac{G_{i, T}(\BFtheta_b)}{G_{j, T}(\BFtheta_b)} =1 \;\; \text{for all } i\neq j, i\neq i^b, j \neq i^b.
        \end{aligned}
    \end{oldequation}
    Suppose $\liminf_{T\rightarrow \infty} \frac{N_{i^b, T}(\BFtheta_b)}{T} = 0$, i.e., $j_1\neq i^b$.  Then,  
    \begin{oldequation}
        \begin{aligned}
        \liminf_{T\rightarrow \infty}\left\{T^{-2}\left\{\frac{N^2_{i^b, T}(\BFtheta_b)}{\lambda_{i^b}^2(\BFtheta_b)} - \sum\nolimits_{i\neq i^b}\frac{N^2_{i, T}(\BFtheta_b)}{\lambda_i^2(\BFtheta_b)}\right\}\right\}&\leq \liminf_{T\rightarrow \infty}\left\{T^{-2}\left\{\frac{N^2_{i^b, T}(\BFtheta_b)}{\lambda_{i^b}^2(\BFtheta_b)} - \frac{N^2_{j_1, T}(\BFtheta_b)}{\lambda_{j_1}^2(\BFtheta_b)}\right\}\right\} \\
        &\leq \liminf_{T\rightarrow \infty} \frac{N_{i_b, T}^2(\BFtheta_b)}{T^2\lambda^2_{i^b}(\BFtheta_b)} - \liminf_{T \rightarrow \infty}\frac{N^2_{j_1, T}(\BFtheta_b)}{T^2\lambda_{j_1}^2(\BFtheta_b)}\\
        & = - \liminf_{T \rightarrow \infty}\frac{N^2_{j_1, T}(\BFtheta_b)}{T^2\lambda_{j_1}^2(\BFtheta_b)}< 0,
        \end{aligned}
    \end{oldequation}
    which contradicts the first condition of~\eqref{eq:two.balance.condition}. The second inequality holds from that $\liminf_{T\rightarrow \infty} a_T + \liminf_{T\rightarrow \infty}b_T \leq \liminf_{T\rightarrow \infty} (a_T+b_T)$ for any two real sequences, $\{a_T\}$ and $\{b_T\}$. This proves $N_{i^b, T}(\BFtheta_b) = \Tas(T)$. 
    
    Next, we show that there exists $j_2 \neq i^b$ such that $\liminf_{T\rightarrow \infty} \frac{N_{j_2, T}(\BFtheta_b)}{T} > 0$. Suppose otherwise. From $N_{i^b, T}(\BFtheta_b)= \Tas(T)$, we have $\liminf_{T\rightarrow\infty}\frac{N_{j, T}(\BFtheta_b)}{N_{i^b, T}(\BFtheta_b)} = 0$ for all $j\neq i^b$ and  $\liminf_{T\rightarrow \infty} \frac{\max_{j\neq i^b}N_{j, T}(\BFtheta_b)}{N_{i^b, T}(\BFtheta_b)} = 0$. Accordingly, there exists subsequence $\{T_\ell\}_{\ell \in \mathbb{N}}$ such that $\lim_{\ell\rightarrow \infty} \frac{\max_{j\neq i^b}N_{j, T_\ell}(\BFtheta_b)}{N_{i^b, T_\ell}(\BFtheta_b)} = 0$, i.e., $\lim_{\ell\rightarrow \infty} \frac{N_{j, T_\ell}(\BFtheta_b)}{N_{i^b, T_\ell}(\BFtheta_b)} = 0$ for all $j\neq i^b$. The first condition of~\eqref{eq:two.balance.condition} stipulates that $\liminf_{\ell\rightarrow \infty}T_\ell^{-2}\left(\frac{N^2_{i^b, T_\ell}(\BFtheta_b)}{\lambda_{i^b}^2(\BFtheta_b)} - \sum\nolimits_{i\neq i^b}\frac{N^2_{i, T_\ell}(\BFtheta_b)}{\lambda_i^2(\BFtheta_b)}\right) = 0$. However,
    \begin{oldequation}
        \begin{aligned}
        \liminf_{\ell\rightarrow \infty}\left\{T_\ell^{-2}\left\{\frac{N^2_{i^b, T_\ell}(\BFtheta_b)}{\lambda_{i^b}^2(\BFtheta_b)} - \sum_{i\neq i^b}\frac{N^2_{i, T_\ell}(\BFtheta_b)}{\lambda_i^2(\BFtheta_b)}\right\}\right\} &\geq \sum_{i\neq i^b}\liminf_{\ell\rightarrow \infty} \left\{\frac{N_{i_b, T_\ell}^2(\BFtheta_b)}{T_\ell^2}\left(\frac{1}{\lambda^2_{i^b}(\BFtheta_b)} -\frac{N^2_{i, T_\ell}(\BFtheta_b)}{N^2_{i^b, T_\ell}(\BFtheta_b)\lambda_{i}^2(\BFtheta_b)}\right)\right\}\\
        & \geq \sum_{i\neq i^b} \liminf_{\ell \rightarrow \infty}\frac{N^2_{i^b, T_\ell}(\BFtheta_b)}{T_\ell^2} \times \liminf_{\ell \rightarrow \infty} \left(\frac{1}{\lambda^2_{i^b}(\BFtheta_b)} -\frac{N^2_{i, T_\ell}(\BFtheta_b)}{N^2_{i^b, T_\ell}(\BFtheta_b)\lambda_{i}^2(\BFtheta_b)}\right)\\
        & = \sum_{i\neq i^b} \frac{1}{\lambda^2_{i^b}(\BFtheta_b)}\liminf_{\ell \rightarrow \infty}\frac{N^2_{i^b, T_\ell}(\BFtheta_b)}{T_\ell^2} > 0,
        \end{aligned}
    \end{oldequation}
    which is a contradiction. The second inequality is derived from a property of limit infimum, $\liminf_{T\rightarrow \infty} a_Tb_T \geq \liminf_{T\rightarrow \infty} a_T \times \liminf_{T\rightarrow \infty} b_T$ for two positive sequences $\{a_T\}$ and $\{b_T\}$.
    
    Finally, we show that $N_j(\BFtheta_b) = \Tas(T)$ for all $j$. Suppose there exist $j_3\notin\{i^b, j_2\}$ such that $\liminf_{T\rightarrow \infty} \frac{N_{j_3, T}(\BFtheta_b)}{T} = 0$. 
    Then, we can observe that
    \begin{oldequation}
        G_{j_3, T}(\BFtheta_b) = \frac{(\mu_{i^b, T}(\BFtheta_b) - \mu_{j_3, T}(\BFtheta_b))^2}{2\left(\frac{\lambda^2_{i^b}(\BFtheta_b)}{N_{i^b, T}(\BFtheta_b)/T} + \frac{\lambda^2_{j_3}(\BFtheta_b)}{N_{j_3, T}(\BFtheta_b)/T}\right)} \leq \frac{(\mu_{i^b, T}(\BFtheta_b) - \mu_{j_3, T}(\BFtheta_b))^2}{2\lambda^2_{j_3}(\BFtheta_b)}\frac{N_{j_3, T}(\BFtheta_b)}{T},
    \end{oldequation}
    which implies $\liminf_{T\rightarrow \infty} G_{j_3, T}(\BFtheta_b) = 0$. Because $N_{j_2, T}(\BFtheta_b) = \Tas(T)$ and $N_{i^b, T}(\BFtheta_b) = \Tas(T)$, $G_{j_2, T}(\BFtheta_b) = \Tas(1)$. Accordingly, we have $\liminf_{T\rightarrow \infty}\frac{G_{j_3, T}(\BFtheta_b)}{G_{j_2, T}(\BFtheta_b)} = 0$, which contradicts the second condition of~\eqref{eq:two.balance.condition}. This concludes the proof. 
    \halmos
    \endproof
    
    \vspace{3mm}

    \noindent{\bf Step 2}. We have $\lim_{T \rightarrow \infty} \bar{\BFK}_{i, T} = \BFK_{i, B+1}$. For all sufficiently large $T$, $\bar{\BFK}_{i, T}$ is positive definite. 

    \proof{\underline{Proof of Step~2}}. 
    Observe that $\BFK_{i, B+1}$ and $\bar{\BFK}_{i, T}$ can be viewed as block matrices by writing them as 
    \begin{oldequation}
        \BFK_{i, B+1} = \left[\begin{array}{ c | c }
        \BFK_{i, B} & \BFu \\
    \hline
    \BFu^\top & K_i(\BFtheta_0, \BFtheta_0) \end{array}\right] \;\; \text{and} \;\; \bar{\BFK}_{i, T} = \left[\begin{array}{ c | c }
        \BFK_{i, B} & \BFu_T \\
    \hline
    \BFu^\top_T & K_i(\hBFtheta_T, \hBFtheta_T) \end{array}\right] 
    \end{oldequation}
  where $\BFK_{i, B} = (K_i(\BFtheta_b, \BFtheta_{b^\prime}))_{1\leq b, b^\prime \leq B}$ is a Gram matrix of $K_i$ at $\Theta^+$, $\BFu = (K_i(\BFtheta_b, \BFtheta_0))_{1\leq b\leq B}$, and $\BFu_T = (K_i(\BFtheta_b, \hBFtheta_T))_{1\leq b\leq B}$. Thanks to the continuity of $K_i$ and the strong consistency of $\hBFtheta_T$, $\lim_{T\rightarrow \infty} \bar{\BFK}_{i, T} = \BFK_{i, B+1}$.
  
  Further, the definition of RHKS implies that $\BFK_{i, B}$ and $\bar{\BFK}_{i, T}$ are positive semidefinite. Thus, to show the positive definiteness of $\bar{\BFK}_{i, T}$, it suffices to show $\text{det}(\bar{\BFK}_{i, T})> 0$ for all sufficiently large $T$. By Lemma~\ref{lem:det.block}, we have $\text{det}(\BFK_{i, B+1}) = \text{det}(\BFK_{i, B})(K_i(\BFtheta_0, \BFtheta_0) - \BFu^\top \BFK_{i, B} \BFu)$ and $\text{det}(\bar{\BFK}_{i, T}) = \text{det}(\BFK_{i, B})(K_i(\hBFtheta_T, \hBFtheta_T) - \BFu_T^\top \BFK_{i, B} \BFu_T)$. The positive definiteness of $\BFK_{i, B+1}$ yields $\text{det}(\BFK_{i, B}) > 0$ and $K_i(\BFtheta_0, \BFtheta_0) - \BFu^\top \BFK_{i, B} \BFu > 0$. From the continuity of $K_i$ and the strong consistency of $\hBFtheta_T$, we have $\lim_{T\rightarrow \infty} K_i(\hBFtheta_T, \hBFtheta_T) - \BFu_T^\top \BFK_{i, B} \BFu_T = K_i(\BFtheta_0, \BFtheta_0) - \BFu^\top \BFK_{i, B} \BFu > 0$. Hence, for all sufficiently large $T$, $\text{det}(\bar{\BFK}_{i, T}) > 0$, as required. \halmos
  \endproof

  \vspace{3mm}

  From Step 2, we can replace the Moore-Penrose inverse in~\eqref{eq:krr_conti_ver} with the exact inverse for all sufficiently large $T$. To show the strong consistency of $(g_{i, T}(\BFtheta_b))_{1\leq b\leq B+1}$, it is enough to verify
  \begin{oldequation}
    \begin{aligned}
      &\lim_{T\rightarrow \infty}(\widetilde{\BFK}_{i, T} \widetilde{\Sigma}_{i, T}^{-1}\widetilde{\BFK}^\top_{i,T} + \kappa \bar{\BFK}_{i, T})^{-1}\widetilde{\BFK}_{i, T}\widetilde{\Sigma}_{i, T}^{-1}\left([\BFmu_{i, T}^\top \;\; \BFY_i^\top]^\top - \gamma_i \BFone_{B + N_{i, T}(\hBFtheta)}\right)
      = \BFK_{i, B+1}^{-1} \left(\BFy_i -\gamma_i \BFone_{B+1}\right),
      \end{aligned}
  \end{oldequation}
  where $\BFy_i := (\eta_i(\BFtheta_b)_{1\leq b\leq B+1})$. We rewrite the left-hand side sequence as
  \begin{oldequation}
      \begin{aligned}
          &(\widetilde{\BFK}_{i, T} \widetilde{\Sigma}_{i, T}^{-1}\widetilde{\BFK}^\top_{i,T} + \kappa \bar{\BFK}_{i, T})^{-1}\widetilde{\BFK}_{i, T}\widetilde{\Sigma}_{i, T}^{-1}\left([\BFmu_{i, T}^\top \;\; \BFY_i^\top]^\top - \gamma_i \BFone_{B + N_{i, T}(\hBFtheta)}\right) \\
          & = (\bar{\Sigma}_{i, T}\bar{\BFK}_{i, T}^{-1}\widetilde{\BFK}_{i, T} \widetilde{\Sigma}_{i, T}^{-1}\widetilde{\BFK}^\top_{i,T} + \kappa \bar{\Sigma}_{i, T})^{-1}\bar{\Sigma}_{i, T}\bar{\BFK}_{i, T}^{-1}\widetilde{\BFK}_{i, T}\widetilde{\Sigma}_{i, T}^{-1}\left([\BFmu_{i, T}^\top \;\; \BFY_i^\top]^\top - \gamma_i \BFone_{B + N_{i, T}(\hBFtheta)}\right) = \BFA_T \BFb_T,
      \end{aligned}
  \end{oldequation}
  where $\bar{\Sigma}_{i, T}$, $\BFA_T$, and $\BFb_T$ are defined as
  \begin{oldequation}
      \begin{aligned}
          \bar{\Sigma}_{i, T} &:= \text{diag}\left(\left\{{\lambda_i^2(\BFtheta_b)}/{N_{i, T}(\BFtheta_b)}\right\}_{1\leq b\leq B+1}\right),\\
          \BFA_T & := (\bar{\Sigma}_{i, T}\bar{\BFK}_{i, T}^{-1}\widetilde{\BFK}_{i, T} \widetilde{\Sigma}_{i, T}^{-1}\widetilde{\BFK}^\top_{i,T} + \kappa \bar{\Sigma}_{i, T})^{-1},\\
          \BFb_T & := \bar{\Sigma}_{i, T}\bar{\BFK}_{i, T}^{-1}\widetilde{\BFK}_{i, T}\widetilde{\Sigma}_{i, T}^{-1}\left([\BFmu_{i, T}^\top \;\; \BFY_i^\top]^\top - \gamma_i \BFone_{B + N_{i, T}(\hBFtheta)}\right).
      \end{aligned}
  \end{oldequation}
  We will analyze the limits of $\BFA_T$ and $\BFb_T$, respectively, in Steps~3 and~4. 

  \vspace{3mm}
    
  \noindent{\bf Step 3}. We have $\lim_{T\rightarrow \infty} \BFb_T = \BFy_i - \gamma_i \BFone_{B+1}$ almost surely. 
  \proof{\underline{Proof of Step~3}}.
  Observe that the $\ell$-th element of $\BFb_T$ is calculated as
  {\small
  \begin{oldequation}
      \begin{aligned}
          (\BFb_T)_\ell &= \sum_{b_1=1}^{B+1}(\bar{\Sigma}_{i, T}\bar{\BFK}_{i, T}^{-1})_{\ell, b_1} \left\{\widetilde{\BFK}_{i, T}\widetilde{\Sigma}_{i, T}^{-1}\left([\BFmu_{i, T}^\top \;\; \BFY_i^\top]^\top - \gamma_i \BFone_{B + N_{i, T}(\hBFtheta)}\right)\right\}_{b_1}\\
          & = \sum_{b_1=1}^{B+1}\frac{\lambda_i^2(\BFtheta_\ell)}{N_{i, T}(\BFtheta_\ell)}(\bar{\BFK}_{i, T}^{-1})_{\ell, b_1}\left\{\sum_{b_2=1}^B \frac{K_i(\BFtheta_{b_1}, \BFtheta_{b_2})N_{i, T}(\BFtheta_{b_2})}{\lambda_i^2(\BFtheta_{b_2})}(\mu_{i, T}(\BFtheta_{b_2})-\gamma_i) + \sum_{r=1}^{N_{i,T}(\hBFtheta)}\frac{K_i(\BFtheta_{b_1}, \BFtheta_{i, r})}{\lambda_i^2(\hBFtheta_{i, r})}(Y_i(\hBFtheta_{i, r})-\gamma_i)\right\}\\
          & = \underbrace{\sum\nolimits_{b_2=1}^{B}\sum\nolimits_{b_1=1}^{B+1}\frac{\lambda_i^2(\BFtheta_\ell)}{N_{i, T}(\BFtheta_\ell)}(\bar{\BFK}_{i, T}^{-1})_{\ell, b_1}\frac{K_i(\BFtheta_{b_1}, \BFtheta_{b_2})N_{i, T}(\BFtheta_{b_2})}{\lambda_i^2(\BFtheta_{b_2})}(\mu_{i, T}(\BFtheta_{b_2})-\gamma_i)}_{:= \text{(I)}}\\
          &\;\; + \underbrace{\sum\nolimits_{r=1}^{N_{i, T}(\hBFtheta)}\sum\nolimits_{b_1 = 1}^{B+1} \frac{\lambda_i^2(\BFtheta_\ell)}{N_{i, T}(\BFtheta_\ell)}(\bar{\BFK}_{i, T}^{-1})_{\ell, b_1}\frac{K_i(\BFtheta_{b_1}, \BFtheta_{i, r})}{\lambda_i^2(\hBFtheta_{i, r})}(Y_i(\hBFtheta_{i, r}) - \gamma_i)}_{:= \text{(II)}}.
      \end{aligned}
    \end{oldequation}
    }
    We can rewrite (I) as
    \begin{oldequation}\label{eq:I}
        \begin{aligned}
            \text{(I)} 
            & = \sum\nolimits_{b_2=1}^{B}\frac{\lambda_i^2(\BFtheta_\ell)\lambda_i^{-2}(\BFtheta_{b_2})}{N_{i, T}(\BFtheta_\ell)N_{i,T}^{-1}(\BFtheta_{b_2})}(\mu_{i, T}(\BFtheta_{b_2}) - \gamma_i)\left(\sum\nolimits_{b_1=1}^{B+1}(\bar{\BFK}_{i, T}^{-1})_{\ell, b_1}K_i(\BFtheta_{b_1}, \BFtheta_{b_2})\right)
        \end{aligned}
    \end{oldequation}
    From Step~2, $\lim_{T \rightarrow \infty} \sum\nolimits_{b_1=1}^{B+1}(\bar{\BFK}_{i, T}^{-1})_{\ell, b_1}K_i(\BFtheta_{b_1}, \BFtheta_{b_2}) = 1\{\ell = b_2\}$. When $\ell \neq b_2$, combining Step~1 with the strong law of large numbers yields
    \begin{oldequation}
        \frac{\lambda_i^2(\BFtheta_\ell)\lambda_i^{-2}(\BFtheta_{b_2})}{N_{i, T}(\BFtheta_\ell)N_{i,T}^{-1}(\BFtheta_{b_2})}(\mu_{i, T}(\BFtheta_{b_2})- \gamma_i)\left(\sum\nolimits_{b_1=1}^{B+1}(\bar{\BFK}_{i, T}^{-1})_{\ell, b_1}K_i(\BFtheta_{b_1}, \BFtheta_{b_2})\right) = \Tas(1)\left(\sum\nolimits_{b_1=1}^{B+1}(\bar{\BFK}_{i, T}^{-1})_{\ell, b_1}K_i(\BFtheta_{b_1}, \BFtheta_{b_2})\right).
    \end{oldequation}
    Thus, $\lim_{T \rightarrow \infty}\frac{\lambda_i^2(\BFtheta_\ell)\lambda_i^{-2}(\BFtheta_{b_2})}{N_{i, T}(\BFtheta_\ell)N_{i,T}^{-1}(\BFtheta_{b_2})}(\mu_{i, T}(\BFtheta_{b_2})- \gamma_i)\left(\sum\nolimits_{b_1=1}^{B+1}(\bar{\BFK}_{i, T}^{-1})_{\ell, b_1}K_i(\BFtheta_{b_1}, \BFtheta_{b_2})\right) = 0$ holds. When $\ell = b_2$, we can derive
    \begin{oldequation}
        \begin{aligned}
        &\frac{\lambda_i^2(\BFtheta_\ell)\lambda_i^{-2}(\BFtheta_{b_2})}{N_{i, T}(\BFtheta_\ell)N_{i,T}^{-1}(\BFtheta_{b_2})}(\mu_{i, T}(\BFtheta_{b_2})- \gamma_i)\left(\sum\nolimits_{b_1=1}^{B+1}(\bar{\BFK}_{i, T}^{-1})_{\ell, b_1}K_i(\BFtheta_{b_1}, \BFtheta_{b_2})\right)\\
        &= (\mu_{i,T}(\BFtheta_{b_2}) - \gamma_i)\left(\sum\nolimits_{b_1=1}^{B+1}(\bar{\BFK}_{i, T}^{-1})_{\ell, b_1}K_i(\BFtheta_{b_1}, \BFtheta_{b_2})\right),
        \end{aligned}
    \end{oldequation}
    which converges to $y_{i}(\BFtheta_{b_2}) - \gamma_i$ almost surely. Hence, we have
    \begin{oldequation}
        \lim_{T \rightarrow \infty} \frac{\lambda_i^2(\BFtheta_\ell)\lambda_i^{-2}(\BFtheta_{b_2})}{N_{i, T}(\BFtheta_\ell)N_{i,T}^{-1}(\BFtheta_{b_2})}(\mu_{i, T}(\BFtheta_{b_2})- \gamma_i)\left(\sum\nolimits_{b_1=1}^{B+1}(\bar{\BFK}_{i, T}^{-1})_{\ell, b_1}K_i(\BFtheta_{b_1}, \BFtheta_{b_2})\right) = (\eta_i(\BFtheta_\ell)-\gamma_i) \mathbf{1}\{b_2 = \ell\}.
    \end{oldequation}
    Accordingly, the limit of~\eqref{eq:I} is given as
    \begin{oldequation}\label{eq:limit_(I)}
        \lim_{T\rightarrow \infty} \text{(I)} = \sum\nolimits_{b_2 = 1}^{B} (\eta_i(\BFtheta_\ell) - \gamma_i) \mathbf{1}\{b_2 = \ell\} = (\eta_{i}(\BFtheta_\ell)-\gamma_i)1\{\ell \leq B\}. 
    \end{oldequation}

    Now, let us show the limit of (II), which can be rewritten as
    \begin{oldequation}
        \begin{aligned}
            \text{(II)} 
            &= \frac{\lambda_i^2(\BFtheta_\ell) N_{i,T}(\hBFtheta)}{N_{i, T}(\BFtheta_\ell)} \frac{1}{N_{i,T}(\hBFtheta)}\sum\nolimits_{r=1}^{N_{i, T}(\hBFtheta)}\frac{Y_i(\hBFtheta_{i, r})-\gamma_i}{\lambda_i^{2}(\hBFtheta_{i, r})}\left(\sum\nolimits_{b_1=1}^{B+1}(\bar{\BFK}_{i, T}^{-1})_{\ell, b_1}K_i(\BFtheta_{b_1}, \hBFtheta_{i, r})\right)
        \end{aligned}
    \end{oldequation}
    Let $X_T = \frac{1}{N_{i,T}(\hBFtheta)}\sum\nolimits_{r=1}^{N_{i, T}(\hBFtheta)}\frac{Y_i(\hBFtheta_{i, r})-\gamma_i}{\lambda_i^{2}(\hBFtheta_{i, r})}\left(\sum\nolimits_{b_1=1}^{B+1}(\bar{\BFK}_{i, T}^{-1})_{\ell, b_1}K_i(\BFtheta_{b_1}, \BFtheta_{i, r})\right)$. Since $Y_i(\hBFtheta_{i, r})$ are normally distributed with mean $\eta_i(\hBFtheta_{i, r})$ and variance $\lambda_i^2(\hBFtheta_{i, r})$, $X_T$ are also normally distributed whose mean and variance are given as
    \begin{oldequation}\label{eq:X_mean_var}
        \begin{aligned}
        \mathrm{E}[X_T] &= {N^{-1}_{i, T}(\hBFtheta)}\sum\nolimits_{r=1}^{N_{i, T}(\hBFtheta)}\frac{\eta_i(\hBFtheta_{i, r})-\gamma_i}{\lambda_i^{2}(\hBFtheta_{i, r})}\left(\sum\nolimits_{b_1=1}^{B+1}(\bar{\BFK}_{i, T}^{-1})_{\ell, b_1}K_i(\BFtheta_{b_1}, \hBFtheta_{i, r})\right)\\
        \text{Var}[X_T] &= {N^{-2}_{i, T}(\hBFtheta)}\sum\nolimits_{r=1}^{N_{i, T}(\hBFtheta)}{\lambda_i^{-2}(\hBFtheta_{i, r})}{\left(\sum\nolimits_{b_1=1}^{B+1}(\bar{\BFK}_{i, T}^{-1})_{\ell, b_1}K_i(\BFtheta_{b_1}, \hBFtheta_{i, r})\right)^2}.
        \end{aligned}
    \end{oldequation}
    Combining $\lim_{r \rightarrow \infty} \hBFtheta_{i, r} = \BFtheta_0$ and Step~2, we have $\lim_{T\rightarrow \infty} \left(\sum\nolimits_{b_1=1}^{B+1}(\bar{\BFK}_{i, T}^{-1})_{\ell, b_1}K_i(\BFtheta_{b_1}, \hBFtheta_{i, r})\right) = \mathbf{1}\{\ell = B+1\}$. From the continuity of $\eta_i$ and $\lambda_i$, we can further obtain
    \begin{oldequation}\label{eq:temp_inner_term}
        \begin{aligned}
            \lim_{r \rightarrow \infty} \frac{\eta_i(\hBFtheta_{i, r}) - \gamma_i}{\lambda_i^{2}(\hBFtheta_{i, r})}\left(\sum\nolimits_{b_1=1}^{B+1}(\bar{\BFK}_{i, T}^{-1})_{\ell, b_1}K_i(\BFtheta_{b_1}, \hBFtheta_{i, r})\right) &= \frac{\eta_i(\BFtheta_0)-\gamma_i}{\lambda_i^2(\BFtheta_0)}1\{\ell = B+1\},\\ 
            \lim_{r \rightarrow \infty} {\lambda_i^{-2}(\hBFtheta_{i, r})}{\left(\sum\nolimits_{b_1=1}^{B+1}(\bar{\BFK}_{i, T}^{-1})_{\ell, b_1}K_i(\BFtheta_{b_1}, \hBFtheta_{i, r})\right)^2} & = \lambda_i^{-2}(\BFtheta_0)1\{\ell = B+1\}.
        \end{aligned}
    \end{oldequation}
    From the Ces{\'a}ro mean property and~\eqref{eq:temp_inner_term}, we can see that
    \begin{oldequation}
        \lim_{T\rightarrow \infty} \mathrm{E}[X_T] = \frac{\eta_i(\BFtheta_0)-\gamma_i}{\lambda_i^2(\BFtheta_0)}1\{\ell = B+1\} \;\; \text{and} \;\; \lim_{T \rightarrow \infty} N_{i, T}(\hBFtheta)\text{Var}[X_T] = \lambda_{i}^{-2}(\BFtheta_0)1\{\ell = B+1\}. 
    \end{oldequation}
    Applying Lemma~\ref{lem:strong.consistency.normal}, $\lim_{T \rightarrow \infty} X_T = \frac{\eta_i(\BFtheta_0)}{\lambda_i^2(\BFtheta_0)}1\{\ell = B+1\}$ holds almost surely. Consequently, we have   
    \begin{oldequation}\label{eq:limit_(II)}
        \lim_{T \rightarrow \infty} \text{(II)} = (\eta_i(\BFtheta_0)-\gamma_i)1\{\ell = B+1\}. 
    \end{oldequation}
    Combining~\eqref{eq:limit_(I)} with~\eqref{eq:limit_(II)} yields $\lim_{T\rightarrow \infty} (\BFb_T)_\ell = \eta_i(\BFtheta_\ell)-\gamma_i$, as desired. \halmos

    \endproof

    \vspace{3mm}

    \noindent{\bf Step~4}. $\lim_{T\rightarrow \infty} \BFA_T = \BFK_{i, B+1}^{-1}$ holds almost surely.

    \proof{\underline{Proof of Step~4}}. Since $\lim_{T \rightarrow \infty} \bar{\Sigma}_{i, T} = \BFzero_{B+1, B+1}$, it suffices to verify  $\lim_{T \rightarrow \infty} \bar{\Sigma}_{i, T}\bar{\BFK}_{i, T}^{-1}\widetilde{\BFK}_{i, T} \widetilde{\Sigma}_{i, T}^{-1}\widetilde{\BFK}^\top_{i,T} = \BFK_{i, B+1}$. For $1\leq b_1, b_2\leq B+1$,
    \begin{oldequation}
        \begin{aligned}
            (\bar{\Sigma}_{i, T}\bar{\BFK}_{i, T}^{-1}\widetilde{\BFK}_{i, T} \widetilde{\Sigma}_{i, T}^{-1}\widetilde{\BFK}^\top_{i,T})_{b_1, b_2}
            & = \sum\nolimits_{\ell = 1}^{B+1} (\bar{\Sigma}_{i, T}\bar{\BFK}_{i, T}^{-1})_{b_1, \ell} (\widetilde{\BFK}_{i, T} \widetilde{\Sigma}_{i, T}^{-1}\widetilde{\BFK}^\top_{i,T})_{\ell, b_2}\\
            & = \sum\nolimits_{\ell = 1}^{B+1} (\bar{\Sigma}_{i, T}\bar{\BFK}_{i, T}^{-1})_{b_1, \ell} \sum\nolimits_{\ell_1=1}^{B+N_{i, T}(\hBFtheta)}(\widetilde{\BFK}_{i, T} \widetilde{\Sigma}_{i, T}^{-1})_{\ell, \ell_1}(\widetilde{\BFK}^\top_{i,T})_{\ell_1, b_2}\\
            & = \underbrace{\sum\nolimits_{\ell = 1}^{B+1} (\bar{\Sigma}_{i, T}\bar{\BFK}_{i, T}^{-1})_{b_1, \ell} \sum\nolimits_{\ell_1=1}^{B}(\widetilde{\BFK}_{i, T} \widetilde{\Sigma}_{i, T}^{-1})_{\ell, \ell_1}(\widetilde{\BFK}^\top_{i,T})_{\ell_1, b_2}}_{:= \text{(III)}} \\
            & \;\; + \underbrace{\sum\nolimits_{\ell = 1}^{B+1} (\bar{\Sigma}_{i, T}\bar{\BFK}_{i, T}^{-1})_{b_1, \ell} \sum\nolimits_{\ell_1=1}^{N_{i, T}(\hBFtheta)}(\widetilde{\BFK}_{i, T} \widetilde{\Sigma}_{i, T}^{-1})_{\ell, \ell_1}(\widetilde{\BFK}^\top_{i,T})_{B+\ell_1, b_2}}_{:=\text{(IV)}}.
        \end{aligned}
    \end{oldequation}
    Let us see (III) first. One can derive that
    \begin{oldequation}
        \begin{aligned}
            \text{(III)}
            & = \sum\nolimits_{\ell = 1}^{B+1} \sum\nolimits_{\ell_1=1}^{B} \frac{\lambda_i^2(\BFtheta_{b_1})}{N_{i, T}(\BFtheta_{b_1})}(\bar{\BFK}_{i, T}^{-1})_{b_1, \ell} K_i(\BFtheta_\ell, \BFtheta_{\ell_1})\frac{N_{i, T}(\BFtheta_{\ell_1})}{\lambda_i^2(\BFtheta_{\ell_1})}K_i(\BFtheta_{\ell_1}, \BFtheta_{b_2})\\
            & = \sum\nolimits_{\ell_1=1}^{B} \frac{\lambda_i^2(\BFtheta_{b_1})\lambda_i^{-2}(\BFtheta_{\ell_1})}{N_{i, T}(\BFtheta_{b_1})N^{-1}_{i, T}(\BFtheta_{\ell_1})}K_i(\BFtheta_{\ell_1}, \BFtheta_{b_2})\left(\sum\nolimits_{\ell = 1}^{B+1} (\bar{\BFK}_{i, T}^{-1})_{b_1, \ell}K_i(\BFtheta_\ell, \BFtheta_{\ell_1})\right)
        \end{aligned}
    \end{oldequation}
    From Step~2 again, we have $\lim_{T\rightarrow \infty} \sum\nolimits_{\ell = 1}^{B+1} (\bar{\BFK}_{i, T}^{-1})_{b_1, \ell}K_i(\BFtheta_\ell, \BFtheta_{\ell_1}) = 1\{b_1 = \ell_1\}$.  From this result, the limit of (III) is given as 
    \begin{oldequation}
        \lim_{T\rightarrow \infty} \text{(III)} = \sum\nolimits_{\ell_1=1}^B K_i(\BFtheta_{\ell_1}, \BFtheta_{b_2}) 1\{b_1 = \ell_1\} = K_i(\BFtheta_{b_1}, \BFtheta_{b_2})1\{b_1\leq B\}.
    \end{oldequation}
    Further, (IV) can be calculated as
    \begin{oldequation}
        \begin{aligned}
            \text{(IV)}
            & = \sum\nolimits_{\ell_1=1}^{N_{i, T}(\hBFtheta)}\sum\nolimits_{\ell = 1}^{B+1}\frac{\lambda_i^2(\BFtheta_{b_1})}{N_{i, T}(\BFtheta_{b_1})}(\bar{\BFK}_{i, T}^{-1})_{b_1, \ell}\frac{K_i(\BFtheta_\ell, \hBFtheta_{i, \ell_1})}{\lambda_i^2(\hBFtheta_{i, \ell_1})}K_i(\BFtheta_{b_2}, \hBFtheta_{i, \ell})\\
            & = \frac{N_{i, T}(\hBFtheta)}{N_{i, T}(\BFtheta_{b_1})}\frac{1}{N_{i, T}(\hBFtheta)}\sum\nolimits_{\ell_1=1}^{N_{i, T}(\hBFtheta)}\frac{\lambda_i^2(\BFtheta_{b_1})}{\lambda_{i}^{2}(\hBFtheta_{i, \ell_1})}K_i(\BFtheta_{b_2}, \hBFtheta_{i, \ell_1})\left(\sum\nolimits_{\ell=1}^{B+1}(\bar{\BFK}_{i, T}^{-1})_{b_1, \ell}K_i(\BFtheta_{\ell},\hBFtheta_{i, \ell_1})\right).
        \end{aligned}
    \end{oldequation}
    From $\lim_{T \rightarrow \infty} \sum\nolimits_{\ell=1}^{B+1}(\bar{\BFK}_{i, T}^{-1})_{b_1, \ell}K_i(\BFtheta_{\ell},\hBFtheta_{i, \ell_1}) = 1\{b_1 = B+1\}$ and the continuity of $\lambda_i^2$, we can see that
    \begin{oldequation}
        \lim_{\ell_1 \rightarrow \infty} \frac{\lambda_i^2(\BFtheta_{b_1})}{\lambda_{i}^{2}(\hBFtheta_{i, \ell_1})}K_i(\BFtheta_{b_2}, \hBFtheta_{i, \ell_1})\left(\sum\nolimits_{\ell=1}^{B+1}(\bar{\BFK}_{i, T}^{-1})_{b_1, \ell}K_i(\BFtheta_{\ell}, \hBFtheta_{i, \ell_1})\right) = \frac{\lambda_i^2(\BFtheta_{b_1})}{\lambda_i^2(\BFtheta_0)}K_i(\BFtheta_{b_2}, \BFtheta_0)1\{b_1 = B+1\}. 
    \end{oldequation}
    Applying the Ces{\'a}ro mean property, one can derive the limit of (IV) as
    \begin{oldequation}
        \lim_{T\rightarrow \infty} \text{(IV)} = K_i(\BFtheta_{b_2}, \BFtheta_0)1\{b_1 = B+1\}.
    \end{oldequation}
    We finally have $\lim_{T \rightarrow \infty} \text{(III)} + \text{(IV)} = K_i(\BFtheta_{b_1}, \BFtheta_{b_2})$, which completes the proof. \halmos   \qed 
    \endproof

\section{Efficient Updates of the Nystr{\"o}m KRR Model}\label{ec:recursive}

In this subsection, we discuss an efficient method to update $\BFc_{i,T}$ discussed in Remark~\ref{rmk:Sherman-Morrsion-OSAR_conti}. Define $\BFA_{i, T} = \widetilde{\BFK}_{i, T} \widetilde{\Sigma}_{i, T}^{-1}\widetilde{\BFK}^\top_{i,T} + \kappa \bar{\BFK}_{i, T}$ and $\BFb_{i, T} = \widetilde{\BFK}_{i, T}\widetilde{\Sigma}_{i, T}^{-1}\BFy_T$, where $\BFy_T = \left([\BFmu_{i, T}^\top \;\; \BFY_i^\top]^\top - \gamma_i \BFone_{B + N_{i, T}(\hBFtheta)}\right)$. From Theorem~\ref{thm:convergence_OSAR+}, Corollary~\ref{cor:least.square} below shows that $\BFc_{i, T}$ is identical to the optimal solution to the linear equation. Namely, we can learn $\BFc_{i, T}$ via the least squares method instead of computing the Moore-Penrose inverse of $\BFA_{i, T}$. 
\begin{corollary}\label{cor:least.square}
    One can view $\BFc_{i, T}$ defined in~\eqref{eq:opt_cit} as the least square solution to 
    \begin{oldequation}\label{eq:least.square.svd}
        \min\nolimits_{\BFx}||\BFA_{i, T}\BFx - \BFb_{i, T}||_2,
\end{oldequation}
where $\BFb_{i, T} = \widetilde{\BFK}_{i, T}\widetilde{\Sigma}_{i, T}^{-1}\BFx$. 
\end{corollary}
Motivated by this, $\BFc_{i, T}$ can be computed by solving~\eqref{eq:least.square.svd}. In this regard, we decompose $\BFA_{i, T}$ via the QR decomposition, $\BFA_{i, T} = \BFQ_{i, T} \BFR_{i, T}$, where $\BFQ_{i, T}$ is the orthogonal matrix (i.e., $\BFQ_{i, T}^\top= \BFQ_{i, T}^{-1}$) and $\BFR_{i, T}$ is the upper triangular matrix. Observe that
\begin{oldequation}
    \begin{aligned}
    ||\BFA_{i, T} \BFx - \BFb_{i, T}||^2_2 &= (\BFA_{i, T} \BFx - \BFb_{i, T})^\top(\BFA_{i, T} \BFx - \BFb_{i, T})\\
    &= (\BFA_{i, T} \BFx - \BFb_{i, T})^\top\BFQ_{i,T}\BFQ_{i,T}^\top(\BFA_{i, T} \BFx - \BFb_{i, T})\\
    &= ||\BFR_{i, T}\BFx - \BFQ_{i,T}^\top \BFb_{i, T}||_2^2,
    \end{aligned}
\end{oldequation}
where the second equality follows from $\BFQ_{i, T}\BFQ_{i, T}^\top = I$. Hence, we have 
\begin{oldequation}\label{def:c_it_QR}
    \BFc_{i, T} = \argmin\nolimits_{\BFx}||\BFR_{i, T}\BFx - \BFQ_{i,T}^\top \BFb_{i, T}||_2.
\end{oldequation}
We will exploit~\eqref{def:c_it_QR} to evaluate $\BFc_{i, T}$ since it is easier to solve than~\eqref{eq:least.square.svd} as $\BFR_{i,T}$ is upper triangular. 

Whenever we collect more simulation data at any $(i, \BFtheta_b)$, $\BFA_{i, T}$ and $\BFb_{i, T}$ are updated. Then, $\BFQ_{i, T}$ and $\BFR_{i, T}$ should be updated accordingly. Proposition~\ref{prop:update_cit} below summarizes the recursive formulae of $\BFA_{i, T}$ and $\BFb_{i, T}$, respectively. From Proposition~\ref{prop:update_cit}, one can confirm that $\BFA_{i, T+1}$ can be written as $\BFA_{i, T+1}=\BFA_{i, T} + \BFu\BFv^\top$ for some $(B+1)$-dimensional vectors, $\BFu$ and $\BFv$. Exploiting this, we can efficiently update $\BFQ_{i, T+1}$ and $\BFR_{i, T+1}$ in a sequential manner using the \textsc{Matlab} built-in function, \textsf{qrupdate}. It is known that the computational complexity of this job is $O(B^2)$, which is cheaper than that of a new QR decomposition, $O(B^3)$~\citep{Golub:MatrixComputation_ec}. 

\begin{proposition}\label{prop:update_cit}
    Assuming that $Y^\text{new}_i$ is the first new observation after pending budget $T$, $\BFA_{i, T}$ and $\BFb_{i, T}$ are updated as
        $
            \BFA_{i, T+1} = \BFA_{i, T} + \lambda_i^{-2}(\BFtheta_b) \BFk_{i, b}^\top\BFk_{i, b}~\text{and}~
            \BFb_{i, T+1} = \BFb_{i, T} + \lambda_{i}^{-2}(\BFtheta_b)(Y_i^\text{new} - \gamma_i)\BFk_{i, b},
        $
    where $ \BFk_{i, b} = \widetilde{\BFK}_{i, T}\BFe_b$.
\end{proposition}

\proof{\underline{Proof of Proposition~\ref{prop:update_cit}}.}
    Notice that $\bar{\BFK}_{i, T+1} = \bar{\BFK}_{i, T}$ and $\widetilde{\BFK}_{i, T+1} = \widetilde{\BFK}_{i, T}$ hold as the MAP is not updated. We consider two cases, (a) $\BFtheta_b \in \Theta^+$ and (b) $\BFtheta_b = \hBFtheta$, separately. 
    
    \noindent (a) Suppose $\BFtheta_b \in \Theta^+$.  Then,
     \begin{oldequation}
        \BFA_{i, T+1} = \widetilde{\BFK}_{i, T+1}\Sigma_{i, T+1}^{-1}\widetilde{\BFK}_{i, T+1}^\top = \widetilde{\BFK}_{i, T}\Sigma_{i, T}^{-1}\widetilde{\BFK}_{i, T}^\top + \lambda_i^{-2}(\BFtheta_b)\BFk_{i, b}\BFk_{i, b}^\top = \BFA_{i, T} + \lambda_i^{-2}(\BFtheta_b)\BFk_{i, b}\BFk_{i, b}^\top,
    \end{oldequation}
    which proves the recursive formula for $\BFA_{i, T+1}$. Recall $\BFy_{T} = \left([\BFmu_{i, T}^\top \;\; \BFY_i^\top]^\top - \gamma_i \BFone_{B + N_{i, T}(\hBFtheta)}\right)$. Let $\widetilde{\mu} = \mu_{i, T+1}(\BFtheta_b) - \mu_{i, T}(\BFtheta_b)$. Then, we have
    \begin{oldequation}
        \begin{aligned}
        \BFb_{i, T+1}
        &=\widetilde{\BFK}_{i, T+1}\widetilde{\Sigma}_{i, T+1}^{-1}\BFy_{T+1}\\
        & = \widetilde{\BFK}_{i, T}(\widetilde{\Sigma}_{i, T}^{-1} + \lambda_i^{-2}(\BFtheta_b)\BFe_b\BFe_b^\top)(\BFy_T + \widetilde{\mu}\BFe_b)\\
        & = \widetilde{\BFK}_{i, T}\left[\widetilde{\Sigma}_{i, T}^{-1}\BFy_T +\left\{\lambda_i^{-2}(\BFtheta_b)\BFe_b^\top\BFy_T+ \lambda_i^{-2}(\BFtheta_b)\widetilde{\mu} + \widetilde{\mu}\left(\frac{N_{i, T}(\BFtheta_b)}{\lambda_{i}^2(\BFtheta_b)}\right)\right\}\BFe_b\right]\\
        & = \widetilde{\BFK}_{i, T}\left[\widetilde{\Sigma}_{i, T}^{-1}\BFy_T +\lambda_i^{-2}(\BFtheta_b)\left\{\mu_{i, T}(\BFtheta_b) - \gamma_i + (N_{i, T}(\BFtheta_b)+1)\widetilde{\mu})\right\}\BFe_b\right]\\
        & = \widetilde{\BFK}_{i, T}\widetilde{\Sigma}_{i, T}^{-1}\BFy_T + \lambda_{i}^{-2}(\BFtheta_b)(Y_i^\text{new} - \gamma_i)\BFk_{i, b} = \BFb_{i, T} + \lambda_{i}^{-2}(\BFtheta_b)(Y_i^\text{new} - \gamma_i)\BFk_{i, b},
        \end{aligned}
    \end{oldequation}
    where the fifth equality follows from $\mu_{i, T}(\BFtheta_b) - \gamma_i + (N_{i, T}(\BFtheta_b)+1)\widetilde{\mu} = Y_i^\text{new} - \gamma_i$. 

    \noindent (b)  Since $\widetilde{\BFK}_{i, T+1} = [\widetilde{\BFK}_{i, T} \;\; \BFk_{i, B+1}]$ and $\widetilde{\Sigma}^{-1}_{i, T+1} =  \left[\begin{array}{ c | c }
       \widetilde{\Sigma}^{-1}_{i, T} & \BFzero_{(B+N_{i, T}(\hBFtheta))\times 1}\\
        \hline
    \BFzero_{(B+N_{i, T}(\hBFtheta))\times 1}^\top & \lambda^{-2}_{i}(\hBFtheta_T)
  \end{array}\right]$, we have
  \begin{oldequation}
      \begin{aligned}
          \BFA_{i, T+1}& = \widetilde{\BFK}_{i, T+1}\widetilde{\Sigma}_{i, T+1}^{-1}\widetilde{\BFK}_{i, T+1}^\top + \kappa \bar{\BFK}_{i, T+1}\\
          & = [\widetilde{\BFK}_{i, T} \;\; \BFk_{i, B+1}]\left[\begin{array}{ c | c }
       \widetilde{\Sigma}^{-1}_{i, T} & \BFzero_{(B+N_{i, T}(\hBFtheta))\times 1}\\
        \hline
    \BFzero_{(B+N_{i, T}(\hBFtheta))\times 1}^\top & \lambda^{-2}_{i}(\hBFtheta_T)
  \end{array}\right][\widetilde{\BFK}_{i, T} \;\; \BFk_{i, B+1}]^\top + \kappa \bar{\BFK}_{i, T} \\
  & = \widetilde{\BFK}_{i, T} \widetilde{\Sigma}_{i, T}^{-1} \widetilde{\BFK}_{i, T}^\top + \lambda_{i}^{-2}(\hBFtheta_T)\BFk_{i, B+1}\BFk_{i, B+1}^\top+ \kappa \bar{\BFK}_{i, T} = \BFA_{i, T} + \lambda_{i}^{-2}(\hBFtheta_T)\BFk_{i, B+1}\BFk_{i, B+1}^\top.
      \end{aligned}
  \end{oldequation}
  To analyze $\BFb_{i, T+1}$, we first observe $\BFy_{T+1} = [\BFy^\top_T\;\;  Y^\text{new}_i-\gamma_i]^\top$ and 
  \begin{oldequation}
    \begin{aligned}
      \widetilde{\BFK}_{i, T+1}\widetilde{\Sigma}_{i, T+1}^{-1}\BFy_{T+1} &= [\widetilde{\BFK}_{i, T} \;\; \BFk_{i, B+1}]\left[\begin{array}{ c | c }
       \widetilde{\Sigma}^{-1}_{i, T} & \BFzero_{(B+N_{i, T}(\hBFtheta))\times 1}\\
        \hline
    \BFzero_{(B+N_{i, T}(\hBFtheta))\times 1}^\top & \lambda^{-2}_{i}(\hBFtheta_T)
  \end{array}\right]\BFy_{T+1}\\
    & = \widetilde{\BFK}_{i, T}\widetilde{\Sigma}_{i, T}^{-1}\BFy_T +\lambda_i^{-2}(\hBFtheta_T)(Y^\text{new}_i-\gamma_i)\BFk_{i, B+1}\\
    &= \BFb_{i, T} + \lambda_i^{-2}(\hBFtheta)(Y^\text{new}_i - \gamma_i) \BFk_{i, B+1},
    \end{aligned}
  \end{oldequation}
  which completes the proof. \qed
\endproof

Corollary~\ref{cor:update_unknown} below presents the revised formulae for $\BFA_{i, T}$ and $\BFb_{i, T}$ when the variances, $\lambda_i^2(\BFtheta_b)$, are unknown.
\begin{corollary}\label{cor:update_unknown}
    Let $\lambda^2_{i, T}(\BFtheta_b)$ be the sample variance at $(i, \BFtheta_b)$ after spending sampling budget $T$. After observing $Y^\text{new}_i$, $\BFA_{i, T}$ and $\BFb_{i, T}$ are updated as follows:
    \begin{enumerate}
        \item [(a)] If $\BFtheta_b \in \Theta^+$, we have
        \begin{oldequation}
            \begin{aligned}
                \BFA_{i, T+1} = \BFA_{i, T} + \widetilde{\lambda} \BFk_{i, b}^\top\BFk_{i, b}~\text{and}~
                \BFb_{i, T+1} = \BFb_{i, T} + \widetilde{\mu}\BFk_{i, b}
            \end{aligned}
        \end{oldequation}
        where 
            $
                \widetilde{\lambda} = \left(\frac{N_{i, T}(\BFtheta_b)+1}{\lambda_{i, T+1}^2(\BFtheta_b)} - \frac{N_{i, T}(\BFtheta_b)}{\lambda_{i, T+1}^2(\BFtheta_b)}\right),
                \widetilde{\mu} = \frac{N_{i, T}(\BFtheta_b) + 1}{\lambda_{i, T+1}^2(\BFtheta_b)}(\mu_{i, T+1}(\BFtheta_b) - \gamma_i) - \frac{N_{i, T}(\BFtheta_b)}{\lambda_{i, T}^2(\BFtheta_b)}(\mu_{i, T}(\BFtheta_b) - \gamma_i),~\text{and}~
                \BFk_{i, b} = \widetilde{\BFK}_{i, T}\BFe_b.
            $

        \item [(b)] If $\BFtheta_b = \hBFtheta$, we can see
        \begin{oldequation}
            \begin{aligned}
                \BFA_{i, T+1} = \BFA_{i, T} + \lambda_i^{-2}(\hBFtheta) \BFk_{i, B+1}^\top\BFk_{i, B+1}~\text{and}~
                \BFb_{i, T+1} = \BFb_{i, T} + \lambda_i^{-2}(\hBFtheta)(Y^\text{new}_i- \gamma_i)\BFk_{i, B+1},
            \end{aligned}
        \end{oldequation}
        where $\BFk_{i, B+1} = (K_i(\hBFtheta, \BFtheta_b))_{1\leq b\leq B+1}$.
    \end{enumerate}
\end{corollary}

\section{Implementation Details of BICO}\label{ec:ungredda}

This section discusses some computational details required to run the Bayesian Information Collection and Optimization (BICO) algorithm suggested by~\cite{ungredda2022_ec} for the examples in Section~\ref{sec:num.exp}. 

Unlike OSAR, BICO  identifies the optimal solution by comparing the expected performance averaged over the joint posterior distribution of simulation error and input parameter. Another difference is that BICO allocates a single replication in each iteration for simulation, however, collects a batch of input data from a source, where the batch size is a user input. We adopt the batch size of $5$ in all our experiments. To determine whether to  collect more input data or simulation output, BICO computes their values of information (VoI).  To compute the VoI of input data, they adopt  Monte Carlo (MC) estimation, which requires two parameters, $R_1$ and $R_2$; the former is the MC sample size  to compute the current expected performance of each solution, and the latter is the MC sample size to compute the improvement in each solution's performance should we sample the batch input data from each input source. Again, these are suggested as user inputs and we adopt $R_1=150$ and $R_2=50$ in our implementations.

\section{Auxiliary Results}\label{ec:aux}

This section contains auxiliary results invoked to show our main results and their proofs.

\begin{lemma}\label{lem:min.property}
Suppose there exists sequence $\{a_{i,n} \}_{n=1}^{\infty}$ such that $a_{i, n} > 0$ and $\lim_{n \rightarrow \infty} -\frac{1}{n}\log a_{i, n} = \lambda_i \geq 0$ for each $i = 1, 2,\ldots, K$. Then, we have $\lim_{n\rightarrow \infty} -\frac{1}{n}\log(\sum_{i=1}^K a_{i, n}) = \min_{1\leq i\leq K} \lambda_i$. 
\end{lemma}
\proof{\underline{Proof.}} Taking $-\frac{1}{n}\log(\cdot)$ operation on all sides of 
    $\max_{1\leq i\leq K} a_{i, n} \leq \sum_{i=1}^K a_{i, n} \leq K \max_{1\leq i\leq K} a_{i, n}$,
we have
\begin{oldequation}\label{ineq.ldr}
    -\frac{1}{n}\log K + \min_{1\leq i\leq K}-\frac{1}{n} \log a_{i, n} \leq -\frac{1}{n}\log\left(\sum\nolimits_{i=1}^K a_{i, n}\right) \leq \min_{1\leq i\leq K}-\frac{1}{n}\log a_{i, n}.
\end{oldequation}
which follows from that $-\log \max_{1\leq i\leq K} a_{i, n} = \min_{1\leq i\leq K} -\log a_{i, n}$. Letting $n \rightarrow \infty$ on both sides of~\eqref{ineq.ldr}, we can obtain the desired result. \qed
\endproof

    \begin{lemma}[Modified version of Theorem 4.5.5 in Durrett (2019)]\label{lem:second_Borel}
        For each $t$, define $Y_t \sim \text{Bin}(\Delta, p_t)$, where $p_t \in \mathcal{F}_{t-1}$ and $\mathcal{F}_{t-1}$ is a filtration generated by the collected data until the $(t-1)$-th iteration. Then, we have
        \begin{oldequation}
            \frac{\sum_{t=1}^T Y_t}{\Delta\sum_{t=1}^T p_t} \rightarrow 1 \;\; \text{a.s.}\;\; \text{conditional on } \{\sum_{t=1}^{\infty} p_t = \infty\}.
        \end{oldequation}
    \end{lemma}

    \proof{\underline{Proof.}}
        Similar to the original proof, we construct a martingale $\{X_t\}_{t\geq 1}$ as $X_0 = 0$ and $X_t - X_{t-1} = \frac{Y_t}{\Delta} - p_t$. This yields that
        \begin{oldequation}
            {\sum\nolimits_{t=1}^T Y_t}/{(\Delta\sum\nolimits_{t=1}^T p_t)} - 1 = {X_T}/{\sum\nolimits_{t=1}^T p_t}. 
        \end{oldequation}
        The increasing process of $X_t$, say $A_t$, is defined as $A_t = \sum_{t_1 = 1}^t \expec[(X_{t_1} - X_{t_1-1})^2|\mathcal{F}_{t_1-1}]$. Then, we have
        \begin{oldequation}
            A_{t} - A_{t-1} = \expec[(X_{t} - X_{t-1})^2|\mathcal{F}_{t-1}] =  \Delta^{-2}{\expec[(Y_t - \Delta p_t)^2|\mathcal{F}_{t-1}]} = \Delta^{-1}p_t(1-p_t) \leq p_t/\Delta.
        \end{oldequation}
        Given $\{A_{\infty} < \infty\}$, Theorem 4.5.2 in~\cite{durrett2019probability_ec} shows that $X_t$ converges to a constant. Hence, we have $\frac{X_T}{\sum_{t=1}^T p_t} \rightarrow 0$ conditional on $\{A_{\infty} < \infty, \sum_{t=1}^\infty p_t = \infty\}$. If $A_{\infty} = \infty$, we have $\{A_{\infty} = \infty\} \subseteq \{\sum_{t=1}^{\infty} p_t = \infty\}$. Applying Theorem 4.5.3 in~\cite{durrett2019probability_ec} with $f(t) = \max(t, 1)$, one can confirm that $X_T/f(A_T) \rightarrow 0$ conditional on $\{A_\infty = \infty\}$. Since $f(A_T) \leq \frac{\sum_{t=1}^T p_t}{\Delta}$, we finally have $X_T/\sum_{t=1}^T p_t \rightarrow 0$ conditional on $\{A_\infty = \infty\} = \{A_{\infty} = \infty, \sum_{t=1}^\infty p_t = \infty\}$. Combining two limit results verifies $\frac{X_T}{\sum_{t=1}^T p_t} \rightarrow 0$ given $\sum_{t=1}^\infty p_t = \infty$, as desired. \qed
    \endproof

    \begin{lemma}\label{lem:det.block}
        Let $A$ be a $B$-by-$B$ invertible matrix. Define $A_1 = \left[\begin{array}{ c | c } A & \BFu \\ \hline \BFv^\top & c \end{array}\right]$ for some $B$-dimensional vectors $\BFu$ and $\BFv$, and constant $c$. Then, we have $\text{det}(A_1) = (c - \BFv^\top A^{-1}\BFu)\text{det}(A)$. 
    \end{lemma}

    \proof{\underline{Proof.}} From a straightforward computation, one can confirm
    \begin{oldequation}
        A_1 = \left[\begin{array}{ c | c } I_B & \mathbf{0}_{B} \\ \hline \BFv^\top A^{-1} & 1 \end{array}\right] \left[\begin{array}{ c | c } A & \mathbf{0}_{B} \\ \hline \mathbf{0}_{B}^\top & c - \BFv^\top A^{-1}\BFu \end{array}\right] \left[\begin{array}{ c | c } I_B & A^{-1}\BFu \\ \hline 0 & 1 \end{array}\right],
    \end{oldequation}
    where $I_B$ is a $B$-by-$B$ identity matrix. Taking a determinant on both sides, we can obtain $\text{det}(A_1) = (c - \BFv^\top A^{-1}\BFu)\text{det}(A)$. \qed
    \endproof

    \begin{lemma}\label{lem:strong.consistency.normal}
        Let $\{Y_n\}_{n \geq 1}$ be a sequence of normal random variables such that $Y_n \sim N(\mu_n, \lambda_n^2)$ for each $n$. If $\limsup_{n \rightarrow \infty}n\lambda^2_n < \infty$ and $\lim_{n \rightarrow \infty}\mu_n = \mu_0$ hold for some $\lambda_0 > 0$ and $\mu_0$, then we have $Y_n \xrightarrow{a.s.} \mu_0$.  
    \end{lemma}

    \proof{\underline{Proof of Lemma~\ref{lem:strong.consistency.normal}.}} Let $Y_n = \mu_n + \epsilon_n$, where $\epsilon_n \sim N(0, \lambda_n^2)$. Since $\lim_{n\rightarrow \infty} \mu_n = \mu_0$, it is sufficient to show $\lim_{n\rightarrow \infty} \epsilon_n = 0$ almost surely. For any $\delta > 0$, we have 
    \begin{oldequation}\label{eq:temp_inf_sum}
        \sum\nolimits_{n=1}^{\infty}\prob\left(\big|\epsilon_n\big| > \delta\right) = 2\sum\nolimits_{n=1}^{\infty}\Phi(-{\delta}/{{\lambda}_n}),
    \end{oldequation}
    where $\Phi(\cdot)$ is a CDF function of standard normal distribution. Combining $\lim_{x\rightarrow \infty} x^4\Phi(-x) = 0$ with $\limsup_{n\rightarrow \infty} n\lambda^2_n < \infty$, one can see that there exists $C, C_1> 0$ such that $\Phi(-{\delta}/\lambda_n) \leq C\lambda_n^4 \leq C_1 n^{-2}$ for all sufficiently large $n$, which in turn implies that~\eqref{eq:temp_inf_sum} is finite. Hence, the Borel-Cantelli lemma yields
    \begin{oldequation}
        \prob\left(|\epsilon_n| > \delta, \text{ infinitely often} \right) = 0, \;\; \text{i.e.,} \;\; \prob\left(\exists N \;\; \text{s.t.} \;\; \big|\epsilon_n\big| \leq \delta, \forall n \geq N\right) = 1.
    \end{oldequation}
    Since $\delta > 0$ is arbitrary, letting $\delta \rightarrow 0$ proves $\prob\left(\lim_{n\rightarrow \infty} \epsilon_n = 0 \right) = 1$, as desired. \qed
    \endproof

\end{document}